\documentclass[%
 aip,
jcp,%
 amsmath,amssymb,
 reprint,%
 groupedaddress,
]{revtex4-1}
\usepackage{graphicx}% Include figure files
\usepackage{dcolumn}% Align table columns on decimal point
\usepackage{bm}% bold math
\usepackage{xcolor}
\usepackage{siunitx}
\usepackage{epstopdf}
\usepackage{sidecap}
\sidecaptionvpos{figure}{c}

\newcommand{\Tr}{\mathrm{Tr}}

\newcommand{\rd}{\mathrm{d}}

\newcommand{\eu}[1]{\mathrm{e}^{#1}}
\usepackage{soul} % provides st for strikethrough

\usepackage{braket}

\usepackage{dcolumn} % D
\newcolumntype{d}[1]{D{.}{.}{#1}} % taXble environment which aligns decimal places

\begin{document}

\title{A partially linearized spin-mapping approach for simulating nonlinear optical spectra}% Force line breaks with \\

\author{Jonathan R. Mannouch}
\email{jonathan.mannouch@phys.chem.ethz.ch}
\author{Jeremy O. Richardson}%
\email{jeremy.richardson@phys.chem.ethz.ch}
\affiliation{Laboratory of Physical Chemistry, ETH Z\"{u}rich, 8093 Z\"{u}rich, Switzerland}

\date{\today}% It is always \today, today,

\begin{abstract}
We present a partially linearized method based on spin mapping for computing both linear and nonlinear optical spectra. As observables are obtained from ensembles of classical trajectories, the approach can be applied to the large condensed-phase systems that undergo photosynthetic light-harvesting processes. In particular, the recently derived spin-PLDM method has been shown to exhibit superior accuracy in computing population dynamics compared to other related classical-trajectory methods. Such a method should also be ideally suited to describing the quantum coherences generated by interaction with light. We demonstrate that this is indeed the case by calculating the nonlinear optical response functions relevant for the  pump--probe and 2D photon-echo spectra for a Frenkel biexciton model and the Fenna--Matthews--Olsen light-harvesting complex. One especially desirable feature of our approach is that the full spectrum can be decomposed into its constituent components associated with the various Liouville-space pathways, offering a greater insight beyond what can be directly obtained from experiment.  
\end{abstract}

\maketitle

\section{\label{sec:intro}Introduction}
Nonlinear optical spectroscopy is a powerful tool for elucidating the exciton dynamics of condensed-phase systems.\cite{MukamelBook} While linear spectroscopic techniques are essentially limited to determining the exciton energies and dephasing rates, nonlinear methods give access to a plethora of extra information.\cite{Brixner2005,Schlaucohen2011} For example, 2D optical photon-echo spectroscopy has previously been used to measure exciton couplings,\cite{Brixner2005,Read2007,Cho2005,*Zigmantas2006} excitonic energy-transfer pathways and timescales,\cite{Brixner2005,Read2007,Cho2005,*Zigmantas2006,Schlaucohen2009,*Myers2010,Engel2007,*Collini2010} exciton coherences\cite{Lee2007,*Schlaucohen2012,*Ishizaki2012,*Ishizaki2012,Engel2007,*Collini2010} and dipole-moment orientations.\cite{Read2008,*Schlaucohen2010} By tuning the polarization of each of the pulses, the relative intensity of peaks within 2D photon-echo spectra can be altered, providing a way to disentangle and assign peaks within crowded regions of the spectra, which are often present for large systems.\cite{Read2007,Schlaucohen2012b} Nonlinear approaches are also useful for removing the effects of inhomogeneous broadening.\cite{MukamelBook} Because of these advantages, 2D optical photon-echo spectroscopy has become the experimental tool of choice for investigating the excitonic energy-transfer processes within photosynthetic light-harvesting systems, where researchers are still trying to understand their impressive ability to harvest sunlight with near perfect quantum efficiency. For theory to help interpret and give further insight to such measurements, there is a need to develop approaches for simulating the processes and calculating spectroscopic quantities which can be applied to systems with a large number of degrees of freedom. 

A common way of computing such spectroscopic quantities is by using the optical response function approach.\cite{MukamelBook} Here the field--matter interaction is treated perturbatively, leading to a set of multi-time correlation functions which involve only the field-free dynamics. % (where we refer to the combination of all the correlation functions associated with a given order of the field--matter interaction as the corresponding optical response function).
Each spectroscopic observable is then typically obtained by various Fourier transforms of the associated optical response functions. Alternatively the so-called explicit light-field approach\cite{Gao2020_lin,Gao2020_nonlin,Provazza2021} can also be used to compute spectroscopic properties, by performing the dynamics associated with the coupling to a classical light field through the use of a time-dependent Hamiltonian. While this alternative approach has the advantage that it can describe processes involving strong light fields, the optical response function approach is nevertheless sufficient for and provides the simplest connection to many experiments. It also has the added advantage that the full signal can be separated into contributions associated with different Liouville-space pathways, offering greater insight beyond what can be directly obtained from experiment. Many methods have been developed to compute linear and nonlinear spectra within this framework.\cite{Fetherolf2017,Tempelaar2013,*Vegte2013,*Jain2019,*Huang2021,Hanna2008,*Hanna2009,*Hanna2011,Bai2014} One of the most commonly used techniques for simulating nonlinear spectra is the Hierarchical equations of motion (HEOM) approach.\cite{Hein2012,Kramer2017,Kramer2018,Tanimura2020HEOM} Although this method enables numerically exact quantum dynamics to be applied to large condensed-phase systems containing a harmonic bath, it is unable to treat more realistic anharmonic problems. A full quantum mechanical approach is also in some sense unnecessary for treating the problems we consider in this paper, as we expect that it should be possible to simulate most relevant biological processes without taking the quantum-mechanical nature of nuclear dynamics into account. We are therefore primarily interested in classical-trajectory techniques, which can in principle be applied to systems with any form for the nuclear potential.

When modelling spectroscopic quantities involving electronic transitions between the ground and a single excited state, there are several different classical-trajectory approximations that can be used; these differ by whether the nuclear degrees of freedom are evolved on the ground-state potential, excited-state potential or some linear combination of both.\cite{Egorov1998vibronic,*Rabani1998vibronic,*Shi2008nonlinear,*Karsten2018vibronic,*McRobbie2009nonlinear,MukamelBook} All such approaches are exact for shifted harmonic models where the ground and excited state have the same frequencies, but give different approximations in general. From previous work, it has been found that the Wigner averaged classical limit (WACL),\cite{Egorov1999,*Shi2004goldenrule,*Shi2005nonadiabatic} which evolves the nuclear degrees of freedom on the time-independent arithmetic mean surface of the ground and a single excited state, generally gives rise to the most accurate results. The forward-backward initial-value representation (FB-IVR),\cite{Makri1998,*Sun1999} which is related to the Herman--Kluk propagator,\cite{Herman+Kluk1984} can additionally exactly reproduce spectra for systems that involve a frequency change under photoexcitation, but this approach cannot easily be converged for high-dimensional condensed-phase problems and the Herman--Kluk prefactor becomes unstable when applied to anharmonic models.\cite{Walton1995} 

While WACL has been relatively successful in accurately obtaining Frank--Condon spectra for certain condensed-phase systems, it cannot describe nonadiabatic transitions between coupled exciton states and so is not an appropriate method for photosynthetic light-harvesting systems. Nonetheless, it is desirable that a nonadiabatic dynamics method reduces to WACL in the case that the excited states are uncoupled. Other features that we would want from such an approach is that it is exact in the static-nuclear limit (when inhomogeneous broadening dominates), is accurate in the high-temperature limit (where classical trajectories are expected to be valid) and can be used to compute not just single-time correlation functions, but also the multi-time correlation functions needed for obtaining nonlinear spectra.

Mapping-based approaches provide a way to accurately include nonadiabatic transitions within a classical trajectory picture. The combined dynamics of the nuclei and excitons are treated as an ensemble of classical trajectories, where the dynamics of the later is described by evolving a set of continuous variables associated with a mapping space. The computational cost of such methods therefore scales with the number of degrees of freedom in the same way as standard classical molecular dynamics, allowing them to be applied to large condensed-phase systems. Originally, most mapping-based approaches employed the so-called Meyer--Miller--Stock--Thoss (MMST) mapping,\cite{Meyer1979nonadiabatic,*Stock1997mapping} where the exciton subsystem is described by the single-excitation space of a set of harmonic oscillators and the dynamics are obtained by evolving the classical phase-space variables associated with this mapping space. Recently, progress has been made in improving the accuracy of such methods through the use of a resolution of the identity,\cite{identity,FMO,linearized} optimized zero-point energy parameters,\cite{Mueller1999pyrazine,Gao2020mapping} the generalized master equation,\cite{Shi2004GQME,*Kelly2016master,*Pfalzgraff2019GQME,*Mulvihill2019LSCGQME}
nonadiabatic ring-polymer molecular dynamics, \cite{mapping,*vibronic,*Ananth2013MVRPMD,*Chowdhury2017CSRPMD}
symmetric windowing (SQC),\cite{Cotton2013SQC,*Cotton2013mapping,*Miller2016Faraday,*Cotton2016_2,*Cotton2016SQC,*Liang2018,*Cotton2019} spin-mapping,\cite{spinmap,multispin} and other alternative classical mapping models.\cite{Liu2016,*xin2019,Kim2014mapping,Miller1986fermions,*Li2012fermions,*Sun2021,Lang2021GDTWA} Some of these advancements have also already been used to obtain both linear and nonlinear optical spectra.\cite{Gao2020_lin,Polley2019,Polley2020vibronic,Polley2021,Braver2021} 
A particularly successful method for computing dynamical observables within exciton systems is the standard partially linearized density matrix (PLDM)\cite{Huo2010,*Huo2011densitymatrix,*Huo2012MolPhys,*Huo2015PLDM,*Huo2012PLDM,*Huo2012_2,*Lee2016,*Castellanos2017,*Mandal2019,*Hsieh2012FBTS,*Hsieh2013FBTS,*Kelly2020,Mannouch2020a,Mannouch2020b,Sun1997} approach, which uses coherent states within the MMST mapping space to describe the dynamics associated with the forward and backward exciton paths separately through the use of two independent sets of mapping variables. Additionally, a procedure for the application of this method to the computation of linear and nonlinear optical spectra has already been developed, which has been used successfully for a range of systems.\cite{Provazza2018_lin,Provazza2018_nonlin} Such an approach seems especially attractive, as it fulfills many of the desirable features that we would like; the approach is exact in the static-nuclear limit, can compute both single- and multi-time correlation functions directly\cite{Provazza2018_lin,Provazza2018_nonlin} and for relatively short propagation times is seen to be extremely accurate in the high-temperature limit.\cite{Mannouch2020a} We note that the approach also reduces to WACL in the absence of diabatic couplings if so-called `focused' initial conditions are used. Despite the many positives, there are however certain aspects of the method that could be improved. 

For standard PLDM, the initial mapping variables can be sampled using either focused initial conditions or a Gaussian sampling approach, which are both defined in Ref.~\onlinecite{Hsieh2013FBTS}. It is in general not obvious which sampling approach is superior. Focused conditions, which limit the initial sampling space of the mapping variables to a region corresponding to the occupation of a single initial excitonic state, have typically been used when computing nonlinear spectra for reasons of improved efficiency.\cite{Provazza2018_nonlin} However it is also known that the accuracy of population dynamics can suffer dramatically by using focused initial conditions instead of a Gaussian sampling approach for the mapping variables.\cite{Huo2012PLDM,Hsieh2013FBTS} Additionally, it has been observed that even in the high-temperature limit, standard PLDM becomes inaccurate for long propagation times even when using Gaussian sampling.\cite{Mannouch2020a} Such an approach could therefore lead to significant errors in the nonlinear spectra when considering long delay times between the pump and probe pulses. One potential way to alleviate some of these problems is by using spin-PLDM,\cite{Mannouch2020a,Mannouch2020b} which is an extension of the standard method that instead utilizes the spin-mapping space reformulated in terms of Stratonovich--Weyl kernels.\cite{spinmap,multispin} For spin-mapping, focused initial conditions are just as rigorous in the sense that this sampling of the mapping variables leads to the product of any two electronic operators still being correctly described by the mapping,\cite{Mannouch2020b} while for focused conditions in the MMST mapping space this is only true for products of population operators. As a result, there is no reason to expect a loss in accuracy when focused initial conditions are used with spin-mapping approaches, as has been observed, and using such initial conditions still gives similar improvements in efficiency as for standard PLDM\@.\cite{spinmap,multispin,Mannouch2020a} Spin-PLDM has also been shown to exhibit greater accuracy than standard PLDM for a range of different problems and in particular it is found to be numerically exact in the high-temperature limit, even for long propagation times.\cite{Mannouch2020a} For these reasons, spin-PLDM is expected to offer a number of key advantages for calculating nonlinear spectra over the standard method and is the only approach of those outlined above that is able to fulfill all of the desirable features that we listed earlier.

After first reviewing how linear optical spectra can be calculated with fully linearized semiclassical (LSC) mapping methods, we introduce our approach for calculating both linear and nonlinear spectra with spin-PLDM\@. The improved accuracy of such an approach is then demonstrated by comparing the optical absorption, fluorescence, pump--probe and two-dimensional (2D) photon-echo spectra for a range of model systems calculated with both spin-PLDM and other commonly used methods. In the literature, a particular Frenkel biexciton model is often used to benchmark approximate methods for computing nonlinear optical spectra. However, we will show that a static-nuclear picture often suffices to obtain reasonably accurate results for this system, and so in this paper we will also test the methods on more difficult parameter regimes. In addition, we compute pump--probe spectra for the seven-state Fenna--Matthews--Olsen complex (FMO) model, which as well as being relevant for studying photosynthetic light-harvesting processes, is also a much more challenging benchmark system. This is in particular because in this model, there are 21 double-exciton states needed for computing nonlinear spectra, whereas biexciton models have only one. To the authors' knowledge, no mapping-based approach has previously tackled the nonlinear spectra of such a large system.

\section{\label{sec:theory}Background Theory}
We consider a molecular system consisting of a single electronic ground state and a manifold of coupled excited states. The Hamiltonian for such a system can be written as:\cite{BardfordBook,KuehnBook}
\begin{equation}
\label{eq:ham}
\hat{H}=\tfrac{1}{2}p^{2}+\hat{V}_{\text{g}}(x)+\hat{V}_{\text{e}}(x)+\hat{V}_{\text{ee}}(x)+\cdots ,
\end{equation}
where both the nuclear configuration $x$ and momentum $p$ are vectors which have been mass weighted such that all degrees of freedom have unit mass.
The ground-state potential operator is $\hat{V}_{\text{g}}(x)=\ket{0}V_{\text{g}}(x)\bra{0}$.%, where $\ket{0}$ is the electronic ground state and $V_g(x)$ is the potential function of this state.
\footnote{In principle, the approach outlined in this paper can also treat systems containing a manifold of ground states, where $\hat{V}_{\text{g}}(x)$ simply becomes a matrix within this subspace.} The remaining potential operators describe all possible excited electronic states associated with a $F$ chromophore system and are partitioned into their constituent subspaces according to the number of excitons present: $\hat{V}_{\text{e}}(x)$ is the $F\times F$ potential operator associated with the single-exciton subspace, $\ket{1}$,\dots,$\ket{F}$, while $\hat{V}_{\text{ee}}(x)$ encodes the contributions from the double-exciton subspace, $\ket{1,2}$,$\ket{1,3}$,\dots,$\ket{F-1,F}$, of dimension $\tfrac{1}{2}F(F-1)$. We set $\hbar=1$ throughout. 

All electronic couplings between the distinct exciton subspaces are neglected such that transitions are only induced when the system interacts with electromagnetic radiation. This field--matter interaction is accounted for by the dipole-moment operator, $\hat{\mu}=\hat{\mu}^{+}+\hat{\mu}^{-}$, which couples subspaces whose exciton numbers differ by one: 
\begin{equation}
\label{eq:dipole_op}
\hat{\mu}^{+}=\sum_{n=1}^{F}\mu_{n}\hat{a}^{\dagger}_{n} ,\qquad\hat{\mu}^{-}=\sum_{n=1}^{F}\mu_{n}\hat{a}_{n} ,
\end{equation}
where $\hat{a}^{\dagger}_{n}$ ($\hat{a}_{n}$) creates (annihilates) an exciton on chromophore $n$ and $\mu_{n}$ is the transition dipole moment associated with this chromophore. 

Spectroscopic quantities of interest can then be calculated from Fourier transforms of the time-dependent polarization, which is induced in the medium by the applied radiation field. One common way to proceed is to perturbatively expand the time-dependent polarization in the field--matter interaction, which leads to the definition of the so-called optical response functions, where $S^{(m)}(t_{1},t_{2},\cdots,t_{m})$, is the the $m$th-order term in the expansion.\cite{MukamelBook} These optical response functions depend solely on the initial density matrix of the system, the dipole-moment operators [Eq.~(\ref{eq:dipole_op})] and the field-free dynamics associated with the Hamiltonian given in Eq.~(\ref{eq:ham}). We begin by considering the linear response in Sec.~\ref{sec:linear} before turning to the nonlinear case in Sec.~\ref{sec:nonlinear}.
\section{Linear Optical Spectroscopy} \label{sec:linear}
The linear optical response can be described though the single-time correlation functions:
\begin{subequations}
\label{eq:dipole_func}
\begin{align}
J_{\text{abs}}(t)&=\Tr\left[\eu{i\hat{H}_{\text{g}}t}\hat{\mu}^{-}\eu{-i\hat{H}_{\text{e}}t}\hat{\mu}^{+}\hat{\rho}_{\text{g}}\right] \label{eq:dipole_abs} , \\
J_{\text{fluor}}(t)&=\Tr\left[\eu{i\hat{H}_{\text{g}}t}\hat{\mu}^{-}\eu{-i\hat{H}_{\text{e}}t}\hat{\rho}_{\text{e}}\,\hat{\mu}^{+}\right] \label{eq:dipole_fluor} ,
\end{align}
\end{subequations}
where the key difference between these two functions is the initial state of the system. $J_{\text{abs}}(t)$ gives rise to the linear absorption spectrum, where the system is initialized in a thermal distribution associated with the electronic ground state, with $\hat{\rho}_{\text{g}}=\eu{-\beta\hat{H}_{\text{g}}}$ and $\hat{H}_{\text{g}}=\ket{0}\tfrac{1}{2}p^{2}\bra{0}+\hat{V}_{\text{g}}(x)$. $J_{\text{fluor}}(t)$ gives rise to the fluorescence spectrum, where the system is initialized in the thermal state associated with the single-exciton subspace, with $\hat{\rho}_{\text{e}}=\eu{-\beta\hat{H}_{\text{e}}}$ and $\hat{H}_{\text{e}}=\sum_{\lambda_{\text{e}}=1}^F\ket{\lambda_{\text{e}}}\tfrac{1}{2}p^{2}\bra{\lambda_{\text{e}}}+\hat{V}_{\text{e}}(x)$. Both of these correlation functions contain a forward ($\eu{-i\hat{H}_\text{e}t}$) and a backward ($\eu{i\hat{H}_\text{g}t}$) propagator, %. Due to the absence of couplings between the various exciton subspaces in the system Hamiltonian [Eq.~(\ref{eq:ham})], the forward (backward) propagator
which correspond to dynamics purely within the single-exciton subspace and electronic ground state respectively.

The linear absorption or fluorescence spectra are then obtained as follows: 
\begin{subequations}
\begin{align}
I(\omega)&=\frac{1}{2\pi}\text{Im}\left[\int_{0}^{\infty}\rd t\,\frac{S^{(1)}(t)}{J(0)}\,\eu{i\omega t}\right] \label{eq:spec_lin} , \\
S^{(1)}(t)&=-2\theta(t)\text{Im}[J(t)],
\end{align}
\end{subequations}
where $\theta(t)$ is the Heaviside step function and $J(t)$ is the corresponding correlation function from one of Eqs.~\ref{eq:dipole_func}. The additional constants included in Eq.~(\ref{eq:spec_lin}) %ensure that the positive frequency part of the linear spectrum is normalized % such that $\int_{0}^{\infty}\rd\omega \, I(\omega)=1$,
%for systems where all the eigenvalues of $\hat{H}_{\text{e}}$ are larger than the thermally accessible states of $\hat{H}_{\text{g}}$.
%(assuming that the excited states are sufficiency higher in energy than the ground state).
normalize the spectrum.

Quasi-classical expressions for single-time correlation functions such as Eqs.~\eqref{eq:dipole_func} can be easily obtained within both fully and partially linearized mapping-based techniques. These offer a practical way of calculating linear spectra for large condensed-phase systems based on phase-space averages over ensembles of classical trajectories.  We will now briefly review these two approaches.
\subsection{Fully Linearized Mapping Methods}\label{sec:linear_fully}
Single-time correlation functions are expressed in a fully linearized mapping approach by representing the initial and final exciton state using the same single set of Cartesian mapping variables.
We represent these as $\mathcal{Z}=\{Z_{0},\cdots,Z_{F}\}$, a set of $N=F+1$ complex numbers whose real and imaginary parts correspond to the usual MMST mapping variables. %(e.g. $N=F$ for the single-exciton space or $N=F+1$ for the combined ground and single-exciton spaces).
Thus the correlation functions given in Eqs.~(\ref{eq:dipole_func}), are approximated by the phase-space averages:
\begin{subequations}
\label{eq:ful_lin}
\begin{align}
J_{\text{abs}}(t)&\approx\Big\langle\mu^{-}(\mathcal{Z}(t))\,\mu^{+}(\mathcal{Z})\rho_{\text{g}}(x,p)\Big\rangle \label{eq:abs_ful_lin}, \\
J_{\text{fluor}}(t)&\approx\Bigg\langle\mu^{-}(\mathcal{Z}(t))\,\rho_{\text{coh}}(x,p,\mathcal{Z})\Bigg\rangle \label{eq:fluor_ful_lin},
\end{align}
\end{subequations}
where $\mu^{+}(\mathcal{Z})=\tfrac{1}{2}\sum_{\lambda_{\text{e}}}Z^{*}_{\lambda_{\text{e}}}\mu_{\lambda_{\text{e}}}Z_{0}$ is the mapping representation of the dipole operator $\hat{\mu}^{+}$, $\ket{0}$ is the electronic ground state and $\ket{\lambda_{\text{e}}}$ is one of the single-exciton states. The dipole operator, $\hat{\mu}^-$, at time $t$ is then represented by the time-evolved mapping variables, $\mu^-(\mathcal{Z}(t))=\mu^+(\mathcal{Z}(t))^*$. %For the fluorescence correlation function [Eq.~(\ref{eq:dipole_fluor})], there are two electronic operators at time $t=0$ ($\hat{\mu}^{+}$ and $\hat{\rho}_{\text{e}}$) which should be combined using quantum-mechanical rules (i.e., $\hat{\rho}_{\text{coh}}(x,p)=\hat{\rho}_{\text{e}}(x,p)\hat{\mu}^{+}$) before they are represented together by mapping variables in the corresponding fully linearized expression [Eq.~(\ref{eq:fluor_ful_lin})].
For the fluorescence correlation function [Eq.~(\ref{eq:dipole_fluor})], there are two electronic operators at time $t=0$ ($\hat{\mu}^{+}$ and $\hat{\rho}_{\text{e}}$), which are represented together by mapping variables in the fully linearized expression [Eq.~(\ref{eq:fluor_ful_lin})] as $\rho_{\text{coh}}(x,p,\mathcal{Z})=\tfrac{1}{2}\sum_{\lambda_{\text{e}}\lambda_{\text{e}}'}Z^{*}_{\lambda_{\text{e}}}\braket{\lambda_{\text{e}}|\hat{\rho}_{\text{e}}(x,p)|\lambda_{\text{e}}'}\mu_{\lambda_{\text{e}}'}Z_{0}$. The excitonic dynamics must be performed in the combined space of $N=F+1$ components, $\mathcal{Z}$, because the correlation functions are represented by a single set of mapping variables within the fully linearized approach. Therefore the dynamics associated with the forward and backward propagators, which occur in the single-exciton subspace and ground state respectively, cannot be treated separately. Additionally, the nuclear operators are described in terms of their Wigner transform. Note that, at each point in the nuclear phase-space, $\hat{\rho}_{\text{e}}(x,p)$ is an $F\times F$ matrix in the single-exciton subspace, while $\rho_{\text{g}}(x,p)$ is a scalar function.

In Eq.~(\ref{eq:ful_lin}), the phase-space average over the initial values for the Cartesian mapping variables, $\mathcal{Z}$, and the nuclear coordinates, $x$ and $p$, is given by:
\begin{equation}
\label{eq:av}
\Braket{\cdots}=\int\rd x\,\rd p\,\rd\mathcal{Z}\cdots\rho_{\text{m}}(\mathcal{Z}) ,
\end{equation}
where $\rho_{\text{m}}(\mathcal{Z})$ is the mapping-variable distribution from which these variables are initially sampled. As outlined above, this fully linearized mapping-based approach for calculating optical absorption spectra has been also used in previous work.\cite{Gao2020_lin} 

In this paper, the fully linearized method we consider is spin-LSC,\cite{multispin} where the excitonic degrees of freedom are represented using spin-mapping on the W-sphere. It has been already shown that such a method is able to reproduce single-time correlation functions to a high accuracy.\cite{spinmap} This is because spin-mapping, unlike for MMST mapping, exactly treats the identity operator within the underlying theory, leading to superior accuracy in calculating identity containing correlation functions (similar to the benefits seen in Refs.~\onlinecite{identity,*FMO,Gao2020mapping,linearized,NRPMDChapter}). Additionally, all configurations in the spin-mapping space correspond to a valid state in the real exciton space, alleviating the problems with MMST mapping associated with classical trajectories being able to `leak' out of the physical mapping subspace.\cite{Stock2005nonadiabatic} For spin-LSC, the mapping-variable distribution corresponds to sampling the Cartesian mapping variables uniformly from the surface of a hypersphere of radius $R_{\text{W}}^{2}=2\sqrt{N+1}$, i.e., $\rho_{\text{m}}(\mathcal{Z})\propto\delta\left(\sum_{\lambda=0}^{F}|Z_{\lambda}|^{2}-R_{\text{W}}^{2}\right)$, which is referred to as full-sphere sampling. The equations of motion for the mapping and nuclear phase-space variables then take the standard form, equivalent to those used in MMST mapping:
\begin{equation}
\dot{x}=p,\qquad\dot{p}=\mathcal{F}(x,\mathcal{Z}),\qquad\dot{Z}_{\lambda}=\sum_{\lambda'=0}^{F}\braket{\lambda|\hat{V}(x)|\lambda'}Z_{\lambda'} ,
\end{equation}
where $\hat{V}(x)=\hat{V}_{\text{g}}(x)+\hat{V}_{\text{e}}(x)$ is the potential matrix associated with the combined ground state and single-exciton subspace (i.e., a direct product) and $\mathcal{F}(x,\mathcal{Z})=-\tfrac{1}{2}\sum^{F}_{\lambda\lambda'=0}\braket{\lambda'|\nabla \hat{V}(x)|\lambda}\left(Z^{*}_{\lambda'}Z_{\lambda}-\gamma\delta_{\lambda'\lambda}\right)$ is the mapping-variable representation of the force operator. Additionally, $\gamma=(R_{\text{W}}^{2}-2)/N$ is the generalized spin-mapping zero-point energy parameter, as derived in Ref.~\onlinecite{multispin}.

A key disadvantage of this fully linearized approach for calculating linear spectra is that for harmonic systems, the method is still not exact even when there are no diabatic couplings between the chromophores. However, in this case, even the much simpler WACL,\cite{Egorov1999,*Shi2004goldenrule,*Shi2005nonadiabatic} for which the nuclear force is associated with the time-independent arithmetic mean of the ground and a single excited state surface (i.e., $\mathcal{F}=-\tfrac{1}{2}(\nabla V_{\text{g}}(x)+\braket{\lambda_{\text{e}}|\nabla\hat{V}_{\text{e}}(x)|\lambda_{\text{e}}})$, where $\ket{\lambda_{\text{e}}}$ is one of the states in the single-exciton subspace), can exactly reproduce linear spectra. We would therefore like to design mapping-based methods which reduce to WACL in this special case. One way of fixing this problem is by `quantizing' the mapping variables in terms of action-angle variables. This approach is employed by the optimized mean-trajectory (OMT) method,\cite{Polley2019,Polley2020vibronic,Polley2021} which initializes half populations in both the ground and one of the excited states. % is exact in the absence of diabatic couplings, because initially setting the classical action variables to $n_{\text{g}}=n_{\lambda_{\text{e}}}=\tfrac{1}{2}$ correctly reproduces the same nuclear force as  WACL in this case.
An alternative is provided by the partially linearized approach, which can also be made to be exact in this limit %by treating the exciton dynamics associated with the forward and backward propagators using different sets of mapping variables. The exciton dynamics associated with the ground and single-exciton subspaces can thus be treated separately,
as we describe in the following.
\subsection{Partially Linearized Mapping Methods}\label{sec:linear_partial}
The main difference between fully and partially linearized approaches is that in the latter,
%Single-time correlation functions are expressed in a partially linearized approach by representing
the electronic forward and backward propagators are represented using separate sets of mapping variables. The nuclear dynamics are however still described using a single set of nuclear variables. For the correlation functions given by Eqs.~(\ref{eq:dipole_func}) only the forward propagator corresponds to coupled exciton--nuclear dynamics within the single-exciton subspace, with the backward propagator involving only nuclear dynamics in the ground state. This means that only one set of mapping variables is actually needed to correctly describe the excitonic dynamics associated with these functions. Therefore expressions for the correlation functions within a partially linearized approach are given by:
\begin{subequations}
\label{eq:part_lin}
\begin{align}
J_{\text{abs}}(t)&\approx\Big\langle\braket{0|\hat{\mu}^{-}\hat{w}_{\text{e}}(\mathcal{Z}_{\text{e}},t)\hat{\mu}^{+}|0}\rho_{\text{g}}(x,p)\Big\rangle , \label{eq:abs_part_lin} \\
J_{\text{fluor}}(t)&\approx\Big\langle\braket{0|\hat{\mu}^{-}\hat{w}_{\text{e}}(\mathcal{Z}_{\text{e}},t)\hat{\rho}_{\text{e}}(x,p)\hat{\mu}^{+}|0}\Big\rangle , \label{eq:fluor_part_lin}
\end{align}
\end{subequations}
where $\mathcal{Z}_{\text{e}}=\{Z_{1},\cdots,Z_{F}\}$ is now a set of Cartesian mapping variables purely within the single-exciton subspace with $N=F$ components. The (partial) Wigner transform of the initial density matrix, $\rho_{\text{g}}(x,p)$ and $\hat{\rho}_{\text{e}}(x,p)$, and all phase-space averages over classical variables are defined equivalently as in the fully linearized approach, % by Eqs.~(\ref{eq:wigner}) and (\ref{eq:av}),
the key difference being that we use the smaller space, $\mathcal{Z}_\text{e}$. 

The time-evolved kernel $\hat{w}_{\text{e}}(\mathcal{Z}_{\text{e}},t)=\hat{U}_{\text{e}}(t)\hat{w}_{\text{e}}(\mathcal{Z}_{\text{e}})$ is a $F\times F$ matrix that is used to represent individual propagators associated with the single-exciton subspace and is evolved in time for each trajectory using the time-ordered propagator, $\hat{U}_{\text{e}}(t)$. At $t=0$, the kernel matrix elements are defined by $\braket{\lambda_{\text{e}}|\hat{w}_{\text{e}}(\mathcal{Z}_{\text{e}})|\lambda'_{\text{e}}}=\tfrac{1}{2}(Z_{\lambda_{\text{e}}}Z_{\lambda'_{\text{e}}}^{*}-\gamma\delta_{\lambda_{\text{e}}\lambda'_{\text{e}}})$, where $\gamma$ is the zero-point energy parameter associated with the mapping. In addition, the time-evolved propagator is defined as:
%\begin{subequations}
\begin{equation}
\label{eq:kernel_evolve}
\hat{U}_{\text{e}}(t)=\eu{-i\hat{V}_{\text{eg}}(x(t))\epsilon}\cdots\eu{-i\hat{V}_{\text{eg}}(x(2\epsilon))\epsilon}\,\eu{-i\hat{V}_{\text{eg}}(x(\epsilon))\epsilon}
, %\\ \label{eq:rel_pot}
%&\hat{V}_{\text{eg}}(x)=\hat{V}_{\text{e}}(x)-V_{\text{g}}(x) ,
\end{equation}
%\end{subequations}
where $\epsilon$ is the time-step, and the potential matrix $\hat{V}_{\text{eg}}(x)=\hat{V}_{\text{e}}(x)-V_{\text{g}}(x)$ corresponds to the potential matrix associated with the single-exciton subspace defined relative to the ground-state potential. Defined in this way, the contribution of the time-evolved kernel to the partially linearized expressions for the correlation functions [Eq.~(\ref{eq:part_lin})] also includes the $\eu{i\int_{0}^{t}\rd t_{1}V_{\text{g}}(x(t_{1}))}$ factor that arises from the quasi-classical expression for the backward ground-state propagator. For excited-state subspaces consisting of multiple excitons (required in Sec.~\ref{sec:nonlinear}), the associated potential matrices used for the exciton dynamics are also defined relative to the ground state potential such that this factor is removed when considering coherences which do not involve the ground state. The mapping variables are evolved under the standard equations of motion and the nuclear force is given as the average force associated with the forward and backward exciton paths:
\begin{subequations}
\begin{align}
\mathcal{F}(x,\mathcal{Z}_{\text{e}})&=-\tfrac{1}{2}\nabla\big(V_{\text{g}}(x)+V_{\text{e}}(\mathcal{Z}_{\text{e}},x)\big), \\
V_{\text{e}}(\mathcal{Z}_{\text{e}},x)&=\tfrac{1}{2}\sum_{\lambda_{\text{e}},\lambda'_{\text{e}}}\braket{\lambda_{\text{e}}|\hat{V}_{\text{e}}(x)|\lambda'_{\text{e}}}\left(Z^{*}_{\lambda_{\text{e}}}Z_{\lambda'_{\text{e}}}-\gamma\delta_{\lambda_{\text{e}}\lambda'_{\text{e}}}\right) ,
\end{align}
\end{subequations}
where $V_{\text{e}}(\mathcal{Z}_{\text{e}},x)$ is the mapping-variable representation of the potential matrix associated with the single-exciton subspace. One of the features that we would like from a nonadiabatic dynamics approach for computing linear optical spectra is that it reduces to WACL in the absence of diabatic couplings between the chromophores and hence is exact for harmonic systems in this limit. The fact that the nuclear force used in WACL is also the average force associated with the single excited state and the ground state suggests that the partially linearized approach for calculating linear spectra can be devised in such a way that it fulfills this requirement. The key advantage of the partially linearized approach over WACL is that it includes the effects of nonadiabatic transitions and exciton energy transfer in systems when the chromophores are coupled.

In order to guarantee that the partially linearized approach does reduce to WACL in this case, the initial values for the mapping variables must be sampled using focused initial conditions in the single-exciton basis $\ket{\lambda_{\text{e}}}$. This means that the mapping-variable distribution in Eq.~(\ref{eq:av}) is given by:
\begin{equation}
\rho_{\text{m}}(\mathcal{Z}_{\text{e}})\propto{\sum_{\lambda_{\text{e}}}\delta(|Z_{\lambda_{\text{e}}}|^{2}-\gamma-2)\prod_{\lambda'_{\text{e}}\neq\lambda_{\text{e}}}}{\delta(|Z_{\lambda'_{\text{e}}}|^{2}-\gamma)} .
\end{equation}
In this paper, we consider two partially linearized techniques both using focused initial conditions: the standard PLDM\cite{Huo2010,*Huo2011densitymatrix,*Huo2012MolPhys,*Huo2015PLDM,*Huo2012PLDM,*Huo2012_2,*Lee2016,*Castellanos2017,*Mandal2019,*Hsieh2012FBTS,*Hsieh2013FBTS,*Kelly2020} approach, which corresponds to $\gamma=0$ and the spin-PLDM\cite{Mannouch2020a,Mannouch2020b} approach, which uses the same W-sphere spin-mapping approach as spin-LSC and therefore corresponds to $\gamma=(R_{\text{W}}^{2}-2)/N$, where $R_{\text{W}}^{2}=2\sqrt{N+1}$ is the W-sphere radius.\cite{multispin} Note, however, that because the size of the exciton space for computing linear spectra is different for spin-LSC and spin-PLDM ($N=F+1$ and $N=F$ respectively), the two methods actually use different values of the spin-radius and zero-point energy parameter. The advantage of treating the electronic ground state and single exciton subspace separately in spin-PLDM is that the most appropriate zero-point energy parameter for each distinct subspace can be used, which is expected to lead to more accurate results.

For standard PLDM, Eq.~(\ref{eq:abs_part_lin}) gives identical results to a previously implemented partially linearized approach for calculating optical absorption spectra.\cite{Provazza2018_lin,Provazza2018_nonlin} However the implementation of our expression for the associated single-time correlation function has been simplified, so that only one set of mapping-variables is used for the simulation and the dimension of this set is reduced to the size of the single-exciton subspace (instead of considering the combined space of the electronic ground state and single-exciton subspace for both forward and backward paths as in the original approach). %Because our partially linearized approach treats the electronic ground state and single-exciton subspace separately unlike in the original approach, the correct zero-point energy parameter for each distinct subspace can again be used and it should therefore lead to more accurate results with spin-PLDM. 
Formally, this has no effect on the accuracy of the standard PLDM method because the focused initial conditions treat the double combined space identically to the separated spaces.
However, as we discussed above, it has important implications for spin-PLDM.

We showed in previous work that using focused initial conditions with spin-PLDM does not significantly affect the accuracy of obtained single-time correlation functions, but does mean that an order-of-magnitude fewer trajectories are needed to reach convergence of the mapping-variable integrals.\cite{Mannouch2020b} In contrast, using focused initial conditions with standard PLDM is known to rapidly degrade the quality of the results with increasing simulation time.\cite{Huo2012PLDM,Hsieh2013FBTS} Therefore one advantage of using spin-PLDM over standard PLDM is that with focused initial conditions, the method 
%both simultaneously reduces to WACL for systems without diabatic couplings and 
gives accurate results for long-time dynamics when couplings exist between the chromophores, while still retaining the connection to WACL. In our previous work, it has also been been observed for a range of model systems that spin-PLDM generally exhibits greater accuracy when calculating single-time correlation functions compared to standard PLDM, even when focused conditions are not implemented. In particular, even though standard PLDM is reasonably accurate at short times, spin-PLDM is observed to be even better.\cite{Mannouch2020a} Because the linear optical response functions generally decay rapidly to zero, spin-PLDM is hence ideally suited to accurately calculate such quantities. 
%Because spin-PLDM also allows the dynamics in the distinct exciton subspaces to be treated separately, unlike for fully linearized approaches, and hence reduces to WACL when applied to a system without diabatic couplings between the chromophores, spin-PLDM therefore appears to be the ultimate quasi-classical mapping-based approach for calculating optical spectra. 
We now apply these techniques to calculate linear spectra for excitonic condensed-phase systems.
\subsection{Results}\label{sec:linear_results}
We consider exciton systems consisting of $F$ chromophores, where each chromophore is linearly coupled to its own independent harmonic bath. The full Hamiltonian is given by: 
\begin{subequations}
\begin{align}
&\hat{H}=\hat{H}_{\text{S}}+\hat{H}_{\text{SB}}+H_{\text{B}} , \\
&\hat{H}_{\text{S}}=\sum_{n=1}^{F}(\omega_{\text{shift}}+\epsilon_{n})\hat{a}^{\dagger}_{n}\hat{a}_{n}+\sum_{n=1}^{F}\sum_{m=1}^{n-1}\Delta_{nm}(\hat{a}^{\dagger}_{n}\hat{a}_{m}+\hat{a}^{\dagger}_{m}\hat{a}_{n}) , \\
&\hat{H}_{\text{SB}}=-\sum_{n=1}^{F}\sum_{j=1}^{f}c_{j}x_{j,n}\hat{a}^{\dagger}_{n}\hat{a}_{n} , \\
&H_{\text{B}}=\sum_{n=1}^{F}\sum_{j=1}^{f}\left(\tfrac{1}{2}p^{2}_{j,n}+\tfrac{1}{2}\omega_{j}^{2}x^{2}_{j,n}\right) ,
\end{align}
\end{subequations}
where $f$ is the number of nuclear degrees of freedom associated with a single chromophore and $\omega_{\text{shift}}$ shifts the chromophore energies such that $\sum_{n=1}^{F}\epsilon_{n}=0$.\footnote{In many cases, we may choose a smaller value of $\omega_{\text{shift}}$, which reduces the oscillation frequency of the correlation functions but in practice is not expected to alter the shape of the spectrum.} From this expression, the Hamiltonian associated with each of the various exciton subspaces can then be obtained. First, the ground-state Hamiltonian corresponds to the bath Hamiltonian projected into the zero exciton subspace: $\hat{H}_{\text{g}}=\ket{0}H_{\text{B}}\bra{0}$. Additionally, the Hamiltonians associated with the other exciton subspaces then contain their corresponding projected bath Hamiltonian, along with an excitonic state matrix containing the associated matrix elements of $\hat{H}_{\text{S}}$ and $\hat{H}_{\text{SB}}$ between all of the basis states of the subspace.

For all the exciton models considered in this paper, the distribution of the nuclear frequencies within each of the baths and their couplings is determined by the Debye spectral density:
\begin{equation}
J_{\text{bath}}(\omega)=2\Lambda\frac{\omega\omega_{\text{c}}}{\omega^{2}+\omega_{\text{c}}^{2}} ,
\end{equation}
where $\Lambda$ is the reorganization energy and $\omega_{\text{c}}$ is the characteristic frequency of the bath. In order to have a finite number of nuclear degrees of freedom for our trajectory simulations, the continuous bath is discretized using the scheme employed in Ref.~\onlinecite{Craig2007condensed}.  

When calculating linear spectra of nonadiabatic systems, quantum master equations are widely used. Of these, Redfield theory\cite{KuehnBook} is the most common, which is derived from a perturbative expansion in the exciton--nuclear coupling and employs a Markovian approximation. %, which guarantees that the required properties of the reduced density matrix are preserved throughout the dynamics, such as being positive-definite.
In this paper, we always apply the secular approximation to Redfield theory, which has the advantage that the obtained dynamical populations are guaranteed to remain positive\cite{KuehnBook} and thus removes the possibility of obtaining unphysical negative peaks within calculated spectra.\cite{Kramer2018} A more sophisticated approach is the second-order time-convolutionless (TCL2) master equation,\cite{OpenQuantum,Berkelbach2012,*Berkelbach2012hybrid,Fetherolf2017} which while still perturbative in nature is based on the second-order cumulant approximation such that it %incorporates some non-Markovian effects and 
is by construction exact for systems with no diabatic couplings interacting with a harmonic bath. Such methods can be easily applied to excitonic systems and provide an interesting comparison with our partially linearized approach, which is the main focus of this paper.   

In order to calculate the linear spectra using mapping-based methods, the (partial) Wigner transform of the Boltzmann operator for the appropriate excitonic subspace must first be calculated. For a harmonic bath, the Wigner transform of the Boltzmann operator in the exciton ground state is given by:\cite{MukamelBook}
\begin{equation}
\rho_{\text{g}}(x,p)=\prod_{n=1}^F\prod_{j=1}^{f}\frac{\alpha_{j}}{\pi}\text{exp}\left[-\frac{2\alpha_{j}}{\omega_{j}}\left(\tfrac{1}{2}p_{j,n}^{2}+\tfrac{1}{2}\omega_{j}^{2}x_{j,n}^{2}\right)\right],
\end{equation}
where $\alpha_{j}=\text{tanh}(\tfrac{1}{2}\beta\omega_{j})$. Due to the presence of exciton--nuclear coupling, the Wigner transform of the Boltzmann operator in the single-exciton subspace is challenging to obtain exactly.\cite{Montoya2017} For simplicity, we approximate this quantity as follows:
\begin{equation}
\label{eq:wigner_approx} 
\frac{\hat{\rho}_{\text{e}}(x,p)}{\rho_{\text{g}}(x,p)}\approx\eu{-\tfrac{\beta}{2}\hat{H}_{\text{S}}}\,\eu{2\sum_{n=1}^{F}\sum_{j=1}^{f}\frac{\alpha_{j}}{\omega_{j}}c_{j}x_{j,n}\hat{a}^{\dagger}_{n}\hat{a}_{n}}\eu{-\tfrac{\beta}{2}\hat{H}_{\text{S}}}, 
\end{equation}
 where this expression is obtained by taking the Wigner transform of the following approximate Trotter splitting for the single-exciton Boltzmann operator: $\hat{\rho}_{\text{e}}\approx\eu{-\tfrac{\beta}{2}\hat{H}_{\text{S}}}\eu{-\beta\left(H_{\text{B}}+\hat{H}_{\text{SB}}\right)}\eu{-\tfrac{\beta}{2}\hat{H}_{\text{S}}}$ and is therefore accurate to second order in the diabatic couplings, $\Delta_{nm}$ and to first order in the exciton--nuclear coupling coefficients, $c_{j}$. In principle, we could go beyond this approximation, but it is found to be sufficient for the cases we have tested. For the quantum master equations, the Boltzmann operator in the single-exciton subspace is approximated using only the purely excitonic Hamiltonian (i.e., $\hat{\rho}_{\text{e}}\approx\eu{-\beta\hat{H}_{\text{S}}}$) in accordance with previous work.\cite{Kramer2018}  
 
 All the mapping-based methods considered here can exactly reproduce the time-dependent quantum dynamics of a bare electronic system. Also for a harmonic bath, our treatment of the bare nuclear dynamics by sampling the initial nuclear coordinates from the exact Wigner density and propagating them using classical equations of motion is also exact. Hence any errors appearing in the calculated optical spectra will solely arise due to the approximate description of the exciton--nuclear coupling. This therefore acts as a good test to compare the relative accuracies of calculating spectra with different mapping-based methods. %Note, however, that a key property of all methods presented in this paper is that they can in principle also be applied to systems with anharmonic potentials.
\subsubsection{The Frenkel Biexciton Model}
We first consider a previously studied Frenkel biexciton model,\cite{Fetherolf2017,Provazza2018_lin,Provazza2018_nonlin,Gao2020_lin,Gao2020_nonlin,Polley2020vibronic,Braver2021} which allows the accuracy of our partially linearized approach to be easily compared to other approaches. We use the same parameter sets considered in Ref.~\onlinecite{Gao2020_lin}, which offer a comprehensive test over the various regimes that the model encompasses, from low- to high-temperature and homogeneous to inhomogeneous broadening. As in previous work, the transition dipole moments associated with the two exciton sites are chosen to be antiparallel with $\mu_{1}=-5\mu_{2}$. Additionally, exact results can be obtained using the HEOM approach, which we calculated using the open source \texttt{pyrho} package.\cite{pyrho}

\begin{figure}
\includegraphics[scale=.55]{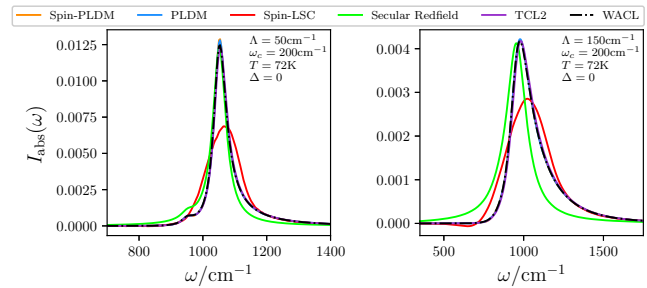}
\caption{The absorption spectra for two different Frenkel biexciton models without diabatic couplings, the parameters for which are given in the figures. Both models have $\epsilon_{1}-\epsilon_{2}=100\,\text{cm}^{-1}$ and $\omega_{\text{shift}}=1050\,\text{cm}^{-1}$. In this case, spin-PLDM, PLDM, TCL2 and WACL all give exact results.}\label{fig:abs-exciton-delta0}
\end{figure}
All results presented in this paper obtained with partially linearized approaches sample the mapping variables using focused initial conditions, for the reasons discussed above. This is illustrated in Fig.~\ref{fig:abs-exciton-delta0}, where the optical absorption spectrum for two low-temperature ($T=72\,\mathrm{K}$) Frenkel biexciton models with no diabatic coupling are reproduced exactly using partially linearized approaches either with the original MMST mapping (PLDM) or spin-mapping (spin-PLDM)\@. Because of the additional linearization approximation applied in the excitonic subspace, fully linearized approaches such as spin-LSC cannot be made exact in this case and exhibit large errors. Figure \ref{fig:abs-exciton-delta0} also illustrates the advantage of using the TCL2 approach over Redfield, as only the former is exact in this case. From a theoretical perspective therefore, partially linearized approaches and TCL2 are better engineered to correctly describe the dynamical coherences when calculating linear spectroscopic quantities of interest. 

\begin{figure*}
\includegraphics[scale=.61]{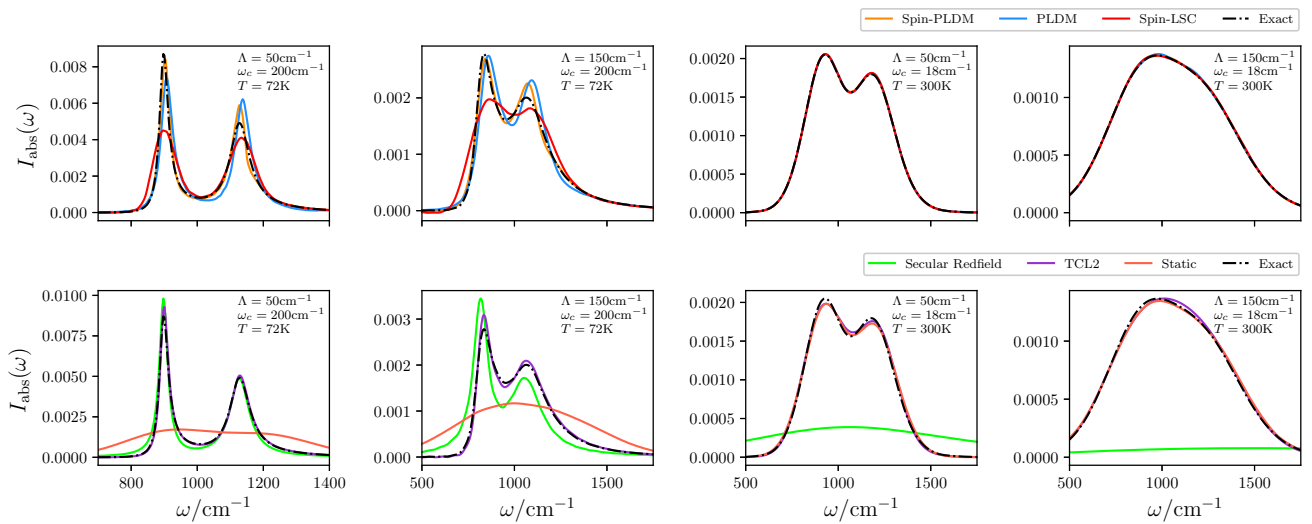}
\caption{The absorption spectra for four different Frenkel biexciton models, the parameters for which are given in the figures. All models have $\epsilon_{1}-\epsilon_{2}=100\,\text{cm}^{-1}$, $\omega_{\text{shift}}=1050\,\text{cm}^{-1}$ and $\Delta_{12}=100\,\text{cm}^{-1}$. Exact HEOM results\cite{pyrho} are given by the dashed black lines. These results may be compared with those calculated using Ehrenfest (Figs. 1 and 3 from Ref.~\onlinecite{Gao2020_lin}), LSC-IVR, PBME and traceless MMST (Figs.~2 and 4 from Ref.~\onlinecite{Gao2020_lin}), SQC (Fig.~1 from Ref.~\onlinecite{Provazza2018_lin}) and OMT (Fig.~3 from Ref.~\onlinecite{Polley2020vibronic}). Our Redfield (Figs.~1 and 3 from Ref.~\onlinecite{Gao2020_lin}), TCL2 (Figs.~2 and 3 from Ref.~\onlinecite{Fetherolf2017}), standard PLDM (Fig.~1 from Ref.~\onlinecite{Provazza2018_lin}) and HEOM (Figs.~2 and 3 from Ref.~\onlinecite{Fetherolf2017} and Figs.~1--4 from Ref.~\onlinecite{Gao2020_lin}) results are in agreement with those already published.}\label{fig:abs-exciton}
\end{figure*}
Even though neither partially linearized approaches nor TCL2 can exactly reproduce the nonadiabatic dynamics of exciton models with diabatic couplings, the advantage of both these methods is also demonstrated by the reasonable accuracy of the numerical results when calculating optical absorption spectra. First we consider the relatively slow-bath and high-temperature models (i.e., small $\omega_{\text{c}}$ and large $T$), where the corresponding absorption spectra calculated using a range of methods are given by the two right-hand columns of Fig.~\ref{fig:abs-exciton}. For both these models, all of the mapping-based methods can almost perfectly reproduce the absorption spectra, due to the following reasons. All such techniques are known to be able to reproduce the short-time dynamics essentially exactly in the high-temperature limit\cite{Mannouch2020a} and because all mapping-based methods exactly describe Rabi oscillations for an isolated excitonic system, they all by construction are able to correctly describe the static-nuclear limit, synonymous with a slow nuclear bath. The fact that a static-nuclear approximation (labelled `Static' in Fig.~\ref{fig:abs-exciton}) can also reproduce the spectrum extremely accurately, where the classical nuclear variables are still sampled from the Wigner distribution, $\rho_{\text{b}}(x,p)$, but are not evolved in time, confirms that the spectrum is almost entirely dominated by inhomogeneous broadening and hence these models do not correspond to a particularly challenging regime. Redfield theory, however, nonetheless completely fails in this case, as its Markovian approximation can only correctly describe the coupling to fast high-frequency modes.\cite{Thoss2001hybrid,Berkelbach2012,*Berkelbach2012hybrid,Fetherolf2017,Gao2020_lin} 

In contrast to the relatively slow-bath models, the static-nuclear approximation cannot reproduce the absorption spectra for the relatively fast-bath and low-temperature Frenkel biexciton models given by the two left-hand columns of Fig.~\ref{fig:abs-exciton}, confirming that homogeneous broadening dominates here and that these models constitute a more challenging regime for accurately obtaining absorption spectra. Of the mapping-based approaches tested, spin-LSC is seen to produce the largest error, in keeping with the fact that fully linearized approaches cannot correctly describe the dynamical coherences generated when measuring spectroscopic quantities and so cannot even reproduce the simple limit of the model without diabatic couplings, as discussed earlier in this section. Similar errors have also been observed for other fully linearized approaches when calculating absorption spectra for these same challenging fast-bath and low-temperature models.\cite{Gao2020_lin} TCL2 also outperforms Redfield for the same reason. Despite this, fully linearized methods are still expected to accurately reproduce linear spectra in the high-temperature and static-nuclear limits, where the associated dynamics are generally observed to be extremely accurate.\cite{spinmap,multispin,Mannouch2020a,linearized} Although both the partially linearized approaches and TCL2 are not able to exactly reproduce the absorption spectra for these relatively fast-bath models given in Fig.~\ref{fig:abs-exciton}, they are able to qualitatively reproduce the important features which is sufficient for most applications. Spin-PLDM is shown to be slightly more accurate than PLDM (although here the difference is not dramatic), which is in line with our previous observations that spin-PLDM  gives consistently more accurate short-time dynamics compared to other mapping-based approaches.\cite{Mannouch2020a}

\begin{figure*}
\includegraphics[scale=.61]{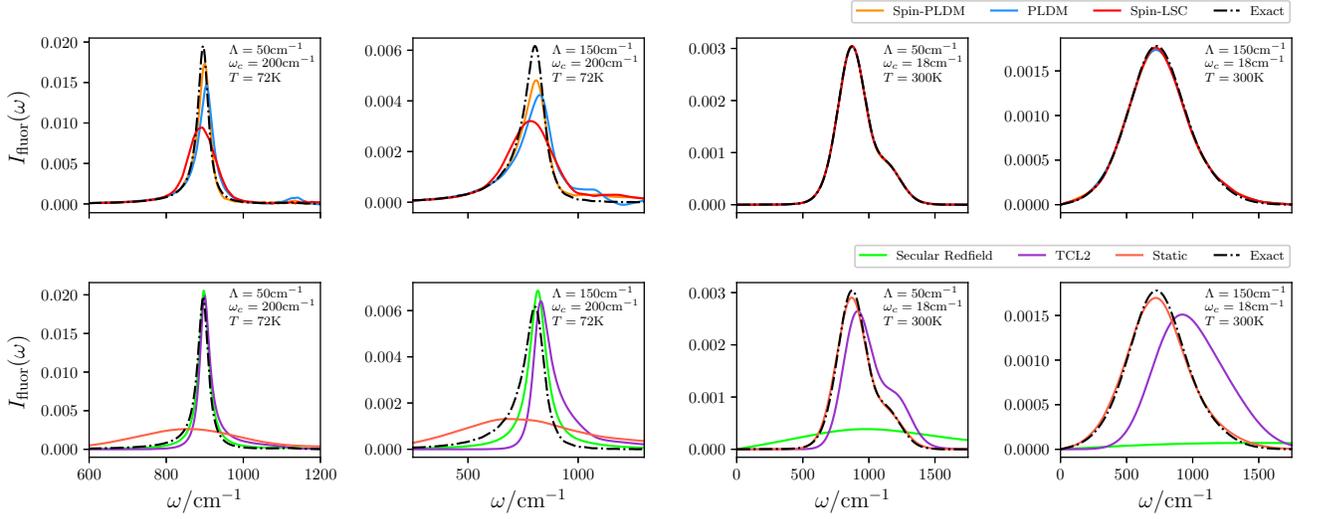}
\caption{The fluorescence spectra for four different Frenkel biexciton models, the parameters for which are given in the figures. All models have $\epsilon_{1}-\epsilon_{2}=100\,\text{cm}^{-1}$, $\omega_{\text{shift}}=1050\,\text{cm}^{-1}$ and $\Delta_{12}=100\,\text{cm}^{-1}$. Exact HEOM results\cite{pyrho} are given by the dashed black lines.}\label{fig:fluor-exciton}
\end{figure*}
Figure \ref{fig:fluor-exciton} gives the fluorescence spectra calculated for a range of different methods for the same Frenkel biexciton models. The quantum master equation approaches (i.e., Redfield and TCL2) are shown to produce much less accurate fluorescence spectra in comparison to the corresponding absorption spectra. This is because for quantum master equations the nuclear degrees of freedom have been integrated out, which means that the Wigner-transformed thermal Boltzmann operator associated with the single-exciton subspace, $\hat{\rho}_{\text{e}}(x,p)$, cannot be accurately described and at best can only be approximated using the purely excitonic part of the Hamiltonian, $\hat{H}_{\text{S}}$. In contrast, as for the absorption spectra, all of the mapping-based approaches can almost exactly reproduce the fluorescence spectra for the relatively slow-bath and high-temperature models (two right-hand columns of Fig.~\ref{fig:fluor-exciton}), because both the dynamics and the approximate expression used for the Wigner-transformed thermal Boltzmann operator associated with the single-exciton subspace [Eq.~(\ref{eq:wigner_approx})] are exact in the high-temperature and static-nuclear limit. The fact that the partially linearized approaches can also qualitatively reproduce the fluorescence spectra for the relatively fast-bath and low-temperature models (two left-hand columns of Fig.~\ref{fig:fluor-exciton}) suggests that Eq.~(\ref{eq:wigner_approx}) is also sufficient for calculating fluorescence spectra away from the high-temperature limit, at least in this case. Again, we observe that spin-PLDM is slightly more accurate at obtaining fluorescence spectra than PLDM.% suggesting that it is the method of choice for calculating linear spectra starting in the single-exciton subspace.  

Using the same methods and approach, the absorption, circular dichroism and fluorescence spectra have also been calculated for a seven-state FMO model\cite{Kramer2018,Gao2020_lin} at a range of temperatures, the results of which are given in the supplementary material. Our partially linearized approach is again able to accurately reproduce the linear spectra associated with this system. While the differing accuracies of the various methods are not so pronounced when applied to the FMO model compared to the Frenkel biexciton models studied above, similar conclusions can nevertheless be drawn.
\section{Nonlinear Optical Spectroscopy}\label{sec:nonlinear}
In an analogous fashion to the linear case, a vast array of nonlinear spectra can be obtained by applying various Fourier transforms to multi-time response functions.\cite{MukamelBook} %, $S^{(m)}(t_{1},t_{2},\cdots,t_{m})$.
In many systems the second-order response vanishes, and we will thus focus on the third-order response function, $S^{(3)}(t_1,t_2,t_3)$.
%It is possible to describe the vast array of distinct nonlinear optical spectra in terms of this function.
%Because numerous optical response functions can be constructed by considering different orders of the field--matter interaction, $m$, and different numbers of delay times over which a Fourier transform may or may not be applied, it is therefore possible to define a vast array of distinct nonlinear spectra, all of which can probe different dynamical processes within a system.\cite{MukamelBook}
This is the basis for 2D optical spectroscopy, which is particularly popular for investigating exciton relaxation, dephasing and transfer processes. Nevertheless the theory presented here can in principle be applied to calculate any nonlinear spectrum of any order through its associated optical response function.

\begin{SCfigure*}
\includegraphics[scale=.5]{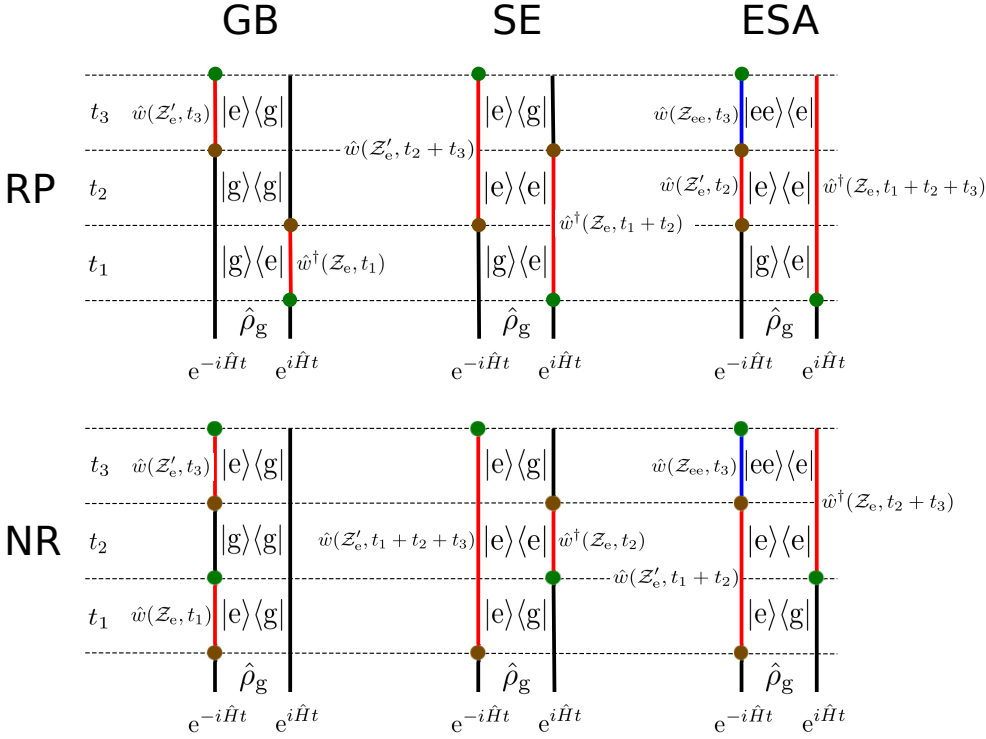}
\caption{The six double-sided Feynman diagrams used to represent the distinct contributions to the response function associated with the 2D optical spectrum. The coloured lines signify the occupied exciton subspace for each propagation path ($\eu{-i\hat{H}t}$ and $\eu{i\hat{H}t}$) during each time interval ($t_{1}$, $t_{2}$ and $t_{3}$), with black corresponding to the ground state, $\ket{\text{g}}$, red to the single-exciton subspace, $\ket{\text{e}}$, and blue to the double-exciton subspace, $\ket{\text{ee}}$. Additionally the coloured circles represent the application of the dipole operator, where the brown circle creates an exciton (i.e., $\hat{\mu}^{+}$), while the green circle removes an exciton (i.e, $\hat{\mu}^{-}$).}\label{fig:feynman}
\end{SCfigure*}

In order to calculate a 2D optical spectrum, the following contributions to the third-order optical response function are needed:\cite{MukamelBook,Jeon2010,Kramer2018,Schlaucohen2011}
\begin{subequations}
\label{eq:nonlinear}
\begin{align}
\begin{split}
&R_{\text{GB,RP}}(t_{1},t_{2},t_{3}) \\
&=\Tr\left[\hat{\mu}^{-}\hat{\mu}^{+}(t_{1})\hat{\mu}^{-}(t_{1}+t_{2}+t_{3})\hat{\mu}^{+}(t_{1}+t_{2})\hat{\rho}_{\text{g}}\right] ,
\end{split} \\
\begin{split}
&R_{\text{SE,RP}}(t_{1},t_{2},t_{3}) \\
&=\Tr\left[\hat{\mu}^{-}\hat{\mu}^{+}(t_{1}+t_{2})\hat{\mu}^{-}(t_{1}+t_{2}+t_{3})\hat{\mu}^{+}(t_{1})\hat{\rho}_{\text{g}}\right] ,
\end{split} \\
\begin{split}
&R_{\text{ESA,RP}}(t_{1},t_{2},t_{3}) \\
&=\Tr\left[\hat{\mu}^{-}\hat{\mu}^{-}(t_{1}+t_{2}+t_{3})\hat{\mu}^{+}(t_{1}+t_{2})\hat{\mu}^{+}(t_{1})\hat{\rho}_{\text{g}}\right] ,
\end{split} \\ 
\begin{split}
&R_{\text{GB,NR}}(t_{1},t_{2},t_{3}) \\
&=\Tr\left[\hat{\mu}^{-}(t_{1}+t_{2}+t_{3})\hat{\mu}^{+}(t_{1}+t_{2})\hat{\mu}^{-}(t_{1})\hat{\mu}^{+}\hat{\rho}_{\text{g}}\right]
\end{split} \\
\begin{split}
&R_{\text{SE,NR}}(t_{1},t_{2},t_{3}) \\
&=\Tr\left[\hat{\mu}^{-}(t_{1})\hat{\mu}^{+}(t_{1}+t_{2})\hat{\mu}^{-}(t_{1}+t_{2}+t_{3})\hat{\mu}^{+}\hat{\rho}_{\text{g}}\right] ,
\end{split} \\
\begin{split}
&R_{\text{ESA,NR}}(t_{1},t_{2},t_{3}) \\
&=\Tr\left[\hat{\mu}^{-}(t_{1})\hat{\mu}^{-}(t_{1}+t_{2}+t_{3})\hat{\mu}^{+}(t_{1}+t_{2})\hat{\mu}^{+}\hat{\rho}_{\text{g}}\right] ,
\end{split}
\end{align}
\end{subequations}
where $\hat{A}(t)=\eu{i\hat{H}t}\hat{A}\eu{-i\hat{H}t}$ is the time-dependent Heisenberg representation of an arbitrary operator $\hat{A}$. Each contribution can then be represented by a double-sided Feynman diagram (Fig.~\ref{fig:feynman}) which illustrates its associated dynamical pathway through the exciton subspaces. %In each diagram, the left and right vertical lines correspond to the forward and backward propagators, $\eu{-i\hat{H}t}$ and $\eu{i\hat{H}t}$, and the circles correspond to the dipole operators applied at the beginning and end of each time interval. 
These diagrams can be generated from their associated mathematical expressions [Eqs.~(\ref{eq:nonlinear})] as follows. First, the propagators for each Heisenberg operator are written explicitly and any propagators directly next to each other are combined. The exciton subspace associated with each combined propagator is then determined by starting at the initial density matrix in the electronic ground state $\hat{\rho}_{\text{g}}$ and noting that $\hat{\mu}^{+}$ ($\hat{\mu}^{-}$) adds (removes) an exciton.  

From Fig.~\ref{fig:feynman}, we see that each of the contributions to the 2D optical spectrum has the following features in common. First, for the initial and final time intervals ($t_{1}$ and $t_{3}$) commonly referred to as the evolution and detection times, the system is in a coherence between two distinct exciton subspaces. Fourier transforms are therefore applied over these times to give the 2D spectrum. The so-called rephasing and nonrephasing diagrams (denoted `RP' and `NR')\cite{Jeon2010,Schlaucohen2011} correspond to the forward or backward propagator occupying the lower energy exciton subspace of the coherence during the evolution time, $t_{1}$. Second, the middle time interval ($t_{2}$) commonly referred to as the delay time, is fixed for a given 2D optical spectrum. The associated dynamics occur within a single-exciton subspace, although the forward and backward paths may however occupy different exciton states. %within that subspace.
The difference between the so-called ground-state bleaching and stimulated emission diagrams (denoted `GB' and `SE' respectively) is that the ground state, $\ket{\text{g}}$, is occupied during the delay time for the former, while the single-exciton subspace, $\ket{\text{e}}$, is occupied for the later. In addition, the excited-state absorption diagrams (denoted `ESA') differ from the stimulated emission diagrams by occupying a coherence state during the detection time which involves the double-exciton subspace, $\ket{\text{ee}}$.

From these contributions, the 2D optical spectrum can be obtained by performing the following Fourier transform:\cite{Schlaucohen2011,Kramer2018}
\begin{subequations}
\begin{align}
\begin{split}
&I_{\text{2D}}(\omega_{1},\omega_{3},t_{2})=-\frac{1}{4\pi}\text{Im}\left[\int_{0}^{\infty}\rd t_{1}\int_{0}^{\infty}\rd t_{3}\,\eu{i\omega_{3}t_{3}}\right. \\
&\left.\times\left(\frac{S^{(3)}_{\text{RP}}(t_{1},t_{2},t_{3})}{R_{\text{RP}}(0,t_{2},0)}\,\eu{-i\omega_{1}t_{1}}+\frac{S^{(3)}_{\text{NR}}(t_{1},t_{2},t_{3})}{R_{\text{NR}}(0,t_{2},0)}\,\eu{i\omega_{1}t_{1}}\right)\right]
\end{split} \\
&S^{(3)}(t_{1},t_{2},t_{3})=2\theta(t_{1})\theta(t_{2})\theta(t_{3})\text{Im}[R(t_{1},t_{2},t_{3})], \label{eq:optical_3} \\
\begin{split}
&R(t_{1},t_{2},t_{3})=R_{\text{GB}}(t_{1},t_{2},t_{3})+R_{\text{SE}}(t_{1},t_{2},t_{3}) \\
&\qquad\qquad\quad\quad-R_{\text{ESA}}(t_{1},t_{2},t_{3}) , \label{eq:rp_nr}
\end{split}
\end{align}
\end{subequations}
where Eqs.~(\ref{eq:optical_3}) and (\ref{eq:rp_nr}) are valid for both the rephasing and nonrephasing terms. Additionally, the impulsive pump--probe spectrum, also known as the transient absorption spectrum, corresponds to the case where the first two light--matter interactions coincide (i.e., when $t_{1}=0$) and hence the spectrum is given by:
\begin{equation}
I_{\text{PP}}(\omega_{3},t_{2})=-\frac{1}{2\pi}\text{Im}\left[\int_{0}^{\infty}\rd t_{3}\,\frac{S^{(3)}_{\text{RP}}(0,t_{2},t_{3})}{R_{\text{RP}}(0,t_{2},0)}\,\eu{i\omega t_{3}}\right] ,
\end{equation}
or equivalently using the nonrephasing diagrams.

In order to compute nonlinear spectra using mapping-based techniques, quasi-classical expressions for multi-time correlation functions, such as those given by Eqs.~(\ref{eq:nonlinear}), must be derived. For fully linearized methods it is not obvious how to accomplish this, because the phase-space integrals over the mapping-variable representations of dipole operators can only at most correctly reproduce the trace of the product of two operators.\cite{spinmap} The OMT method has however been used to compute multi-time correlation functions by incorporating discontinuous jumps in the mapping variables at the end of each time interval to mimic the transitions induced by the field--matter interaction.\cite{Polley2019,Polley2020vibronic} 
While this offers a viable approach for computing nonlinear spectra using fully-linearized approaches, we note that it does not reduce to WACL when the system has no diabatic couplings between the chromophores. It was also found that for the pump--probe spectrum with $t_{2}=0$, the OMT approach can give different results for the GB and SE contributions to the full spectrum, even though the associated quantum correlation functions for each are identical in this limit.\cite{Polley2020vibronic}

Fully linearized mapping based techniques have additionally been used to obtain nonlinear spectra using a explicit light-field approach, where the excitonic mapping variables are evolved in time under the field--matter interaction.\cite{Gao2020_nonlin} This is in principle more general than the optical response function approach, and is thus able to compute spectra in strong electromagnetic fields. However, when the experiment is well described by the perturbation theory, it is cleaner to directly compute the nonlinear multi-time response functions. For example, using the explicit light-field approach to obtain the weak-field limit of the 2D optical spectra is more computationally expensive, because the simulation must be performed with twelve different pulse phases, in order to implement the phase-matching condition required in order to obtain the desired signal. Additionally, because the explicit light-field approach closely mirrors the experimental procedure, it is difficult to analyze the different contributions to the spectrum beyond what can be done from experiment. In contrast, because within the response function approach, each individual contribution to the 2D optical spectrum [Eqs.~(\ref{eq:nonlinear})] can be computed independently, the full signal can then be decomposed into the underlying parts associated with the corresponding processes. 

The partially linearized approach overcomes many of the problems of the fully linearized approaches, in particular because it uses a different set of mapping variables to represent each propagator and can therefore be directly applied to calculating multi-time correlation functions. In Sec.~\ref{sec:nonlinear_partial}, we use such an approach to generate quasi-classical expressions for the multi-time correlation functions given in Eqs.~(\ref{eq:nonlinear}).  
\subsection{Partially Linearized Mapping Methods}\label{sec:nonlinear_partial}
In order to generate partially linearized mapping based expressions for the third-order correlation functions, we employ an analogous approach to that outlined in Sec.~\ref{sec:linear_partial} for linear spectroscopy. We first consider the correlation functions which do not involve the double-exciton subspace. As before, the propagators associated with the single-exciton subspace are each represented with a time-evolved kernel containing an independent set of mapping variables purely within this subspace. While the propagators associated with the ground state do not need to be represented by mapping variables, their effect is still accounted for by propagating the single-exciton subspace kernels using the associated potential matrix defined relative to the ground state potential, as described in Eq.~(\ref{eq:kernel_evolve}). Finally, the nuclear force associated with each time interval is given as the average of the forces associated with the forward and backward exciton paths, as before. Hence using the Feynman diagrams associated with the contributions to the third-order optical response function (Fig.~\ref{fig:feynman}), which explicitly show the forward and backward exciton paths during each time interval, the partially linearized expressions can be easily generated. For example, the stimulated emission rephasing correlation function can be computed by:
\begin{subequations}
\begin{align}
\begin{split}
&R_{\text{SE,RP}}(t_{1},t_{2},t_{3})\approx \\
&\Big\langle\braket{0|\hat{\mu}^{-}\hat{w}_{\text{e}}^{\dagger}(\mathcal{Z}_{\text{e}},t_{1}+t_{2})\hat{\mu}^{+}\hat{\mu}^{-}\hat{w}_{\text{e}}(\mathcal{Z}'_{\text{e}},t_{3}+t_{2})\hat{\mu}^{+}|0}\rho_{\text{g}}(x,p)\Big\rangle
\end{split} \label{eq:part_se} \\
\begin{split}
&\mathcal{F}=
\begin{cases}
     -\tfrac{1}{2}\nabla\left(V_{\text{g}}(x)+V_{\text{e}}(\mathcal{Z}_{\text{e}},x)\right), \qquad & \text{for}\ 0\le t<t_{1} \\
      -\tfrac{1}{2}\nabla\left(V_{\text{e}}(\mathcal{Z}_{\text{e}},x)+V_{\text{e}}(\mathcal{Z}_{\text{e}}',x)\right), &  \text{for}\ t_{1}\le t<t_{1}+t_{2} \\
      -\tfrac{1}{2}\nabla\left(V_{\text{g}}(x)+V_{\text{e}}(\mathcal{Z}_{\text{e}}',x)\right), & \text{for}\ t \ge t_{1}+t_{2}.
    \end{cases} 
\end{split} \label{eq:force_se}
\end{align}
\end{subequations}

In practice, the expression for this three-time correlation function can be evaluated as follows. First, the nuclear phase-space variables are sampled from the initial Wigner density, $\rho_{\text{g}}(x,p)$ and the mapping variables $\mathcal{Z}_{\text{e}}$ from focused initial conditions. %, and the associated Stratonovich--Weyl kernel, $\hat{w}_{\text{e}}(\mathcal{Z}_{\text{e}})$, is generated.
All these classical variables, along with the associated Stratonovich--Weyl kernel, $\hat{w}_{\text{e}}(\mathcal{Z}_{\text{e}})$, are then propagated for the maximum considered $t_{1}$ time using the nuclear force given by the first line of Eq.~\eqref{eq:force_se} and the values of these quantities are retained for each intermediate time-step. Second, a new set of mapping variables, $\mathcal{Z}'_{\text{e}}$, are sampled and a new kernel, $\hat{w}_{\text{e}}(\mathcal{Z}'_{\text{e}})$, is generated. For each intermediate $t_{1}$ time, the saved classical coordinates ($x(t_{1})$, $p(t_{1})$, $\mathcal{Z}_{\text{e}}(t_{1})$ and $\mathcal{Z}'_{\text{e}}$) along with both kernels ($\hat{w}_{\text{e}}(\mathcal{Z}_{\text{e}},t_{1})$ and $\hat{w}_{\text{e}}(\mathcal{Z}'_{\text{e}})$) are propagated for the $t_{2}$ delay time using the nuclear force given by the second line of Eq.~\eqref{eq:force_se}. Finally, the classical coordinates at the end of the $t_{2}$ time-interval, along with the kernel $\hat{w}_{\text{e}}(\mathcal{Z}'_{\text{e}},t_{2})$ are propagated for the maximum considered $t_{3}$ time using the nuclear force given by the final line of Eq.~\eqref{eq:force_se} and the matrix elements of the kernel are retained at each intermediate time-step. The contribution to this correlation function [Eq.~\eqref{eq:part_se}] is then calculated explicitly for each intermediate $t_{1}$ and $t_{3}$ using the corresponding time-evolved kernels. 
This data is then Fourier-transformed to generate the required spectrum.

For the excited-state absorption correlation functions, an additional time-evolved kernel containing mapping variables purely within the double-exciton subspace, $\hat{w}_{\text{ee}}(\mathcal{Z}_{\text{ee}})$, is required. In the same way as before, the partially linearized expression for the excited-state absorption correlation functions and the associated nuclear forces can be generated from their associated Feynman diagrams (Fig.~\ref{fig:feynman}), which for the rephasing diagram gives rise to:
\begin{subequations}
\begin{align}
\begin{split}
&R_{\text{ESA,RP}}(t_{1},t_{2},t_{3})\approx \\
&\!\!\!\!\Big\langle\braket{0|\hat{\mu}^{-}\hat{w}_{\text{e}}^{\dagger}(\mathcal{Z}_{\text{e}},t_{1}+t_{2}+t_{3})\hat{\mu}^{-}\hat{w}_{\text{ee}}(\mathcal{Z}_{\text{ee}},t_{3})\hat{\mu}^{+}\hat{w}_{\text{e}}(\mathcal{Z}'_{\text{e}},t_{2})\hat{\mu}^{+}|0}\Big\rangle 
\end{split} \label{eq:part_esarp} \\
\begin{split}
&\mathcal{F}= 
\begin{cases}
     -\tfrac{1}{2}\nabla\left(V_{\text{g}}(x)+V_{\text{e}}(\mathcal{Z}_{\text{e}},x)\right), \qquad & \text{for}\ t<t_{1} \\
      -\tfrac{1}{2}\nabla\left(V_{\text{e}}(\mathcal{Z}_{\text{e}},x)+V_{\text{e}}(\mathcal{Z}_{\text{e}}',x)\right), & \text{for}\ t_{1}<t<t_{1}+t_{2} \\
      -\tfrac{1}{2}\nabla\left(V_{\text{e}}(\mathcal{Z}_{\text{e}},x)+V_{\text{ee}}(\mathcal{Z}_{\text{ee}},x)\right), & \text{for}\ t \ge t_{1}+t_{2}.
      \end{cases} 
\end{split}
\end{align}
\end{subequations}
The implementation for this correlation function is similar as for the stimulated emission correlation function considered previously, except that in general a third set of mapping variables $\mathcal{Z}_\text{ee}$ is sampled after the $t_2$ delay time.

While a partially linearized approach has already been used to calculate third-order optical response functions,\cite{Provazza2018_nonlin} our approach differs in the following ways. First, spin-mapping instead of MMST mapping is used to describe the excitonic dynamics. Not only is spin-PLDM seen to offer an improvement over standard PLDM for linear spectra (Sec.~\ref{sec:linear}), it has also previously been observed to give rise to superior accuracy in obtaining population dynamics.\cite{Mannouch2020a} This difference is particularly pronounced when focused initial conditions are implemented, because they are known to significantly degrade the accuracy of standard PLDM.\cite{Huo2012PLDM,Hsieh2013FBTS} On the other hand, using focused initial conditions in a partially linearized approach is particularly advantageous, as by construction the method will correctly reduce to WACL in the absence of exciton couplings within each distinct subspace. Focused spin-PLDM is therefore the method of choice, as it can simultaneously obtain correlation functions extremely accurately, while still retaining this connection to WACL. Second, we calculate each contribution to the third-order optical response function separately using its own specific mapping-based expression, which allows the dynamics in each of the distinct excitonic subspaces to be treated separately. While this makes no difference to the obtained results when using focused standard PLDM, the different dimensions of the considered exciton spaces for the two approaches will result in different values for the zero-point energy parameters when using spin-PLDM. An advantage of our approach therefore is that by using the most appropriate zero-point energy for each of the distinct exciton subspaces with spin-PLDM should lead to more accurate results. Finally, in the original PLDM approach of Ref.~\onlinecite{Provazza2018_nonlin}, the mapping variables are resampled at the beginning of each time-interval, whereas in our approach the mapping variables are only resampled when a new exciton is created by a dipole operator. Because the resampling of mapping variables in a method generally makes the results more difficult to converge, we hence choose to do this only when it is absolutely necessary.  
%While resampling the mapping variables more often should in principle lead to more accurate results, it also makes the approach more computationally expensive. 
%to resample the mapping variables only when a new exciton is created by a dipole operator in order to qualitatively reproduce the important features of the nonlinear spectra for the models considered in this paper, particularly when spin-PLDM is used. 
Because of this difference, our standard PLDM approach for nonlinear spectra is in principle subtly different from that presented in Ref.~\onlinecite{Provazza2018_nonlin}, although in practice the obtained results are seen to be essentially identical for the systems treated in this paper.

%Additionally, fully linearized mapping-based approaches have also been designed for computing nonlinear spectra. However because these approaches cannot describe the excitonic dynamics associated with the forward and backward paths separately, they must include the pulses explicitly within the dynamics, such that the calculation that must be performed is identical to the experimental setup. This results in a more computationally expensive approach, because the simulation must be performed multiple times with different pulse phases, in order to implement the phase-matching condition required in order to obtain the desired signal. Additionally, such an approach can also not reduce the full third-order signal into its constituent components, given by the Feynman diagrams in Fig.~\ref{fig:feynman}. Because of this, our partially linearized approach leads to further insight into the various dynamical pathways that contribute to the full nonlinear spectrum, beyond what can be achieved directly from experiment.

\subsection{Results}
We now apply our partially linearized approach outlined in Sec.~\ref{sec:nonlinear_partial} to calculate pump--probe and 2D optical spectra. We study some of the same Frenkel biexciton models for which linear spectra were computed in Sec.~\ref{sec:linear_results} as well as a seven-state FMO model. Specific details associated with these models, as well as implementation details of the various methods, can be found in Sec.~\ref{sec:linear_results} and the supplementary material. Because all of these models are harmonic, exact results can again be obtained using the Hierarchical equations of motion (HEOM) approach, which we calculated using the open-source \texttt{pyrho} package.\cite{pyrho}
\subsubsection{The Frenkel Biexciton Model}
An interesting aspect of the nonlinear spectra we consider is that they also probe the double-exciton subspace. For the Frenkel biexciton model, this subspace consists of a single state where both chromophores are occupied, such that mapping variables are not needed to describe the double-exciton dynamics. The corresponding Stratonovich--Weyl kernel in the partially linearized expressions for the ESA response functions [for example, Eq.~(\ref{eq:part_esarp})] can thus be replaced as follows:
\begin{equation}
\hat{w}_{\text{ee}}(\mathcal{Z}_{\text{ee}},t_{3})\rightarrow\eu{-i\int^{t_{3}}_{0}\rd t\left[V_{\text{ee}}(x(t))-V_{\text{g}}(x(t))\right]} , 
\end{equation}
which is essentially the single state analogue of Eq.~(\ref{eq:kernel_evolve}). 

\begin{figure*}
\includegraphics[scale=.81]{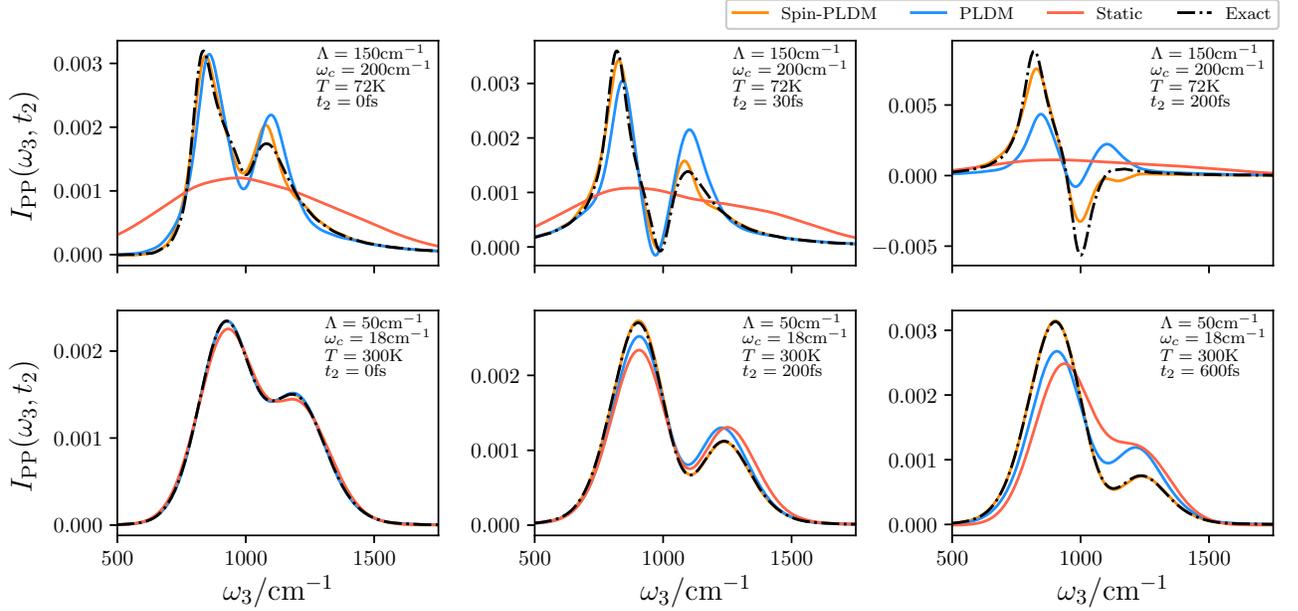}
\caption{The pump--probe spectra for two different Frenkel biexciton models, calculated for various $t_{2}$ delay times as indicated. Both models have $\epsilon_{1}-\epsilon_{2}=100\,\text{cm}^{-1}$, $\omega_{\text{shift}}=1050\,\text{cm}^{-1}$ and $\Delta_{12}=100\,\text{cm}^{-1}$, with  reorganization energy $\Lambda$ and cut-off frequency $\omega_c$ as specified. Exact HEOM results\cite{pyrho} are given by the dashed black lines. The results given in the first row of figures may be compared with those calculated using Ehrenfest, various flavours of MMST (Fig.~6 from Ref.~\onlinecite{Gao2020_nonlin}) and TCL2 (Fig.~6 from Ref.~\onlinecite{Fetherolf2017}). Our HEOM results are in agreement with those already published (Fig.~6 from Ref.~\onlinecite{Fetherolf2017}, Fig.~6 from Ref.~\onlinecite{Gao2020_nonlin}).}\label{fig:pump-exciton}
\end{figure*}
Figure \ref{fig:pump-exciton} presents the pump--probe spectra for both a low- and high-temperature Frenkel biexciton model for various $t_{2}$ delay times. Only the methods from Sec.~\ref{sec:linear_results} which are able to calculate nonlinear spectra within the optical response function approach are used. %The fact that fully linearized methods cannot be used to calculate multi-time correlation functions shows that their partially linearized counterparts have a particular advantage in computing spectroscopic observables of interest.
In particular, we do not present results of fully linearized methods as these cannot directly calculate multi-time correlation functions, although the reader may wish to compare our results with the fully linearized method of Refs.~\onlinecite{Gao2020_lin} and \onlinecite{Gao2020_nonlin} obtained using the explicit light-field approach.
For the relatively slow-bath and high-temperature model (i.e., small $\omega_{\text{c}}$ and large $T$), given by the bottom row of Fig.~\ref{fig:pump-exciton}, the spin-PLDM results are essentially indistinguishable from the numerically exact results, in keeping with what was also observed for linear spectra. While the standard PLDM approach is also able to accurately reproduce the high-temperature pump--probe spectra at $t_{2}=0$, the error in the results increases for longer delay times. This is essentially consistent with earlier work,\cite{Mannouch2020a} where standard PLDM was found to exhibit significant errors when computing long-time population dynamics, even at high-temperature. In this case, even though standard PLDM still captures the correct qualitative features of the pump--probe spectra at non-zero delay times, the error is not much better than that exhibited by far simpler approaches, such as computing the exciton dynamics with static nuclei (labelled `Static'). The fact that the static-nuclear approximation deviates from the exact results for these pump--probe spectra at non-zero $t_{2}$ times also illustrates that homogeneous broadening effects can still be present in nonlinear spectra even for models with relatively slow baths. 

For pump--probe spectra associated with the relatively fast-bath and low-temperature Frenkel biexciton model, given by the top row of Fig.~\ref{fig:pump-exciton}, all methods now exhibit errors compared to the numerically exact results. This is to be expected because for this relatively fast bath homogeneous broadening dominates, as illustrated by the fact that the static-nuclear approximation fails dramatically, and hence this system poses a greater challenge for accurately obtaining the associated nonlinear spectra. Spin-PLDM however is still able to qualitatively capture the correct features of the spectrum, even at large $t_{2}$ delay times %when standard PLDM starts to incorrectly predict an additional peak which is not present in the exact results.
and is significantly more accurate than standard PLDM\@.

\begin{figure*}
\includegraphics[scale=.81]{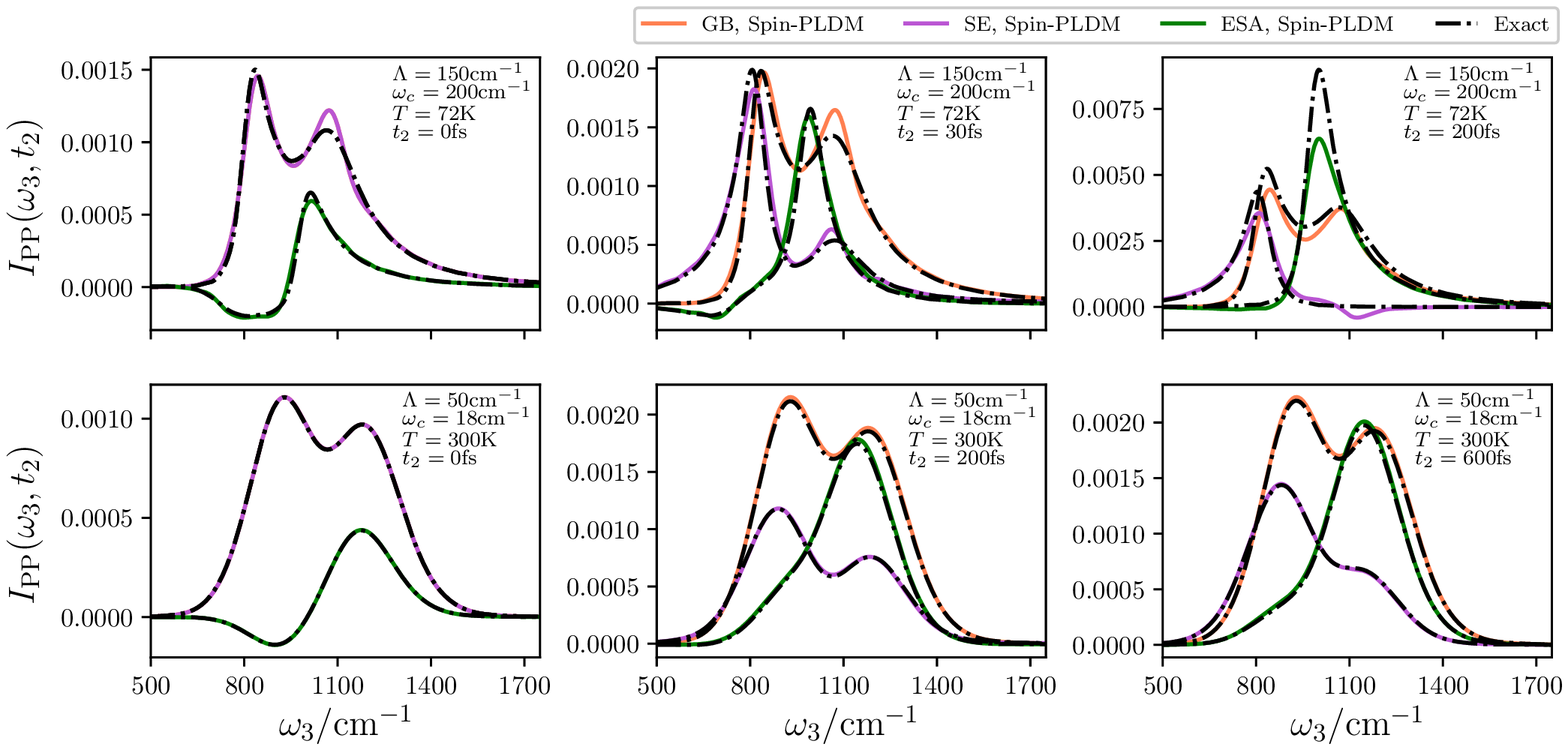}
\caption{The individual contributions to the pump--probe spectra for two different Frenkel biexciton models, calculated with spin-PLDM for various $t_{2}$ delay times as indicated. Both models have $\epsilon_{1}-\epsilon_{2}=100\,\text{cm}^{-1}$, $\omega_{\text{shift}}=1050\,\text{cm}^{-1}$ and $\Delta_{12}=100\,\text{cm}^{-1}$, with the other model specific parameters given in the figures. Exact HEOM results\cite{pyrho} are given by the dashed black lines.}\label{fig:comp-exciton}
\end{figure*}
Another advantage of the optical response function approach for calculating nonlinear spectra is that the full signal can be decomposed into its constituent parts associated with different dynamical pathways, giving greater insight beyond what can be directly obtained from experiment or the explicit light-field approach. Figure \ref{fig:comp-exciton} gives the three distinct contributions to the pump--probe spectra (GB, SE and ESA) for the two Frenkel-biexciton models considered in Fig.~\ref{fig:pump-exciton} at various $t_{2}$ delay times, calculated using spin-PLDM (coloured lines) and the numerically exact HEOM approach (dashed black lines). As discussed before, the spin-PLDM approach is able to qualitatively reproduce the main features of the various contributions to and thus the full pump--probe spectra in both the high-temperature and low-temperature regimes. Considering each of these contributions in turn allows the full pump--probe spectra given in Fig.~\ref{fig:pump-exciton} to be better understood. First, the GB signal corresponds to the linear absorption spectrum for all $t_{2}$ delay times, because the initial nuclear density matrix, $\hat{\rho}_{\text{g}}$, is conserved by the ground state dynamics during the $t_{2}$ delay time. Second, the SE signal becomes the fluorescence spectrum in the limit of $t_{2}\rightarrow\infty$, as the system relaxes to $\hat{\rho}_{\text{e}}$ due to the dynamics in the single-exciton subspace during the $t_{2}$ delay time. This explains why the lowest-energy peak within the SE signal increases in intensity with increasing $t_{2}$ delay time, in accordance with Kasha's rule. Finally, the highest-energy peak in the ESA signal increases in intensity with increasing $t_{2}$ delay time, because absorption to the double-exciton subspace occurs from the remaining unoccupied state within the single-exciton subspace. 

\begin{figure*}
\includegraphics[scale=.8]{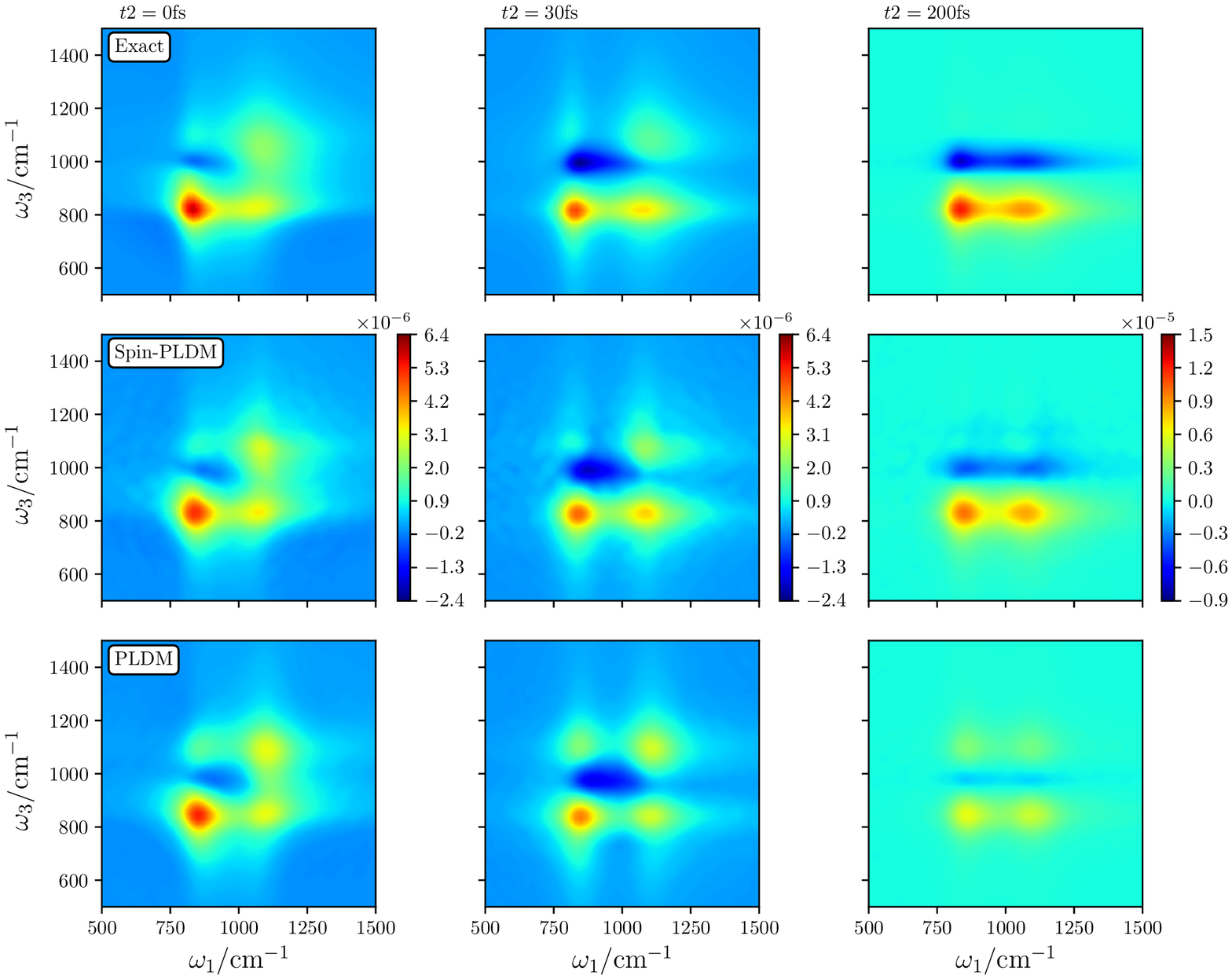}
\caption{The 2D optical spectra for a relatively fast-bath and low-temperature Frenkel biexciton model ($\epsilon_{1}-\epsilon_{2}=100\,\text{cm}^{-1}$, $\omega_{\text{shift}}=1050\,\text{cm}^{-1}$, $\Delta_{12}=100\,\text{cm}^{-1}$, $\Lambda=150\,\text{cm}^{-1}$, $\omega_{c}=200\,\text{cm}^{-1}$ and $T=72\,$K), calculated for various $t_{2}$ delay times as indicated. Exact HEOM results\cite{pyrho} correspond to the first row of figures. Note that a different colour scale is used for the $t_{2}=200\,$fs results (right-hand column) compared to the others.}\label{fig:2d-exciton}
\end{figure*}
The 2D optical spectrum can also be calculated using the same approach. This is more computationally demanding than pump--probe spectra, as now the contributions to the optical response function must be computed on a ($t_{1}$,$t_{3}$) two-dimensional grid. The 2D optical spectra for the relatively slow-bath and high-temperature Frenkel biexciton model for various $t_{2}$ delay times is given in the supplementary material, which is the main benchmark used in previous work for nonlinear optical spectra.\cite{Provazza2018_nonlin,Gao2020_nonlin,Fetherolf2017,Polley2020vibronic} As for the pump--probe spectra, the spin-PLDM 2D optical spectra for this regime are essentially indistinguishable from the exact results, while even though the standard PLDM results are qualitatively accurate, they do exhibit small but noticeable errors for large $t_{2}$ delay times. We however note that although the static-nuclear approximation does deviate from the exact results for this parameter set, it is nevertheless able to qualitatively reproduce the important features of the spectra, illustrating that this model does not pose a %rigorous
challenging test for our method. We therefore consider the 2D optical spectra for a more challenging relatively fast-bath and low-temperature regime of the model, where homogeneous broadening effects dominate and the static-nuclear approximation fails. In this regime, shown in Fig.~\ref{fig:2d-exciton}, both partially linearized approaches are able to qualitatively reproduce the most important features. However, for non-zero $t_{2}$ delay times, spin-PLDM is again more accurate than standard PLDM, in particular for the $t_{2}=200$ fs case. % where standard PLDM predicts peaks at $\omega_{3}\approx 1100\,\text{cm}^{-1}$, which are absent in the exact results.
\subsubsection{The Fenna--Matthews--Olsen complex}
An example of a biologically relevant system for which it is interesting to compute nonlinear spectra is the seven-state FMO model. All the Hamiltonian parameters for this model can be found in Ref.~\onlinecite{Kramer2018}. This system contains $F=7$ single-exciton states and $F(F-1)/2=21$ double-exciton states and hence requires a set of mapping variables to describe the associated exciton dynamics within each of these subspaces. The excitonic space of this system is thus much larger than that of the Frenkel biexciton model considered previously (which contains two single-exciton states and only one double-exciton state), therefore acting as a good test for the viability of these methods to calculate nonlinear spectra in realistic condensed-phase systems. 
%To the authors' knowledge, no mapping-based approach has previously tackled the nonlinear spectra of such a large system. 
To compute the pump--probe spectra, we assign dipole-moment orientations to each chromophore following Ref.~\onlinecite{Kramer2018} and consider the case where all laser pulses are equally polarized, such that the rotational averaging of these spectra can be performed by averaging the multi-time correlation functions over 10 representative electric-field directions (corresponding to the vertices of a dodecahedron) as described in Ref.~\onlinecite{Hein2012}. %This corresponds to averaging over a set of ten multi-time correlation functions, that each contain a different orientation for the dipole operators.
This averaging can be performed as a post-processing step after each trajectory is computed, to determine its contribution to the ensemble average. The same approach could also be applied to more complicated polarization sequences, where at most 21 electric field directions have to be considered.\cite{Gelin2017}

\begin{figure*}
\includegraphics[scale=.81]{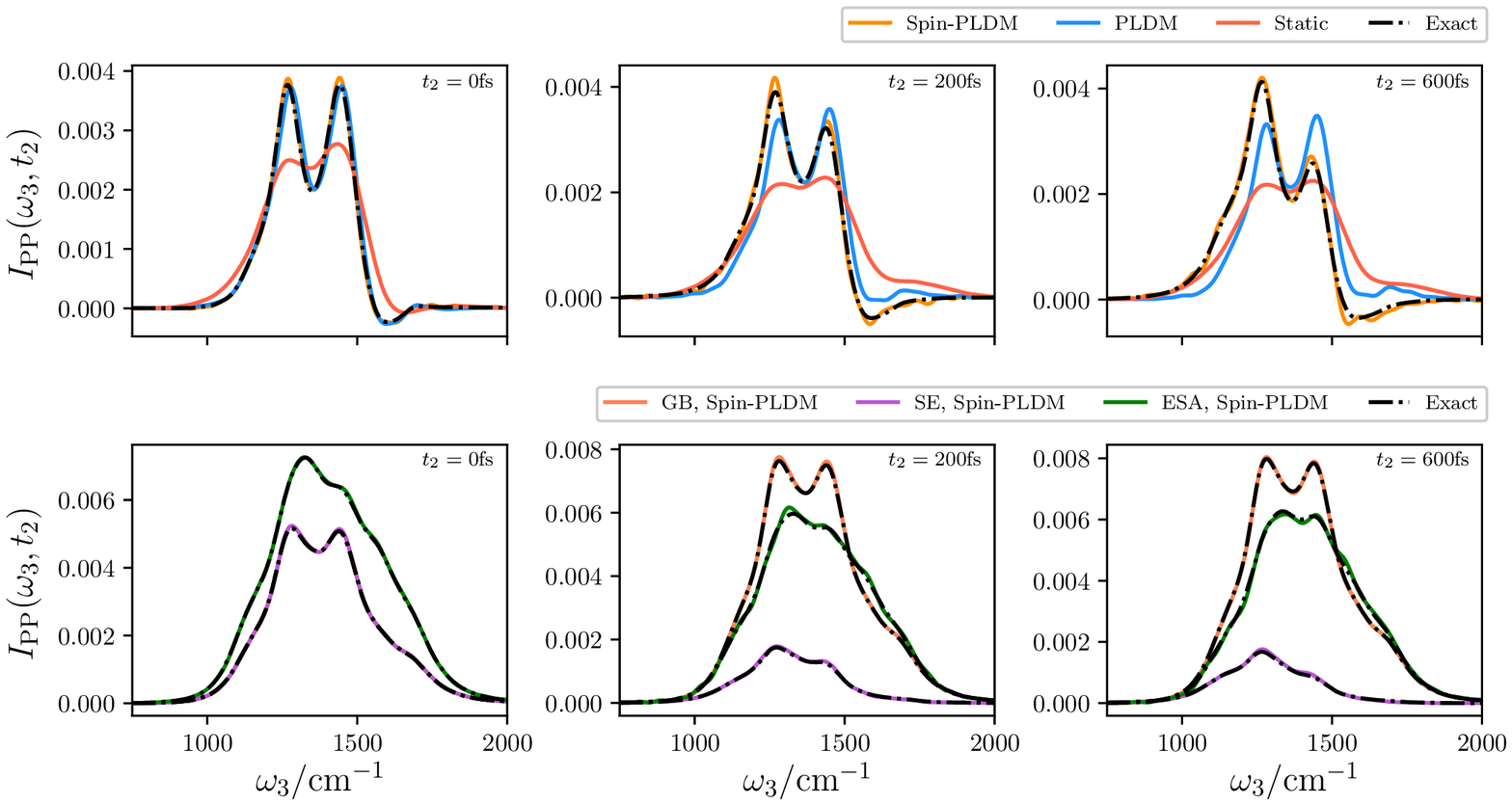}
\caption{The pump--probe spectra for a seven-state FMO model at $T=300\,$K, calculated for various $t_{2}$ delay times as indicated. We use $\omega_{\text{shift}}=1430\,\text{cm}^{-1}$, with all other model parameters given in Ref.~\onlinecite{Kramer2018}. The top row of figures gives the full pump--probe spectra, whereas the bottom row gives the individual contributions calculated with spin-PLDM\@. Exact HEOM results\cite{pyrho} are given by the dashed black lines.}\label{fig:pump-fmo}
\end{figure*}
Figure \ref{fig:pump-fmo} gives the pump--probe spectra for the FMO model at $T=300$ K, for three different $t_{2}$ delay times. As for the other high-temperature models, the spin-PLDM results are essentially indistinguishable from the benchmark, for both the components and the full spectra, with any errors being numerical in nature rather than systematic. As was also found for the Frenkel biexciton models, standard PLDM is accurate for the high-temperature pump--probe spectra at $t_{2}=0$ fs, but the error associated with the method increases for longer $t_{2}$ delay times. Spin-PLDM therefore clearly exhibits superior accuracy in computing nonlinear optical spectra than standard PLDM.  

It seems that the only disadvantage of the spin-PLDM approach relative to standard PLDM is that for large exciton subspaces, the method requires a greater number of trajectories to converge. For the relatively small exciton subspaces of the Frenkel-biexciton models, there is no issue and both standard PLDM and spin-PLDM require about $10^{6}$ trajectories to reach full convergence on the scale of the plot when calculating the multi-time correlation functions needed for nonlinear spectra. However when calculating the excited-state absorption correlation functions of the FMO model, for which the very large 21 state double-exciton subspace must be considered, we used $1\times10^{9}$ spin-PLDM trajectories to reach convergence compared to $2\times 10^{7}$ trajectories when using standard PLDM. This is in contrast to the $1\times 10^{7}$ and $4\times 10^{7}$ spin-PLDM trajectories used to obtain the ground-state bleaching and stimulated emission correlation functions respectively and the $1\times 10^{6}$ trajectories needed to obtain the linear absorption and fluorescence spectra.

We have, however, found a simple way to dramatically reduce the cost of the spin-PLDM approach for calculating these excited-state absorption correlation functions, while still retaining the superior accuracy of the method. For the double-exciton subspace, we choose to use $\gamma=0$ for the focused initial conditions of the mapping variables, $\mathcal{Z}_{\text{ee}}$, the associated kernel, $\hat{w}_{\text{ee}}(\mathcal{Z}_{\text{ee}})$ and the mapping variable expression for the force operator, $\nabla V_{\text{ee}}(\mathcal{Z}_{\text{ee}},x)$, while the spin-PLDM expression for the zero-point energy parameter is still used for the single-exciton subspace. This essentially corresponds to a standard PLDM treatment of the double-exciton subspace, which is a reasonable approximation because the correlation functions decay rapidly as a function of the $t_{3}$ time and it is known that standard PLDM is also accurate for relatively short propagation times. For example, a maximum $t_{3}$ time of only 500\,fs was required to correctly obtain the pump--probe spectra for this FMO model at $T=300$\,K\@. Hence treating the largest double-exciton subspace using standard PLDM should drastically reduce the number of trajectories needed for convergence, without significantly affecting the accuracy of the method. Fig.~S4 in the supplementary material gives the pump--probe spectra for our modified spin-PLDM approach, where we only used $4\times 10^{7}$ trajectories to obtain the excited-state absorption correlation function. The accuracy of these results can be seen to be almost equal to the full spin-PLDM method, but now obtained at a significantly reduced computational cost similar to the standard PLDM approach. % In practice with this modified approach, the excited-state absorption correlation function can now be calculated within 24 hours using 1440 cores on a standard cluster. 
This significant improvement in efficiently should make it possible to compute the 2D spectra of FMO with our partially linearized approach, which requires additional scanning over the $t_{1}$ time.
We will address this in future work.
%While using such a large number of trajectories would surely be impractical for \emph{ab initio} simulations, the sampling of the initial mapping variables within spin-PLDM has yet to be optimized and we suspect that more sophisticated schemes should enable a drastic reduction in the number of trajectories required for convergence. 
\section{Conclusions}
In this paper we have shown how classical trajectory mapping-based methods can be used to compute optical spectra for nonadiabatic systems through the constituent response functions. In particular we use a partially linearized approach, which represents each individual propagator within any correlation function using independent sets of mapping variables. This means that single- and multi-time correlation functions, used to compute linear and nonlinear spectra respectively, can all be calculated on an equal footing. We have demonstrated that this approach can accurately and consistently reproduce the important features of absorption, fluorescence, pump--probe and 2D optical spectra for a range of systems and parameter regimes.
%In contrast, it is not at all obvious how such an extension could be correctly achieved for fully linearized approaches. 
%and methods based on the quantum master equation such as TCL2\@.

%[what's also better about partial vs linear - more subtle things that we wanted our method to be able to do]
One way our approach differs from others is by requiring that the expressions for the correlation functions reduce to WACL when there are no diabatic couplings between the chromophores, such that the method is exact in this limit for systems containing harmonic nuclear degrees of freedom. While such a requirement can be satisfied using a partially linearized approach with focused initial conditions, 
%and we have initially shown that such methods are as a result able to accurately reproduce optical absorption and fluorescence spectra consistently for systems across a wide range of parameter regimes.
it cannot be satisfied with fully linearized methods, often resulting in inaccurate linear spectra for systems in more extreme and challenging parameter regimes and generally giving rise to overbroadened peaks. 
%While quantum master equations, like TCL2, can accurately compute linear absorption spectra, they cannot describe an initial correlated electron-nuclear state and are therefore unable to correctly reproduce fluorescence spectra.
Another distinction of our approach is that we treat the excitonic dynamics separately within the distinct subspaces (something which also cannot be done within a fully linearized approach); this eliminates unphysical forces which arise from mapping-variable contributions to exciton configurations that cannot occur and also means that the most appropriate spin-mapping zero-point energy parameter for each subspace can be used, which is expected to result in more accurate dynamics. Overall, these points illustrate how partially linearized approaches are much more suited to computing spectroscopic quantities than their fully linearized counterparts.

%For multi-time correlation functions, our approach still reduces to WACL in the absence of diabatic couplings between the chromophores and allows the dynamics associated with distinct excitonic subspaces to be treated separately, as desired; 
% 
We have considered two partially linearized mapping-based methods, standard PLDM and spin-PLDM, and have demonstrated that spin-PLDM consistently gives rise to the most accurate spectra. For high-temperature systems, the spin-PLDM results are essentially numerically exact, while we found that the method can still qualitatively reproduce the important spectroscopic features in more challenging low-temperature regimes. %This was to be expected, as it was already known that spin-PLDM is able to extremely accurately reproduce the relatively short time behaviour of single-time correlation functions.\cite{Mannouch2020a} 
In contrast, standard PLDM was shown to exhibit errors in the nonlinear optical spectra at non-zero delay times, even in the high-temperature regime.
% and gave rise to incorrect spectroscopic features, such as additional peaks, for more challenging systems.
One of the reasons that standard PLDM is less accurate than spin-PLDM is because focused initial conditions in standard PLDM are known to rapidly degrade the quality of the results, whereas focused spin-PLDM is observed to be just as accurate as the full-sphere sampling variant.\cite{spinmap,multispin,Mannouch2020b} Therefore one of the advantages of spin-PLDM over standard PLDM is that focused initial conditions can be used to ensure that the method reduces to WACL in the absence of diabatic couplings, without compromising the accuracy of the obtained results for systems with coupled chromophores. 

The only drawback of the spin-mapping version of PLDM is that the method is computationally expensive for systems with large exciton subspaces. To alleviate this problem, we introduced a modified spin-PLDM approach which treats the double-exciton subspace using the standard PLDM approach.
%, while still describing the single-exciton subspace using spin-PLDM\@. 
Because the double-exciton subspace dynamics only occurs during the $t_{3}$ time, over which the multi-time correlation function rapidly decays, the relatively good short-time accuracy of standard PLDM should be sufficient to still accurately describe this. This modification was found to significantly reduce the number of trajectories needed to reach convergence, such that the new approach has a comparable computational cost to standard PLDM,
%@. However, it was also found that the modified method 
while still retaining the superior accuracy of the original spin-PLDM approach.
%in calculating multi-time correlation functions.
This therefore offers a practical way of computing nonlinear spectra in light-harvesting systems.

All our approaches are built on the perturbative response of the exciton--nuclear dynamics due to a weak coupling to the light field.
%There are certain aspects of our approach which could be extended upon in the future.
However, for experiments involving intense laser fields, the effect of the classical electromagnetic radiation on the coupled exciton--nuclear dynamics must be included explicitly via a time-dependent Hamiltonian. In principle, it it will be possible to develop a partially linearized approach for this scenario which reduces to that already outlined in this paper in the weak-field limit. Because of this requirement, such a method will therefore differ from other previously considered mapping-based approaches to this problem.\cite{Gao2020_lin,Gao2020_nonlin,Provazza2021} Additionally for experiments involving extremely low intensity radiation, such as those performed within a cavity, a quantum description of light is needed to correctly describe the induced exciton--nuclear dynamics. This could be achieved within a partially linearized approach by treating each cavity mode as an additional harmonic degree of freedom within the system and then treating the dynamics of these modes classically.\cite{Miller1978radiation}

Because spin-PLDM can in principle be applied to anharmonic problems, it also offers a route to computing linear and nonlinear spectra for more complex systems which cannot be treated by HEOM\@. Hence the importance of anharmonic effects on excitonic energy-transfer in photosynthetic light-harvesting systems could be investigated by performing atomistic simulations based on classical force fields,\cite{Kim2012FMO}
%Additionally better mechanistic insight could be obtained through analysis of the individual trajectories,
leading to a greater overall understanding of the mechanism that gives rise to such highly efficient energy transfer in light-harvesting systems.

\section*{Supplementary}
See the supplementary material for further results and for the data files of calculated optical response functions for spin-LSC, standard PLDM and spin-PLDM, which can be used to generate all the associated spectra presented in this paper.

\begin{acknowledgments}
The authors would like to acknowledge the support from the Swiss National Science Foundation through the NCCR MUST (Molecular Ultrafast Science and Technology) Network. We also thank Graziano Amati, Johan Runeson and Joseph Lawrence for helpful discussions and comments. 
\end{acknowledgments}

\section*{Data Availability}
The data that supports the findings of this study are available within the article and its supplementary material.

\bibliography{jonathan,references}

\begin{thebibliography}{13}%
\makeatletter
\providecommand \@ifxundefined [1]{%
 \@ifx{#1\undefined}
}%
\providecommand \@ifnum [1]{%
 \ifnum #1\expandafter \@firstoftwo
 \else \expandafter \@secondoftwo
 \fi
}%
\providecommand \@ifx [1]{%
 \ifx #1\expandafter \@firstoftwo
 \else \expandafter \@secondoftwo
 \fi
}%
\providecommand \natexlab [1]{#1}%
\providecommand \enquote  [1]{``#1''}%
\providecommand \bibnamefont  [1]{#1}%
\providecommand \bibfnamefont [1]{#1}%
\providecommand \citenamefont [1]{#1}%
\providecommand \href@noop [0]{\@secondoftwo}%
\providecommand \href [0]{\begingroup \@sanitize@url \@href}%
\providecommand \@href[1]{\@@startlink{#1}\@@href}%
\providecommand \@@href[1]{\endgroup#1\@@endlink}%
\providecommand \@sanitize@url [0]{\catcode `\\12\catcode `\$12\catcode
  `\&12\catcode `\#12\catcode `\^12\catcode `\_12\catcode `\%12\relax}%
\providecommand \@@startlink[1]{}%
\providecommand \@@endlink[0]{}%
\providecommand \url  [0]{\begingroup\@sanitize@url \@url }%
\providecommand \@url [1]{\endgroup\@href {#1}{\urlprefix }}%
\providecommand \urlprefix  [0]{URL }%
\providecommand \Eprint [0]{\href }%
\providecommand \doibase [0]{http://dx.doi.org/}%
\providecommand \selectlanguage [0]{\@gobble}%
\providecommand \bibinfo  [0]{\@secondoftwo}%
\providecommand \bibfield  [0]{\@secondoftwo}%
\providecommand \translation [1]{[#1]}%
\providecommand \BibitemOpen [0]{}%
\providecommand \bibitemStop [0]{}%
\providecommand \bibitemNoStop [0]{.\EOS\space}%
\providecommand \EOS [0]{\spacefactor3000\relax}%
\providecommand \BibitemShut  [1]{\csname bibitem#1\endcsname}%
\let\auto@bib@innerbib\@empty
%</preamble>
\bibitem [{\citenamefont {Bosnich}(1969)}]{Bosnich1969}%
  \BibitemOpen
  \bibfield  {author} {\bibinfo {author} {\bibfnamefont {B.}~\bibnamefont
  {Bosnich}},\ }\href@noop {} {\bibfield  {journal} {\bibinfo  {journal} {Acc.
  Chem. Res.}\ }\textbf {\bibinfo {volume} {2}},\ \bibinfo {pages} {266}
  (\bibinfo {year} {1969})}\BibitemShut {NoStop}%
\bibitem [{\citenamefont {Kramer}\ \emph {et~al.}(2018)\citenamefont {Kramer},
  \citenamefont {Noack}, \citenamefont {Reinefeld}, \citenamefont
  {Rodríguez},\ and\ \citenamefont {Zelinskyy}}]{Kramer2018}%
  \BibitemOpen
  \bibfield  {author} {\bibinfo {author} {\bibfnamefont {T.}~\bibnamefont
  {Kramer}}, \bibinfo {author} {\bibfnamefont {M.}~\bibnamefont {Noack}},
  \bibinfo {author} {\bibfnamefont {A.}~\bibnamefont {Reinefeld}}, \bibinfo
  {author} {\bibfnamefont {M.}~\bibnamefont {Rodríguez}}, \ and\ \bibinfo
  {author} {\bibfnamefont {Y.}~\bibnamefont {Zelinskyy}},\ }\href@noop {}
  {\bibfield  {journal} {\bibinfo  {journal} {J. Comput. Chem.}\ }\textbf
  {\bibinfo {volume} {39}},\ \bibinfo {pages} {1779} (\bibinfo {year}
  {2018})}\BibitemShut {NoStop}%
\bibitem [{\citenamefont {Hein}\ \emph {et~al.}(2012)\citenamefont {Hein},
  \citenamefont {Kreisbeck}, \citenamefont {Kramer},\ and\ \citenamefont
  {Rodr{\'{\i}}guez}}]{Hein2012}%
  \BibitemOpen
  \bibfield  {author} {\bibinfo {author} {\bibfnamefont {B.}~\bibnamefont
  {Hein}}, \bibinfo {author} {\bibfnamefont {C.}~\bibnamefont {Kreisbeck}},
  \bibinfo {author} {\bibfnamefont {T.}~\bibnamefont {Kramer}}, \ and\ \bibinfo
  {author} {\bibfnamefont {M.}~\bibnamefont {Rodr{\'{\i}}guez}},\ }\href@noop
  {} {\bibfield  {journal} {\bibinfo  {journal} {New J. Phys.}\ }\textbf
  {\bibinfo {volume} {14}},\ \bibinfo {pages} {023018} (\bibinfo {year}
  {2012})}\BibitemShut {NoStop}%
\bibitem [{\citenamefont {Berkelbach}()}]{pyrho}%
  \BibitemOpen
  \bibfield  {author} {\bibinfo {author} {\bibfnamefont {T.~C.}\ \bibnamefont
  {Berkelbach}},\ }\href@noop {} {\enquote {\bibinfo {title} {Pyrho: A python
  package for reduced density matrix techniques},}\ }\bibinfo {howpublished}
  {\url{https://github.com/berkelbach-group/pyrho}}\BibitemShut {NoStop}%
\bibitem [{\citenamefont {Gao}, \citenamefont {Lai},\ and\ \citenamefont
  {Geva}(2020)}]{Gao2020_lin}%
  \BibitemOpen
  \bibfield  {author} {\bibinfo {author} {\bibfnamefont {X.}~\bibnamefont
  {Gao}}, \bibinfo {author} {\bibfnamefont {Y.}~\bibnamefont {Lai}}, \ and\
  \bibinfo {author} {\bibfnamefont {E.}~\bibnamefont {Geva}},\ }\href@noop {}
  {\bibfield  {journal} {\bibinfo  {journal} {J. Chem. Theory Comput.}\
  }\textbf {\bibinfo {volume} {16}},\ \bibinfo {pages} {6465} (\bibinfo {year}
  {2020})}\BibitemShut {NoStop}%
\bibitem [{\citenamefont {Runeson}\ and\ \citenamefont
  {Richardson}(2019)}]{spinmap}%
  \BibitemOpen
  \bibfield  {author} {\bibinfo {author} {\bibfnamefont {J.~E.}\ \bibnamefont
  {Runeson}}\ and\ \bibinfo {author} {\bibfnamefont {J.~O.}\ \bibnamefont
  {Richardson}},\ }\href {\doibase 10.1063/1.5100506} {\bibfield  {journal}
  {\bibinfo  {journal} {J. Chem. Phys.}\ }\textbf {\bibinfo {volume} {151}},\
  \bibinfo {pages} {044119} (\bibinfo {year} {2019})},\ \Eprint
  {http://arxiv.org/abs/1904.08293} {arXiv:1904.08293 [physics.chem-ph]}
  \BibitemShut {NoStop}%
\bibitem [{\citenamefont {Runeson}\ and\ \citenamefont
  {Richardson}(2020)}]{multispin}%
  \BibitemOpen
  \bibfield  {author} {\bibinfo {author} {\bibfnamefont {J.~E.}\ \bibnamefont
  {Runeson}}\ and\ \bibinfo {author} {\bibfnamefont {J.~O.}\ \bibnamefont
  {Richardson}},\ }\href {\doibase 10.1063/1.5143412} {\bibfield  {journal}
  {\bibinfo  {journal} {J. Chem. Phys.}\ }\textbf {\bibinfo {volume} {152}},\
  \bibinfo {pages} {084110} (\bibinfo {year} {2020})},\ \Eprint
  {http://arxiv.org/abs/1912.10906} {arXiv:1912.10906 [physics.chem-ph]}
  \BibitemShut {NoStop}%
\bibitem [{\citenamefont {Mannouch}\ and\ \citenamefont
  {Richardson}(2020{\natexlab{a}})}]{Mannouch2020a}%
  \BibitemOpen
  \bibfield  {author} {\bibinfo {author} {\bibfnamefont {J.~R.}\ \bibnamefont
  {Mannouch}}\ and\ \bibinfo {author} {\bibfnamefont {J.~O.}\ \bibnamefont
  {Richardson}},\ }\href@noop {} {\bibfield  {journal} {\bibinfo  {journal} {J.
  Chem. Phys.}\ }\textbf {\bibinfo {volume} {153}},\ \bibinfo {pages} {194109}
  (\bibinfo {year} {2020}{\natexlab{a}})}\BibitemShut {NoStop}%
\bibitem [{\citenamefont {Fetherolf}\ and\ \citenamefont
  {Berkelbach}(2017)}]{Fetherolf2017}%
  \BibitemOpen
  \bibfield  {author} {\bibinfo {author} {\bibfnamefont {J.~H.}\ \bibnamefont
  {Fetherolf}}\ and\ \bibinfo {author} {\bibfnamefont {T.~C.}\ \bibnamefont
  {Berkelbach}},\ }\href {\doibase 10.1063/1.5006824} {\bibfield  {journal}
  {\bibinfo  {journal} {J. Chem. Phys.}\ }\textbf {\bibinfo {volume} {147}},\
  \bibinfo {pages} {244109} (\bibinfo {year} {2017})}\BibitemShut {NoStop}%
\bibitem [{\citenamefont {Gao}\ and\ \citenamefont
  {Geva}(2020)}]{Gao2020_nonlin}%
  \BibitemOpen
  \bibfield  {author} {\bibinfo {author} {\bibfnamefont {X.}~\bibnamefont
  {Gao}}\ and\ \bibinfo {author} {\bibfnamefont {E.}~\bibnamefont {Geva}},\
  }\href@noop {} {\bibfield  {journal} {\bibinfo  {journal} {J. Chem. Theory
  Comput.}\ }\textbf {\bibinfo {volume} {16}},\ \bibinfo {pages} {6491}
  (\bibinfo {year} {2020})}\BibitemShut {NoStop}%
\bibitem [{\citenamefont {Provazza}\ \emph {et~al.}(2018)\citenamefont
  {Provazza}, \citenamefont {Segatta}, \citenamefont {Garavelli},\ and\
  \citenamefont {Coker}}]{Provazza2018_nonlin}%
  \BibitemOpen
  \bibfield  {author} {\bibinfo {author} {\bibfnamefont {J.}~\bibnamefont
  {Provazza}}, \bibinfo {author} {\bibfnamefont {F.}~\bibnamefont {Segatta}},
  \bibinfo {author} {\bibfnamefont {M.}~\bibnamefont {Garavelli}}, \ and\
  \bibinfo {author} {\bibfnamefont {D.~F.}\ \bibnamefont {Coker}},\ }\href@noop
  {} {\bibfield  {journal} {\bibinfo  {journal} {J. Chem. Theory Comput.}\
  }\textbf {\bibinfo {volume} {14}},\ \bibinfo {pages} {856} (\bibinfo {year}
  {2018})}\BibitemShut {NoStop}%
\bibitem [{\citenamefont {Mannouch}\ and\ \citenamefont
  {Richardson}(2020{\natexlab{b}})}]{Mannouch2020b}%
  \BibitemOpen
  \bibfield  {author} {\bibinfo {author} {\bibfnamefont {J.~R.}\ \bibnamefont
  {Mannouch}}\ and\ \bibinfo {author} {\bibfnamefont {J.~O.}\ \bibnamefont
  {Richardson}},\ }\href@noop {} {\bibfield  {journal} {\bibinfo  {journal} {J.
  Chem. Phys.}\ }\textbf {\bibinfo {volume} {153}},\ \bibinfo {pages} {194110}
  (\bibinfo {year} {2020}{\natexlab{b}})}\BibitemShut {NoStop}%
\bibitem [{\citenamefont {Provazza}\ and\ \citenamefont
  {Coker}(2018)}]{Provazza2018_lin}%
  \BibitemOpen
  \bibfield  {author} {\bibinfo {author} {\bibfnamefont {J.}~\bibnamefont
  {Provazza}}\ and\ \bibinfo {author} {\bibfnamefont {D.~F.}\ \bibnamefont
  {Coker}},\ }\href@noop {} {\bibfield  {journal} {\bibinfo  {journal} {J.
  Chem. Phys.}\ }\textbf {\bibinfo {volume} {148}},\ \bibinfo {pages} {181102}
  (\bibinfo {year} {2018})}\BibitemShut {NoStop}%
\end{thebibliography}%


\begin{thebibliography}{113}%
\makeatletter
\providecommand \@ifxundefined [1]{%
 \@ifx{#1\undefined}
}%
\providecommand \@ifnum [1]{%
 \ifnum #1\expandafter \@firstoftwo
 \else \expandafter \@secondoftwo
 \fi
}%
\providecommand \@ifx [1]{%
 \ifx #1\expandafter \@firstoftwo
 \else \expandafter \@secondoftwo
 \fi
}%
\providecommand \natexlab [1]{#1}%
\providecommand \enquote  [1]{``#1''}%
\providecommand \bibnamefont  [1]{#1}%
\providecommand \bibfnamefont [1]{#1}%
\providecommand \citenamefont [1]{#1}%
\providecommand \href@noop [0]{\@secondoftwo}%
\providecommand \href [0]{\begingroup \@sanitize@url \@href}%
\providecommand \@href[1]{\@@startlink{#1}\@@href}%
\providecommand \@@href[1]{\endgroup#1\@@endlink}%
\providecommand \@sanitize@url [0]{\catcode `\\12\catcode `\$12\catcode
  `\&12\catcode `\#12\catcode `\^12\catcode `\_12\catcode `\%12\relax}%
\providecommand \@@startlink[1]{}%
\providecommand \@@endlink[0]{}%
\providecommand \url  [0]{\begingroup\@sanitize@url \@url }%
\providecommand \@url [1]{\endgroup\@href {#1}{\urlprefix }}%
\providecommand \urlprefix  [0]{URL }%
\providecommand \Eprint [0]{\href }%
\providecommand \doibase [0]{http://dx.doi.org/}%
\providecommand \selectlanguage [0]{\@gobble}%
\providecommand \bibinfo  [0]{\@secondoftwo}%
\providecommand \bibfield  [0]{\@secondoftwo}%
\providecommand \translation [1]{[#1]}%
\providecommand \BibitemOpen [0]{}%
\providecommand \bibitemStop [0]{}%
\providecommand \bibitemNoStop [0]{.\EOS\space}%
\providecommand \EOS [0]{\spacefactor3000\relax}%
\providecommand \BibitemShut  [1]{\csname bibitem#1\endcsname}%
\let\auto@bib@innerbib\@empty
%</preamble>
\bibitem [{\citenamefont {Mukamel}(1995)}]{MukamelBook}%
  \BibitemOpen
  \bibfield  {author} {\bibinfo {author} {\bibfnamefont {S.}~\bibnamefont
  {Mukamel}},\ }\href@noop {} {\emph {\bibinfo {title} {Principles of Nonlinear
  Optical Spectroscopy}}}\ (\bibinfo  {publisher} {Oxford University Press},\
  \bibinfo {year} {1995})\BibitemShut {NoStop}%
\bibitem [{\citenamefont {Brixner}\ \emph {et~al.}(2005)\citenamefont
  {Brixner}, \citenamefont {Stenger}, \citenamefont {Vaswani}, \citenamefont
  {Cho}, \citenamefont {Blankenship},\ and\ \citenamefont
  {Fleming}}]{Brixner2005}%
  \BibitemOpen
  \bibfield  {author} {\bibinfo {author} {\bibfnamefont {T.}~\bibnamefont
  {Brixner}}, \bibinfo {author} {\bibfnamefont {J.}~\bibnamefont {Stenger}},
  \bibinfo {author} {\bibfnamefont {H.~M.}\ \bibnamefont {Vaswani}}, \bibinfo
  {author} {\bibfnamefont {M.}~\bibnamefont {Cho}}, \bibinfo {author}
  {\bibfnamefont {R.~E.}\ \bibnamefont {Blankenship}}, \ and\ \bibinfo {author}
  {\bibfnamefont {G.~R.}\ \bibnamefont {Fleming}},\ }\href@noop {} {\bibfield
  {journal} {\bibinfo  {journal} {Nature}\ }\textbf {\bibinfo {volume} {434}},\
  \bibinfo {pages} {625} (\bibinfo {year} {2005})}\BibitemShut {NoStop}%
\bibitem [{\citenamefont {Schlau-Cohen}, \citenamefont {Ishizaki},\ and\
  \citenamefont {Fleming}(2011)}]{Schlaucohen2011}%
  \BibitemOpen
  \bibfield  {author} {\bibinfo {author} {\bibfnamefont {G.~S.}\ \bibnamefont
  {Schlau-Cohen}}, \bibinfo {author} {\bibfnamefont {A.}~\bibnamefont
  {Ishizaki}}, \ and\ \bibinfo {author} {\bibfnamefont {G.~R.}\ \bibnamefont
  {Fleming}},\ }\href@noop {} {\bibfield  {journal} {\bibinfo  {journal} {Chem.
  Phys.}\ }\textbf {\bibinfo {volume} {386}},\ \bibinfo {pages} {1} (\bibinfo
  {year} {2011})}\BibitemShut {NoStop}%
\bibitem [{\citenamefont {Read}\ \emph {et~al.}(2007)\citenamefont {Read},
  \citenamefont {Engel}, \citenamefont {Calhoun}, \citenamefont {Man{\v c}al},
  \citenamefont {Ahn}, \citenamefont {Blankenship},\ and\ \citenamefont
  {Fleming}}]{Read2007}%
  \BibitemOpen
  \bibfield  {author} {\bibinfo {author} {\bibfnamefont {E.~L.}\ \bibnamefont
  {Read}}, \bibinfo {author} {\bibfnamefont {G.~S.}\ \bibnamefont {Engel}},
  \bibinfo {author} {\bibfnamefont {T.~R.}\ \bibnamefont {Calhoun}}, \bibinfo
  {author} {\bibfnamefont {T.}~\bibnamefont {Man{\v c}al}}, \bibinfo {author}
  {\bibfnamefont {T.~K.}\ \bibnamefont {Ahn}}, \bibinfo {author} {\bibfnamefont
  {R.~E.}\ \bibnamefont {Blankenship}}, \ and\ \bibinfo {author} {\bibfnamefont
  {G.~R.}\ \bibnamefont {Fleming}},\ }\href@noop {} {\bibfield  {journal}
  {\bibinfo  {journal} {Proc. Natl. Acad. Sci. USA}\ }\textbf {\bibinfo
  {volume} {104}},\ \bibinfo {pages} {14203} (\bibinfo {year}
  {2007})}\BibitemShut {NoStop}%
\bibitem [{\citenamefont {Cho}\ \emph {et~al.}(2005)\citenamefont {Cho},
  \citenamefont {Vaswani}, \citenamefont {Brixner}, \citenamefont {Stenger},\
  and\ \citenamefont {Fleming}}]{Cho2005}%
  \BibitemOpen
  \bibfield  {author} {\bibinfo {author} {\bibfnamefont {M.}~\bibnamefont
  {Cho}}, \bibinfo {author} {\bibfnamefont {H.~M.}\ \bibnamefont {Vaswani}},
  \bibinfo {author} {\bibfnamefont {T.}~\bibnamefont {Brixner}}, \bibinfo
  {author} {\bibfnamefont {J.}~\bibnamefont {Stenger}}, \ and\ \bibinfo
  {author} {\bibfnamefont {G.~R.}\ \bibnamefont {Fleming}},\ }\href@noop {}
  {\bibfield  {journal} {\bibinfo  {journal} {J. Phys. Chem. B}\ }\textbf
  {\bibinfo {volume} {109}},\ \bibinfo {pages} {10542} (\bibinfo {year}
  {2005})}\BibitemShut {NoStop}%
\bibitem [{\citenamefont {Zigmantas}\ \emph {et~al.}(2006)\citenamefont
  {Zigmantas}, \citenamefont {Read}, \citenamefont {Man{\v c}al}, \citenamefont
  {Brixner}, \citenamefont {Gardiner}, \citenamefont {Cogdell},\ and\
  \citenamefont {Fleming}}]{Zigmantas2006}%
  \BibitemOpen
  \bibfield  {author} {\bibinfo {author} {\bibfnamefont {D.}~\bibnamefont
  {Zigmantas}}, \bibinfo {author} {\bibfnamefont {E.~L.}\ \bibnamefont {Read}},
  \bibinfo {author} {\bibfnamefont {T.}~\bibnamefont {Man{\v c}al}}, \bibinfo
  {author} {\bibfnamefont {T.}~\bibnamefont {Brixner}}, \bibinfo {author}
  {\bibfnamefont {A.~T.}\ \bibnamefont {Gardiner}}, \bibinfo {author}
  {\bibfnamefont {R.~J.}\ \bibnamefont {Cogdell}}, \ and\ \bibinfo {author}
  {\bibfnamefont {G.~R.}\ \bibnamefont {Fleming}},\ }\href@noop {} {\bibfield
  {journal} {\bibinfo  {journal} {Proc. Natl. Acad. Sci. USA}\ }\textbf
  {\bibinfo {volume} {103}},\ \bibinfo {pages} {12672} (\bibinfo {year}
  {2006})}\BibitemShut {NoStop}%
\bibitem [{\citenamefont {Schlau-Cohen}\ \emph {et~al.}(2009)\citenamefont
  {Schlau-Cohen}, \citenamefont {Calhoun}, \citenamefont {Ginsberg},
  \citenamefont {Read}, \citenamefont {Ballottari}, \citenamefont {Bassi},
  \citenamefont {van Grondelle},\ and\ \citenamefont
  {Fleming}}]{Schlaucohen2009}%
  \BibitemOpen
  \bibfield  {author} {\bibinfo {author} {\bibfnamefont {G.~S.}\ \bibnamefont
  {Schlau-Cohen}}, \bibinfo {author} {\bibfnamefont {T.~R.}\ \bibnamefont
  {Calhoun}}, \bibinfo {author} {\bibfnamefont {N.~S.}\ \bibnamefont
  {Ginsberg}}, \bibinfo {author} {\bibfnamefont {E.~L.}\ \bibnamefont {Read}},
  \bibinfo {author} {\bibfnamefont {M.}~\bibnamefont {Ballottari}}, \bibinfo
  {author} {\bibfnamefont {R.}~\bibnamefont {Bassi}}, \bibinfo {author}
  {\bibfnamefont {R.}~\bibnamefont {van Grondelle}}, \ and\ \bibinfo {author}
  {\bibfnamefont {G.~R.}\ \bibnamefont {Fleming}},\ }\href@noop {} {\bibfield
  {journal} {\bibinfo  {journal} {J. Phys. Chem. B}\ }\textbf {\bibinfo
  {volume} {113}},\ \bibinfo {pages} {15352} (\bibinfo {year}
  {2009})}\BibitemShut {NoStop}%
\bibitem [{\citenamefont {Myers}\ \emph {et~al.}(2010)\citenamefont {Myers},
  \citenamefont {Lewis}, \citenamefont {Fuller}, \citenamefont {Tekavec},
  \citenamefont {Yocum},\ and\ \citenamefont {Ogilvie}}]{Myers2010}%
  \BibitemOpen
  \bibfield  {author} {\bibinfo {author} {\bibfnamefont {J.~A.}\ \bibnamefont
  {Myers}}, \bibinfo {author} {\bibfnamefont {K.~L.~M.}\ \bibnamefont {Lewis}},
  \bibinfo {author} {\bibfnamefont {F.~D.}\ \bibnamefont {Fuller}}, \bibinfo
  {author} {\bibfnamefont {P.~F.}\ \bibnamefont {Tekavec}}, \bibinfo {author}
  {\bibfnamefont {C.~F.}\ \bibnamefont {Yocum}}, \ and\ \bibinfo {author}
  {\bibfnamefont {J.~P.}\ \bibnamefont {Ogilvie}},\ }\href@noop {} {\bibfield
  {journal} {\bibinfo  {journal} {J. Phys. Chem. Lett.}\ }\textbf {\bibinfo
  {volume} {1}},\ \bibinfo {pages} {2774} (\bibinfo {year} {2010})}\BibitemShut
  {NoStop}%
\bibitem [{\citenamefont {Engel}\ \emph {et~al.}(2007)\citenamefont {Engel},
  \citenamefont {Calhoun}, \citenamefont {Read}, \citenamefont {Ahn},
  \citenamefont {Man{\v c}al}, \citenamefont {Cheng}, \citenamefont
  {Blankenship},\ and\ \citenamefont {Fleming}}]{Engel2007}%
  \BibitemOpen
  \bibfield  {author} {\bibinfo {author} {\bibfnamefont {G.~S.}\ \bibnamefont
  {Engel}}, \bibinfo {author} {\bibfnamefont {T.~R.}\ \bibnamefont {Calhoun}},
  \bibinfo {author} {\bibfnamefont {E.~L.}\ \bibnamefont {Read}}, \bibinfo
  {author} {\bibfnamefont {T.-K.}\ \bibnamefont {Ahn}}, \bibinfo {author}
  {\bibfnamefont {T.}~\bibnamefont {Man{\v c}al}}, \bibinfo {author}
  {\bibfnamefont {Y.-C.}\ \bibnamefont {Cheng}}, \bibinfo {author}
  {\bibfnamefont {R.~E.}\ \bibnamefont {Blankenship}}, \ and\ \bibinfo {author}
  {\bibfnamefont {G.~R.}\ \bibnamefont {Fleming}},\ }\href@noop {} {\bibfield
  {journal} {\bibinfo  {journal} {Nature}\ }\textbf {\bibinfo {volume} {446}},\
  \bibinfo {pages} {782} (\bibinfo {year} {2007})}\BibitemShut {NoStop}%
\bibitem [{\citenamefont {Collini}\ \emph {et~al.}(2010)\citenamefont
  {Collini}, \citenamefont {Wong}, \citenamefont {Wilk}, \citenamefont {Curmi},
  \citenamefont {Brumer},\ and\ \citenamefont {Scholes}}]{Collini2010}%
  \BibitemOpen
  \bibfield  {author} {\bibinfo {author} {\bibfnamefont {E.}~\bibnamefont
  {Collini}}, \bibinfo {author} {\bibfnamefont {C.~Y.}\ \bibnamefont {Wong}},
  \bibinfo {author} {\bibfnamefont {K.~E.}\ \bibnamefont {Wilk}}, \bibinfo
  {author} {\bibfnamefont {P.~M.~G.}\ \bibnamefont {Curmi}}, \bibinfo {author}
  {\bibfnamefont {P.}~\bibnamefont {Brumer}}, \ and\ \bibinfo {author}
  {\bibfnamefont {G.~D.}\ \bibnamefont {Scholes}},\ }\href@noop {} {\bibfield
  {journal} {\bibinfo  {journal} {Nature}\ }\textbf {\bibinfo {volume} {463}},\
  \bibinfo {pages} {644} (\bibinfo {year} {2010})}\BibitemShut {NoStop}%
\bibitem [{\citenamefont {Lee}, \citenamefont {Cheng},\ and\ \citenamefont
  {Fleming}(2007)}]{Lee2007}%
  \BibitemOpen
  \bibfield  {author} {\bibinfo {author} {\bibfnamefont {H.}~\bibnamefont
  {Lee}}, \bibinfo {author} {\bibfnamefont {Y.-C.}\ \bibnamefont {Cheng}}, \
  and\ \bibinfo {author} {\bibfnamefont {G.~R.}\ \bibnamefont {Fleming}},\
  }\href@noop {} {\bibfield  {journal} {\bibinfo  {journal} {Science}\ }\textbf
  {\bibinfo {volume} {316}},\ \bibinfo {pages} {1462} (\bibinfo {year}
  {2007})}\BibitemShut {NoStop}%
\bibitem [{\citenamefont {Schlau-Cohen}\ \emph
  {et~al.}(2012{\natexlab{a}})\citenamefont {Schlau-Cohen}, \citenamefont
  {Ishizaki}, \citenamefont {Calhoun}, \citenamefont {Ginsberg}, \citenamefont
  {Ballottari}, \citenamefont {Bassi},\ and\ \citenamefont
  {Fleming}}]{Schlaucohen2012}%
  \BibitemOpen
  \bibfield  {author} {\bibinfo {author} {\bibfnamefont {G.~S.}\ \bibnamefont
  {Schlau-Cohen}}, \bibinfo {author} {\bibfnamefont {A.}~\bibnamefont
  {Ishizaki}}, \bibinfo {author} {\bibfnamefont {T.~R.}\ \bibnamefont
  {Calhoun}}, \bibinfo {author} {\bibfnamefont {N.~S.}\ \bibnamefont
  {Ginsberg}}, \bibinfo {author} {\bibfnamefont {M.}~\bibnamefont
  {Ballottari}}, \bibinfo {author} {\bibfnamefont {R.}~\bibnamefont {Bassi}}, \
  and\ \bibinfo {author} {\bibfnamefont {G.~R.}\ \bibnamefont {Fleming}},\
  }\href@noop {} {\bibfield  {journal} {\bibinfo  {journal} {Nature Chem.}\
  }\textbf {\bibinfo {volume} {4}},\ \bibinfo {pages} {389} (\bibinfo {year}
  {2012}{\natexlab{a}})}\BibitemShut {NoStop}%
\bibitem [{\citenamefont {Ishizaki}\ and\ \citenamefont
  {Fleming}(2012)}]{Ishizaki2012}%
  \BibitemOpen
  \bibfield  {author} {\bibinfo {author} {\bibfnamefont {A.}~\bibnamefont
  {Ishizaki}}\ and\ \bibinfo {author} {\bibfnamefont {G.~R.}\ \bibnamefont
  {Fleming}},\ }\href@noop {} {\bibfield  {journal} {\bibinfo  {journal} {Annu.
  Rev. Conden. Matter Phys.}\ }\textbf {\bibinfo {volume} {3}},\ \bibinfo
  {pages} {333} (\bibinfo {year} {2012})}\BibitemShut {NoStop}%
\bibitem [{\citenamefont {Read}\ \emph {et~al.}(2008)\citenamefont {Read},
  \citenamefont {Schlau-Cohen}, \citenamefont {Engel}, \citenamefont {Wen},
  \citenamefont {Blankenship},\ and\ \citenamefont {Fleming}}]{Read2008}%
  \BibitemOpen
  \bibfield  {author} {\bibinfo {author} {\bibfnamefont {E.~L.}\ \bibnamefont
  {Read}}, \bibinfo {author} {\bibfnamefont {G.~S.}\ \bibnamefont
  {Schlau-Cohen}}, \bibinfo {author} {\bibfnamefont {G.~S.}\ \bibnamefont
  {Engel}}, \bibinfo {author} {\bibfnamefont {J.}~\bibnamefont {Wen}}, \bibinfo
  {author} {\bibfnamefont {R.~E.}\ \bibnamefont {Blankenship}}, \ and\ \bibinfo
  {author} {\bibfnamefont {G.~R.}\ \bibnamefont {Fleming}},\ }\href@noop {}
  {\bibfield  {journal} {\bibinfo  {journal} {Biophysical J.}\ }\textbf
  {\bibinfo {volume} {95}},\ \bibinfo {pages} {847} (\bibinfo {year}
  {2008})}\BibitemShut {NoStop}%
\bibitem [{\citenamefont {Schlau-Cohen}\ \emph {et~al.}(2010)\citenamefont
  {Schlau-Cohen}, \citenamefont {Calhoun}, \citenamefont {Ginsberg},
  \citenamefont {Ballottari}, \citenamefont {Bassi},\ and\ \citenamefont
  {Fleming}}]{Schlaucohen2010}%
  \BibitemOpen
  \bibfield  {author} {\bibinfo {author} {\bibfnamefont {G.~S.}\ \bibnamefont
  {Schlau-Cohen}}, \bibinfo {author} {\bibfnamefont {T.~R.}\ \bibnamefont
  {Calhoun}}, \bibinfo {author} {\bibfnamefont {N.~S.}\ \bibnamefont
  {Ginsberg}}, \bibinfo {author} {\bibfnamefont {M.}~\bibnamefont
  {Ballottari}}, \bibinfo {author} {\bibfnamefont {R.}~\bibnamefont {Bassi}}, \
  and\ \bibinfo {author} {\bibfnamefont {G.~R.}\ \bibnamefont {Fleming}},\
  }\href@noop {} {\bibfield  {journal} {\bibinfo  {journal} {Proc. Natl. Acad.
  Sci. USA}\ }\textbf {\bibinfo {volume} {107}},\ \bibinfo {pages} {13276}
  (\bibinfo {year} {2010})}\BibitemShut {NoStop}%
\bibitem [{\citenamefont {Schlau-Cohen}\ \emph
  {et~al.}(2012{\natexlab{b}})\citenamefont {Schlau-Cohen}, \citenamefont
  {De~Re}, \citenamefont {Cogdell},\ and\ \citenamefont
  {Fleming}}]{Schlaucohen2012b}%
  \BibitemOpen
  \bibfield  {author} {\bibinfo {author} {\bibfnamefont {G.~S.}\ \bibnamefont
  {Schlau-Cohen}}, \bibinfo {author} {\bibfnamefont {E.}~\bibnamefont {De~Re}},
  \bibinfo {author} {\bibfnamefont {R.~J.}\ \bibnamefont {Cogdell}}, \ and\
  \bibinfo {author} {\bibfnamefont {G.~R.}\ \bibnamefont {Fleming}},\
  }\href@noop {} {\bibfield  {journal} {\bibinfo  {journal} {J Phys. Chem.
  Lett.}\ }\textbf {\bibinfo {volume} {3}},\ \bibinfo {pages} {2487} (\bibinfo
  {year} {2012}{\natexlab{b}})}\BibitemShut {NoStop}%
\bibitem [{\citenamefont {Gao}, \citenamefont {Lai},\ and\ \citenamefont
  {Geva}(2020)}]{Gao2020_lin}%
  \BibitemOpen
  \bibfield  {author} {\bibinfo {author} {\bibfnamefont {X.}~\bibnamefont
  {Gao}}, \bibinfo {author} {\bibfnamefont {Y.}~\bibnamefont {Lai}}, \ and\
  \bibinfo {author} {\bibfnamefont {E.}~\bibnamefont {Geva}},\ }\href@noop {}
  {\bibfield  {journal} {\bibinfo  {journal} {J. Chem. Theory Comput.}\
  }\textbf {\bibinfo {volume} {16}},\ \bibinfo {pages} {6465} (\bibinfo {year}
  {2020})}\BibitemShut {NoStop}%
\bibitem [{\citenamefont {Gao}\ and\ \citenamefont
  {Geva}(2020{\natexlab{a}})}]{Gao2020_nonlin}%
  \BibitemOpen
  \bibfield  {author} {\bibinfo {author} {\bibfnamefont {X.}~\bibnamefont
  {Gao}}\ and\ \bibinfo {author} {\bibfnamefont {E.}~\bibnamefont {Geva}},\
  }\href@noop {} {\bibfield  {journal} {\bibinfo  {journal} {J. Chem. Theory
  Comput.}\ }\textbf {\bibinfo {volume} {16}},\ \bibinfo {pages} {6491}
  (\bibinfo {year} {2020}{\natexlab{a}})}\BibitemShut {NoStop}%
\bibitem [{\citenamefont {Provazza}, \citenamefont {Segatta},\ and\
  \citenamefont {Coker}(2021)}]{Provazza2021}%
  \BibitemOpen
  \bibfield  {author} {\bibinfo {author} {\bibfnamefont {J.}~\bibnamefont
  {Provazza}}, \bibinfo {author} {\bibfnamefont {F.}~\bibnamefont {Segatta}}, \
  and\ \bibinfo {author} {\bibfnamefont {D.~F.}\ \bibnamefont {Coker}},\
  }\href@noop {} {\bibfield  {journal} {\bibinfo  {journal} {J. Chem. Theory
  Comput.}\ }\textbf {\bibinfo {volume} {17}},\ \bibinfo {pages} {29} (\bibinfo
  {year} {2021})}\BibitemShut {NoStop}%
\bibitem [{\citenamefont {Fetherolf}\ and\ \citenamefont
  {Berkelbach}(2017)}]{Fetherolf2017}%
  \BibitemOpen
  \bibfield  {author} {\bibinfo {author} {\bibfnamefont {J.~H.}\ \bibnamefont
  {Fetherolf}}\ and\ \bibinfo {author} {\bibfnamefont {T.~C.}\ \bibnamefont
  {Berkelbach}},\ }\href {\doibase 10.1063/1.5006824} {\bibfield  {journal}
  {\bibinfo  {journal} {J. Chem. Phys.}\ }\textbf {\bibinfo {volume} {147}},\
  \bibinfo {pages} {244109} (\bibinfo {year} {2017})}\BibitemShut {NoStop}%
\bibitem [{\citenamefont {Tempelaar}\ \emph {et~al.}(2013)\citenamefont
  {Tempelaar}, \citenamefont {van~der Vegte}, \citenamefont {Knoester},\ and\
  \citenamefont {Jansen}}]{Tempelaar2013}%
  \BibitemOpen
  \bibfield  {author} {\bibinfo {author} {\bibfnamefont {R.}~\bibnamefont
  {Tempelaar}}, \bibinfo {author} {\bibfnamefont {C.~P.}\ \bibnamefont {van~der
  Vegte}}, \bibinfo {author} {\bibfnamefont {J.}~\bibnamefont {Knoester}}, \
  and\ \bibinfo {author} {\bibfnamefont {T.~L.~C.}\ \bibnamefont {Jansen}},\
  }\href@noop {} {\bibfield  {journal} {\bibinfo  {journal} {J. Chem. Phys.}\
  }\textbf {\bibinfo {volume} {138}},\ \bibinfo {pages} {164106} (\bibinfo
  {year} {2013})}\BibitemShut {NoStop}%
\bibitem [{\citenamefont {van~der Vegte}\ \emph {et~al.}(2013)\citenamefont
  {van~der Vegte}, \citenamefont {Dijkstra}, \citenamefont {Knoester},\ and\
  \citenamefont {Jansen}}]{Vegte2013}%
  \BibitemOpen
  \bibfield  {author} {\bibinfo {author} {\bibfnamefont {C.~P.}\ \bibnamefont
  {van~der Vegte}}, \bibinfo {author} {\bibfnamefont {A.~G.}\ \bibnamefont
  {Dijkstra}}, \bibinfo {author} {\bibfnamefont {J.}~\bibnamefont {Knoester}},
  \ and\ \bibinfo {author} {\bibfnamefont {T.~L.~C.}\ \bibnamefont {Jansen}},\
  }\href@noop {} {\bibfield  {journal} {\bibinfo  {journal} {J. Phys. Chem. A}\
  }\textbf {\bibinfo {volume} {117}},\ \bibinfo {pages} {5970} (\bibinfo {year}
  {2013})}\BibitemShut {NoStop}%
\bibitem [{\citenamefont {Jain}\ \emph {et~al.}(2019)\citenamefont {Jain},
  \citenamefont {Petit}, \citenamefont {Anna},\ and\ \citenamefont
  {Subotnik}}]{Jain2019}%
  \BibitemOpen
  \bibfield  {author} {\bibinfo {author} {\bibfnamefont {A.}~\bibnamefont
  {Jain}}, \bibinfo {author} {\bibfnamefont {A.~S.}\ \bibnamefont {Petit}},
  \bibinfo {author} {\bibfnamefont {J.~M.}\ \bibnamefont {Anna}}, \ and\
  \bibinfo {author} {\bibfnamefont {J.~E.}\ \bibnamefont {Subotnik}},\
  }\href@noop {} {\bibfield  {journal} {\bibinfo  {journal} {J. Phys. Chem. B}\
  }\textbf {\bibinfo {volume} {123}},\ \bibinfo {pages} {1602} (\bibinfo {year}
  {2019})}\BibitemShut {NoStop}%
\bibitem [{\citenamefont {Huang}\ \emph {et~al.}(2021)\citenamefont {Huang},
  \citenamefont {Xie}, \citenamefont {Došlić}, \citenamefont {Gelin},\ and\
  \citenamefont {Domcke}}]{Huang2021}%
  \BibitemOpen
  \bibfield  {author} {\bibinfo {author} {\bibfnamefont {X.}~\bibnamefont
  {Huang}}, \bibinfo {author} {\bibfnamefont {W.}~\bibnamefont {Xie}}, \bibinfo
  {author} {\bibfnamefont {N.}~\bibnamefont {Došlić}}, \bibinfo {author}
  {\bibfnamefont {M.~F.}\ \bibnamefont {Gelin}}, \ and\ \bibinfo {author}
  {\bibfnamefont {W.}~\bibnamefont {Domcke}},\ }\href@noop {} {\bibfield
  {journal} {\bibinfo  {journal} {J. Phys. Chem. Lett.}\ }\textbf {\bibinfo
  {volume} {12}},\ \bibinfo {pages} {11736} (\bibinfo {year}
  {2021})}\BibitemShut {NoStop}%
\bibitem [{\citenamefont {Hanna}\ and\ \citenamefont {Geva}(2008)}]{Hanna2008}%
  \BibitemOpen
  \bibfield  {author} {\bibinfo {author} {\bibfnamefont {G.}~\bibnamefont
  {Hanna}}\ and\ \bibinfo {author} {\bibfnamefont {E.}~\bibnamefont {Geva}},\
  }\href@noop {} {\bibfield  {journal} {\bibinfo  {journal} {J. Phys. Chem. B}\
  }\textbf {\bibinfo {volume} {112}},\ \bibinfo {pages} {15793} (\bibinfo
  {year} {2008})}\BibitemShut {NoStop}%
\bibitem [{\citenamefont {Hanna}\ and\ \citenamefont {Geva}(2009)}]{Hanna2009}%
  \BibitemOpen
  \bibfield  {author} {\bibinfo {author} {\bibfnamefont {G.}~\bibnamefont
  {Hanna}}\ and\ \bibinfo {author} {\bibfnamefont {E.}~\bibnamefont {Geva}},\
  }\href@noop {} {\bibfield  {journal} {\bibinfo  {journal} {J. Phys. Chem. B}\
  }\textbf {\bibinfo {volume} {113}},\ \bibinfo {pages} {9278} (\bibinfo {year}
  {2009})}\BibitemShut {NoStop}%
\bibitem [{\citenamefont {Hanna}\ and\ \citenamefont {Geva}(2011)}]{Hanna2011}%
  \BibitemOpen
  \bibfield  {author} {\bibinfo {author} {\bibfnamefont {G.}~\bibnamefont
  {Hanna}}\ and\ \bibinfo {author} {\bibfnamefont {E.}~\bibnamefont {Geva}},\
  }\href@noop {} {\bibfield  {journal} {\bibinfo  {journal} {J. Phys. Chem. B}\
  }\textbf {\bibinfo {volume} {115}},\ \bibinfo {pages} {5191} (\bibinfo {year}
  {2011})}\BibitemShut {NoStop}%
\bibitem [{\citenamefont {Bai}\ \emph {et~al.}(2014)\citenamefont {Bai},
  \citenamefont {Xie}, \citenamefont {Zhu},\ and\ \citenamefont
  {Shi}}]{Bai2014}%
  \BibitemOpen
  \bibfield  {author} {\bibinfo {author} {\bibfnamefont {S.}~\bibnamefont
  {Bai}}, \bibinfo {author} {\bibfnamefont {W.}~\bibnamefont {Xie}}, \bibinfo
  {author} {\bibfnamefont {L.}~\bibnamefont {Zhu}}, \ and\ \bibinfo {author}
  {\bibfnamefont {Q.}~\bibnamefont {Shi}},\ }\href@noop {} {\bibfield
  {journal} {\bibinfo  {journal} {J. Chem. Phys.}\ }\textbf {\bibinfo {volume}
  {140}},\ \bibinfo {pages} {084105} (\bibinfo {year} {2014})}\BibitemShut
  {NoStop}%
\bibitem [{\citenamefont {Hein}\ \emph {et~al.}(2012)\citenamefont {Hein},
  \citenamefont {Kreisbeck}, \citenamefont {Kramer},\ and\ \citenamefont
  {Rodr{\'{\i}}guez}}]{Hein2012}%
  \BibitemOpen
  \bibfield  {author} {\bibinfo {author} {\bibfnamefont {B.}~\bibnamefont
  {Hein}}, \bibinfo {author} {\bibfnamefont {C.}~\bibnamefont {Kreisbeck}},
  \bibinfo {author} {\bibfnamefont {T.}~\bibnamefont {Kramer}}, \ and\ \bibinfo
  {author} {\bibfnamefont {M.}~\bibnamefont {Rodr{\'{\i}}guez}},\ }\href@noop
  {} {\bibfield  {journal} {\bibinfo  {journal} {New J. Phys.}\ }\textbf
  {\bibinfo {volume} {14}},\ \bibinfo {pages} {023018} (\bibinfo {year}
  {2012})}\BibitemShut {NoStop}%
\bibitem [{\citenamefont {Kramer}, \citenamefont {Rodríguez},\ and\
  \citenamefont {Zelinskyy}(2017)}]{Kramer2017}%
  \BibitemOpen
  \bibfield  {author} {\bibinfo {author} {\bibfnamefont {T.}~\bibnamefont
  {Kramer}}, \bibinfo {author} {\bibfnamefont {M.}~\bibnamefont {Rodríguez}},
  \ and\ \bibinfo {author} {\bibfnamefont {Y.}~\bibnamefont {Zelinskyy}},\
  }\href@noop {} {\bibfield  {journal} {\bibinfo  {journal} {J. Phys. Chem. B}\
  }\textbf {\bibinfo {volume} {121}},\ \bibinfo {pages} {463} (\bibinfo {year}
  {2017})}\BibitemShut {NoStop}%
\bibitem [{\citenamefont {Kramer}\ \emph {et~al.}(2018)\citenamefont {Kramer},
  \citenamefont {Noack}, \citenamefont {Reinefeld}, \citenamefont
  {Rodríguez},\ and\ \citenamefont {Zelinskyy}}]{Kramer2018}%
  \BibitemOpen
  \bibfield  {author} {\bibinfo {author} {\bibfnamefont {T.}~\bibnamefont
  {Kramer}}, \bibinfo {author} {\bibfnamefont {M.}~\bibnamefont {Noack}},
  \bibinfo {author} {\bibfnamefont {A.}~\bibnamefont {Reinefeld}}, \bibinfo
  {author} {\bibfnamefont {M.}~\bibnamefont {Rodríguez}}, \ and\ \bibinfo
  {author} {\bibfnamefont {Y.}~\bibnamefont {Zelinskyy}},\ }\href@noop {}
  {\bibfield  {journal} {\bibinfo  {journal} {J. Comput. Chem.}\ }\textbf
  {\bibinfo {volume} {39}},\ \bibinfo {pages} {1779} (\bibinfo {year}
  {2018})}\BibitemShut {NoStop}%
\bibitem [{\citenamefont {Tanimura}(2020)}]{Tanimura2020HEOM}%
  \BibitemOpen
  \bibfield  {author} {\bibinfo {author} {\bibfnamefont {Y.}~\bibnamefont
  {Tanimura}},\ }\href@noop {} {\bibfield  {journal} {\bibinfo  {journal} {J.
  Chem. Phys.}\ }\textbf {\bibinfo {volume} {153}},\ \bibinfo {pages} {020901}
  (\bibinfo {year} {2020})}\BibitemShut {NoStop}%
\bibitem [{\citenamefont {Egorov}, \citenamefont {Rabani},\ and\ \citenamefont
  {Berne}(1998)}]{Egorov1998vibronic}%
  \BibitemOpen
  \bibfield  {author} {\bibinfo {author} {\bibfnamefont {S.~A.}\ \bibnamefont
  {Egorov}}, \bibinfo {author} {\bibfnamefont {E.}~\bibnamefont {Rabani}}, \
  and\ \bibinfo {author} {\bibfnamefont {B.~J.}\ \bibnamefont {Berne}},\ }\href
  {\doibase 10.1063/1.475512} {\bibfield  {journal} {\bibinfo  {journal}
  {J.~Chem. Phys.}\ }\textbf {\bibinfo {volume} {108}},\ \bibinfo {pages}
  {1407} (\bibinfo {year} {1998})}\BibitemShut {NoStop}%
\bibitem [{\citenamefont {Rabani}, \citenamefont {Egorov},\ and\ \citenamefont
  {Berne}(1998)}]{Rabani1998vibronic}%
  \BibitemOpen
  \bibfield  {author} {\bibinfo {author} {\bibfnamefont {E.}~\bibnamefont
  {Rabani}}, \bibinfo {author} {\bibfnamefont {S.~A.}\ \bibnamefont {Egorov}},
  \ and\ \bibinfo {author} {\bibfnamefont {B.~J.}\ \bibnamefont {Berne}},\
  }\href {\doibase 10.1063/1.477280} {\bibfield  {journal} {\bibinfo  {journal}
  {J.~Chem. Phys.}\ }\textbf {\bibinfo {volume} {109}},\ \bibinfo {pages}
  {6376} (\bibinfo {year} {1998})}\BibitemShut {NoStop}%
\bibitem [{\citenamefont {Shi}\ and\ \citenamefont
  {Geva}(2008)}]{Shi2008nonlinear}%
  \BibitemOpen
  \bibfield  {author} {\bibinfo {author} {\bibfnamefont {Q.}~\bibnamefont
  {Shi}}\ and\ \bibinfo {author} {\bibfnamefont {E.}~\bibnamefont {Geva}},\
  }\href@noop {} {\bibfield  {journal} {\bibinfo  {journal} {J.~Chem. Phys.}\
  }\textbf {\bibinfo {volume} {129}},\ \bibinfo {pages} {124505} (\bibinfo
  {year} {2008})}\BibitemShut {NoStop}%
\bibitem [{\citenamefont {Karsten}\ \emph {et~al.}(2018)\citenamefont
  {Karsten}, \citenamefont {Ivanov}, \citenamefont {Bokarev},\ and\
  \citenamefont {K{\"u}hn}}]{Karsten2018vibronic}%
  \BibitemOpen
  \bibfield  {author} {\bibinfo {author} {\bibfnamefont {S.}~\bibnamefont
  {Karsten}}, \bibinfo {author} {\bibfnamefont {S.~D.}\ \bibnamefont {Ivanov}},
  \bibinfo {author} {\bibfnamefont {S.~I.}\ \bibnamefont {Bokarev}}, \ and\
  \bibinfo {author} {\bibfnamefont {O.}~\bibnamefont {K{\"u}hn}},\ }\href
  {\doibase 10.1063/1.5011764} {\bibfield  {journal} {\bibinfo  {journal}
  {J.~Chem. Phys.}\ }\textbf {\bibinfo {volume} {148}},\ \bibinfo {pages}
  {102337} (\bibinfo {year} {2018})}\BibitemShut {NoStop}%
\bibitem [{\citenamefont {McRobbie}\ and\ \citenamefont
  {Geva}(2009)}]{McRobbie2009nonlinear}%
  \BibitemOpen
  \bibfield  {author} {\bibinfo {author} {\bibfnamefont {P.~L.}\ \bibnamefont
  {McRobbie}}\ and\ \bibinfo {author} {\bibfnamefont {E.}~\bibnamefont
  {Geva}},\ }\href@noop {} {\bibfield  {journal} {\bibinfo  {journal} {J.~Phys.
  Chem.~A}\ }\textbf {\bibinfo {volume} {113}},\ \bibinfo {pages} {10425}
  (\bibinfo {year} {2009})}\BibitemShut {NoStop}%
\bibitem [{\citenamefont {Egorov}, \citenamefont {Rabani},\ and\ \citenamefont
  {Berne}(1999)}]{Egorov1999}%
  \BibitemOpen
  \bibfield  {author} {\bibinfo {author} {\bibfnamefont {S.~A.}\ \bibnamefont
  {Egorov}}, \bibinfo {author} {\bibfnamefont {E.}~\bibnamefont {Rabani}}, \
  and\ \bibinfo {author} {\bibfnamefont {B.~J.}\ \bibnamefont {Berne}},\
  }\href@noop {} {\bibfield  {journal} {\bibinfo  {journal} {J. Chem. Phys.}\
  }\textbf {\bibinfo {volume} {110}},\ \bibinfo {pages} {5238} (\bibinfo {year}
  {1999})}\BibitemShut {NoStop}%
\bibitem [{\citenamefont {Shi}\ and\ \citenamefont
  {Geva}(2004{\natexlab{a}})}]{Shi2004goldenrule}%
  \BibitemOpen
  \bibfield  {author} {\bibinfo {author} {\bibfnamefont {Q.}~\bibnamefont
  {Shi}}\ and\ \bibinfo {author} {\bibfnamefont {E.}~\bibnamefont {Geva}},\
  }\href {\doibase 10.1021/jp049547g} {\bibfield  {journal} {\bibinfo
  {journal} {J.~Phys. Chem.~A}\ }\textbf {\bibinfo {volume} {108}},\ \bibinfo
  {pages} {6109} (\bibinfo {year} {2004}{\natexlab{a}})}\BibitemShut {NoStop}%
\bibitem [{\citenamefont {Shi}\ and\ \citenamefont
  {Geva}(2005)}]{Shi2005nonadiabatic}%
  \BibitemOpen
  \bibfield  {author} {\bibinfo {author} {\bibfnamefont {Q.}~\bibnamefont
  {Shi}}\ and\ \bibinfo {author} {\bibfnamefont {E.}~\bibnamefont {Geva}},\
  }\href@noop {} {\bibfield  {journal} {\bibinfo  {journal} {J.~Chem. Phys.}\
  }\textbf {\bibinfo {volume} {122}},\ \bibinfo {pages} {064506} (\bibinfo
  {year} {2005})}\BibitemShut {NoStop}%
\bibitem [{\citenamefont {Makri}\ and\ \citenamefont
  {Thompson}(1998)}]{Makri1998}%
  \BibitemOpen
  \bibfield  {author} {\bibinfo {author} {\bibfnamefont {N.}~\bibnamefont
  {Makri}}\ and\ \bibinfo {author} {\bibfnamefont {K.}~\bibnamefont
  {Thompson}},\ }\href@noop {} {\bibfield  {journal} {\bibinfo  {journal}
  {Chem. Phys. Lett.}\ }\textbf {\bibinfo {volume} {291}},\ \bibinfo {pages}
  {101} (\bibinfo {year} {1998})}\BibitemShut {NoStop}%
\bibitem [{\citenamefont {Sun}\ and\ \citenamefont {Miller}(1999)}]{Sun1999}%
  \BibitemOpen
  \bibfield  {author} {\bibinfo {author} {\bibfnamefont {X.}~\bibnamefont
  {Sun}}\ and\ \bibinfo {author} {\bibfnamefont {W.~H.}\ \bibnamefont
  {Miller}},\ }\href@noop {} {\bibfield  {journal} {\bibinfo  {journal} {J.
  Chem. Phys.}\ }\textbf {\bibinfo {volume} {110}},\ \bibinfo {pages} {6635}
  (\bibinfo {year} {1999})}\BibitemShut {NoStop}%
\bibitem [{\citenamefont {Herman}\ and\ \citenamefont
  {Kluk}(1984)}]{Herman+Kluk1984}%
  \BibitemOpen
  \bibfield  {author} {\bibinfo {author} {\bibfnamefont {M.~F.}\ \bibnamefont
  {Herman}}\ and\ \bibinfo {author} {\bibfnamefont {E.}~\bibnamefont {Kluk}},\
  }\href@noop {} {\bibfield  {journal} {\bibinfo  {journal} {Chem. Phys.}\
  }\textbf {\bibinfo {volume} {91}},\ \bibinfo {pages} {27} (\bibinfo {year}
  {1984})}\BibitemShut {NoStop}%
\bibitem [{\citenamefont {Walton}\ and\ \citenamefont
  {Manolopoulos}(1995)}]{Walton1995}%
  \BibitemOpen
  \bibfield  {author} {\bibinfo {author} {\bibfnamefont {A.~R.}\ \bibnamefont
  {Walton}}\ and\ \bibinfo {author} {\bibfnamefont {D.~E.}\ \bibnamefont
  {Manolopoulos}},\ }\href@noop {} {\bibfield  {journal} {\bibinfo  {journal}
  {Chem. Phys. Lett.}\ }\textbf {\bibinfo {volume} {244}},\ \bibinfo {pages}
  {448} (\bibinfo {year} {1995})}\BibitemShut {NoStop}%
\bibitem [{\citenamefont {Meyer}\ and\ \citenamefont
  {Miller}(1979)}]{Meyer1979nonadiabatic}%
  \BibitemOpen
  \bibfield  {author} {\bibinfo {author} {\bibfnamefont {H.-D.}\ \bibnamefont
  {Meyer}}\ and\ \bibinfo {author} {\bibfnamefont {W.~H.}\ \bibnamefont
  {Miller}},\ }\href {\doibase 10.1063/1.437910} {\bibfield  {journal}
  {\bibinfo  {journal} {J.~Chem. Phys.}\ }\textbf {\bibinfo {volume} {70}},\
  \bibinfo {pages} {3214} (\bibinfo {year} {1979})}\BibitemShut {NoStop}%
\bibitem [{\citenamefont {Stock}\ and\ \citenamefont
  {Thoss}(1997)}]{Stock1997mapping}%
  \BibitemOpen
  \bibfield  {author} {\bibinfo {author} {\bibfnamefont {G.}~\bibnamefont
  {Stock}}\ and\ \bibinfo {author} {\bibfnamefont {M.}~\bibnamefont {Thoss}},\
  }\href {\doibase 10.1103/PhysRevLett.78.578} {\bibfield  {journal} {\bibinfo
  {journal} {Phys. Rev. Lett.}\ }\textbf {\bibinfo {volume} {78}},\ \bibinfo
  {pages} {578} (\bibinfo {year} {1997})}\BibitemShut {NoStop}%
\bibitem [{\citenamefont {Saller}, \citenamefont {Kelly},\ and\ \citenamefont
  {Richardson}(2019)}]{identity}%
  \BibitemOpen
  \bibfield  {author} {\bibinfo {author} {\bibfnamefont {M.~A.~C.}\
  \bibnamefont {Saller}}, \bibinfo {author} {\bibfnamefont {A.}~\bibnamefont
  {Kelly}}, \ and\ \bibinfo {author} {\bibfnamefont {J.~O.}\ \bibnamefont
  {Richardson}},\ }\href {\doibase 10.1063/1.5082596} {\bibfield  {journal}
  {\bibinfo  {journal} {J. Chem. Phys.}\ }\textbf {\bibinfo {volume} {150}},\
  \bibinfo {pages} {071101} (\bibinfo {year} {2019})},\ \Eprint
  {http://arxiv.org/abs/1811.08830} {arXiv:1811.08830 [physics.chem-ph]}
  \BibitemShut {NoStop}%
\bibitem [{\citenamefont {Saller}, \citenamefont {Kelly},\ and\ \citenamefont
  {Richardson}(2020)}]{FMO}%
  \BibitemOpen
  \bibfield  {author} {\bibinfo {author} {\bibfnamefont {M.~A.~C.}\
  \bibnamefont {Saller}}, \bibinfo {author} {\bibfnamefont {A.}~\bibnamefont
  {Kelly}}, \ and\ \bibinfo {author} {\bibfnamefont {J.~O.}\ \bibnamefont
  {Richardson}},\ }\href {\doibase 10.1039/C9FD00050J} {\bibfield  {journal}
  {\bibinfo  {journal} {Faraday Discuss.}\ }\textbf {\bibinfo {volume} {221}},\
  \bibinfo {pages} {150} (\bibinfo {year} {2020})},\ \Eprint
  {http://arxiv.org/abs/1904.11847} {arXiv:1904.11847 [physics.chem-ph]}
  \BibitemShut {NoStop}%
\bibitem [{\citenamefont {Gao}\ \emph {et~al.}(2020)\citenamefont {Gao},
  \citenamefont {Saller}, \citenamefont {Liu}, \citenamefont {Kelly},
  \citenamefont {Richardson},\ and\ \citenamefont {Geva}}]{linearized}%
  \BibitemOpen
  \bibfield  {author} {\bibinfo {author} {\bibfnamefont {X.}~\bibnamefont
  {Gao}}, \bibinfo {author} {\bibfnamefont {M.~A.~C.}\ \bibnamefont {Saller}},
  \bibinfo {author} {\bibfnamefont {Y.}~\bibnamefont {Liu}}, \bibinfo {author}
  {\bibfnamefont {A.}~\bibnamefont {Kelly}}, \bibinfo {author} {\bibfnamefont
  {J.~O.}\ \bibnamefont {Richardson}}, \ and\ \bibinfo {author} {\bibfnamefont
  {E.}~\bibnamefont {Geva}},\ }\href {\doibase 10.1021/acs.jctc.9b01267}
  {\bibfield  {journal} {\bibinfo  {journal} {J.~Chem.~Theory Comput.}\
  }\textbf {\bibinfo {volume} {16}},\ \bibinfo {pages} {2883} (\bibinfo {year}
  {2020})}\BibitemShut {NoStop}%
\bibitem [{\citenamefont {M{\"u}ller}\ and\ \citenamefont
  {Stock}(1999)}]{Mueller1999pyrazine}%
  \BibitemOpen
  \bibfield  {author} {\bibinfo {author} {\bibfnamefont {U.}~\bibnamefont
  {M{\"u}ller}}\ and\ \bibinfo {author} {\bibfnamefont {G.}~\bibnamefont
  {Stock}},\ }\href {\doibase 10.1063/1.479255} {\bibfield  {journal} {\bibinfo
   {journal} {J.~Chem. Phys.}\ }\textbf {\bibinfo {volume} {111}},\ \bibinfo
  {pages} {77} (\bibinfo {year} {1999})}\BibitemShut {NoStop}%
\bibitem [{\citenamefont {Gao}\ and\ \citenamefont
  {Geva}(2020{\natexlab{b}})}]{Gao2020mapping}%
  \BibitemOpen
  \bibfield  {author} {\bibinfo {author} {\bibfnamefont {X.}~\bibnamefont
  {Gao}}\ and\ \bibinfo {author} {\bibfnamefont {E.}~\bibnamefont {Geva}},\
  }\href@noop {} {\bibfield  {journal} {\bibinfo  {journal} {J. Phys. Chem. A}\
  } (\bibinfo {year} {2020}{\natexlab{b}})}\BibitemShut {NoStop}%
\bibitem [{\citenamefont {Shi}\ and\ \citenamefont
  {Geva}(2004{\natexlab{b}})}]{Shi2004GQME}%
  \BibitemOpen
  \bibfield  {author} {\bibinfo {author} {\bibfnamefont {Q.}~\bibnamefont
  {Shi}}\ and\ \bibinfo {author} {\bibfnamefont {E.}~\bibnamefont {Geva}},\
  }\href@noop {} {\bibfield  {journal} {\bibinfo  {journal} {J.~Chem. Phys.}\
  }\textbf {\bibinfo {volume} {120}},\ \bibinfo {pages} {10647} (\bibinfo
  {year} {2004}{\natexlab{b}})}\BibitemShut {NoStop}%
\bibitem [{\citenamefont {Kelly}\ \emph {et~al.}(2016)\citenamefont {Kelly},
  \citenamefont {Montoya-Castillo}, \citenamefont {Wang},\ and\ \citenamefont
  {Markland}}]{Kelly2016master}%
  \BibitemOpen
  \bibfield  {author} {\bibinfo {author} {\bibfnamefont {A.}~\bibnamefont
  {Kelly}}, \bibinfo {author} {\bibfnamefont {A.}~\bibnamefont
  {Montoya-Castillo}}, \bibinfo {author} {\bibfnamefont {L.}~\bibnamefont
  {Wang}}, \ and\ \bibinfo {author} {\bibfnamefont {T.~E.}\ \bibnamefont
  {Markland}},\ }\href {\doibase 10.1063/1.4948612} {\bibfield  {journal}
  {\bibinfo  {journal} {J.~Chem. Phys.}\ }\textbf {\bibinfo {volume} {144}},\
  \bibinfo {pages} {184105} (\bibinfo {year} {2016})}\BibitemShut {NoStop}%
\bibitem [{\citenamefont {Pfalzgraff}\ \emph {et~al.}(2019)\citenamefont
  {Pfalzgraff}, \citenamefont {Montoya-Castillo}, \citenamefont {Kelly},\ and\
  \citenamefont {Markland}}]{Pfalzgraff2019GQME}%
  \BibitemOpen
  \bibfield  {author} {\bibinfo {author} {\bibfnamefont {W.~C.}\ \bibnamefont
  {Pfalzgraff}}, \bibinfo {author} {\bibfnamefont {A.}~\bibnamefont
  {Montoya-Castillo}}, \bibinfo {author} {\bibfnamefont {A.}~\bibnamefont
  {Kelly}}, \ and\ \bibinfo {author} {\bibfnamefont {T.~E.}\ \bibnamefont
  {Markland}},\ }\href@noop {} {\bibfield  {journal} {\bibinfo  {journal}
  {J.~Chem. Phys.}\ }\textbf {\bibinfo {volume} {150}},\ \bibinfo {pages}
  {244109} (\bibinfo {year} {2019})}\BibitemShut {NoStop}%
\bibitem [{\citenamefont {Mulvihill}\ \emph {et~al.}(2019)\citenamefont
  {Mulvihill}, \citenamefont {Gao}, \citenamefont {Liu}, \citenamefont
  {Schubert}, \citenamefont {Dunietz},\ and\ \citenamefont
  {Geva}}]{Mulvihill2019LSCGQME}%
  \BibitemOpen
  \bibfield  {author} {\bibinfo {author} {\bibfnamefont {E.}~\bibnamefont
  {Mulvihill}}, \bibinfo {author} {\bibfnamefont {X.}~\bibnamefont {Gao}},
  \bibinfo {author} {\bibfnamefont {Y.}~\bibnamefont {Liu}}, \bibinfo {author}
  {\bibfnamefont {A.}~\bibnamefont {Schubert}}, \bibinfo {author}
  {\bibfnamefont {B.~D.}\ \bibnamefont {Dunietz}}, \ and\ \bibinfo {author}
  {\bibfnamefont {E.}~\bibnamefont {Geva}},\ }\href@noop {} {\bibfield
  {journal} {\bibinfo  {journal} {J.~Chem. Phys.}\ }\textbf {\bibinfo {volume}
  {151}},\ \bibinfo {pages} {074103} (\bibinfo {year} {2019})}\BibitemShut
  {NoStop}%
\bibitem [{\citenamefont {Richardson}\ and\ \citenamefont
  {Thoss}(2013)}]{mapping}%
  \BibitemOpen
  \bibfield  {author} {\bibinfo {author} {\bibfnamefont {J.~O.}\ \bibnamefont
  {Richardson}}\ and\ \bibinfo {author} {\bibfnamefont {M.}~\bibnamefont
  {Thoss}},\ }\href {\doibase 10.1063/1.4816124} {\bibfield  {journal}
  {\bibinfo  {journal} {J.~Chem. Phys.}\ }\textbf {\bibinfo {volume} {139}},\
  \bibinfo {pages} {031102} (\bibinfo {year} {2013})}\BibitemShut {NoStop}%
\bibitem [{\citenamefont {Richardson}\ \emph {et~al.}(2017)\citenamefont
  {Richardson}, \citenamefont {Meyer}, \citenamefont {Pleinert},\ and\
  \citenamefont {Thoss}}]{vibronic}%
  \BibitemOpen
  \bibfield  {author} {\bibinfo {author} {\bibfnamefont {J.~O.}\ \bibnamefont
  {Richardson}}, \bibinfo {author} {\bibfnamefont {P.}~\bibnamefont {Meyer}},
  \bibinfo {author} {\bibfnamefont {M.-O.}\ \bibnamefont {Pleinert}}, \ and\
  \bibinfo {author} {\bibfnamefont {M.}~\bibnamefont {Thoss}},\ }\href
  {\doibase 10.1016/j.chemphys.2016.09.036} {\bibfield  {journal} {\bibinfo
  {journal} {Chem. Phys.}\ }\textbf {\bibinfo {volume} {482}},\ \bibinfo
  {pages} {124} (\bibinfo {year} {2017})},\ \Eprint
  {http://arxiv.org/abs/1609.00644} {arXiv:1609.00644 [physics.chem-ph]}
  \BibitemShut {NoStop}%
\bibitem [{\citenamefont {Ananth}(2013)}]{Ananth2013MVRPMD}%
  \BibitemOpen
  \bibfield  {author} {\bibinfo {author} {\bibfnamefont {N.}~\bibnamefont
  {Ananth}},\ }\href {\doibase 10.1063/1.4821590} {\bibfield  {journal}
  {\bibinfo  {journal} {J.~Chem. Phys.}\ }\textbf {\bibinfo {volume} {139}},\
  \bibinfo {pages} {124102} (\bibinfo {year} {2013})}\BibitemShut {NoStop}%
\bibitem [{\citenamefont {Chowdhury}\ and\ \citenamefont
  {Huo}(2017)}]{Chowdhury2017CSRPMD}%
  \BibitemOpen
  \bibfield  {author} {\bibinfo {author} {\bibfnamefont {S.~N.}\ \bibnamefont
  {Chowdhury}}\ and\ \bibinfo {author} {\bibfnamefont {P.}~\bibnamefont
  {Huo}},\ }\href {\doibase 10.1063/1.4995616} {\bibfield  {journal} {\bibinfo
  {journal} {J.~Chem. Phys.}\ }\textbf {\bibinfo {volume} {147}},\ \bibinfo
  {pages} {214109} (\bibinfo {year} {2017})},\ \Eprint
  {http://arxiv.org/abs/1706.08403} {arXiv:1706.08403 [physics.chem-ph]}
  \BibitemShut {NoStop}%
\bibitem [{\citenamefont {Cotton}\ and\ \citenamefont
  {Miller}(2013{\natexlab{a}})}]{Cotton2013SQC}%
  \BibitemOpen
  \bibfield  {author} {\bibinfo {author} {\bibfnamefont {S.~J.}\ \bibnamefont
  {Cotton}}\ and\ \bibinfo {author} {\bibfnamefont {W.~H.}\ \bibnamefont
  {Miller}},\ }\href@noop {} {\bibfield  {journal} {\bibinfo  {journal}
  {J.~Phys. Chem.~A}\ }\textbf {\bibinfo {volume} {117}},\ \bibinfo {pages}
  {7190} (\bibinfo {year} {2013}{\natexlab{a}})}\BibitemShut {NoStop}%
\bibitem [{\citenamefont {Cotton}\ and\ \citenamefont
  {Miller}(2013{\natexlab{b}})}]{Cotton2013mapping}%
  \BibitemOpen
  \bibfield  {author} {\bibinfo {author} {\bibfnamefont {S.~J.}\ \bibnamefont
  {Cotton}}\ and\ \bibinfo {author} {\bibfnamefont {W.~H.}\ \bibnamefont
  {Miller}},\ }\href {\doibase 10.1063/1.4845235} {\bibfield  {journal}
  {\bibinfo  {journal} {J.~Chem. Phys.}\ }\textbf {\bibinfo {volume} {139}},\
  \bibinfo {pages} {234112} (\bibinfo {year} {2013}{\natexlab{b}})}\BibitemShut
  {NoStop}%
\bibitem [{\citenamefont {Miller}\ and\ \citenamefont
  {Cotton}(2016)}]{Miller2016Faraday}%
  \BibitemOpen
  \bibfield  {author} {\bibinfo {author} {\bibfnamefont {W.~H.}\ \bibnamefont
  {Miller}}\ and\ \bibinfo {author} {\bibfnamefont {S.~J.}\ \bibnamefont
  {Cotton}},\ }\href {\doibase 10.1039/C6FD00181E} {\bibfield  {journal}
  {\bibinfo  {journal} {Faraday Discuss.}\ }\textbf {\bibinfo {volume} {195}},\
  \bibinfo {pages} {9} (\bibinfo {year} {2016})}\BibitemShut {NoStop}%
\bibitem [{\citenamefont {Cotton}\ and\ \citenamefont
  {Miller}(2016{\natexlab{a}})}]{Cotton2016_2}%
  \BibitemOpen
  \bibfield  {author} {\bibinfo {author} {\bibfnamefont {S.~J.}\ \bibnamefont
  {Cotton}}\ and\ \bibinfo {author} {\bibfnamefont {W.~H.}\ \bibnamefont
  {Miller}},\ }\href@noop {} {\bibfield  {journal} {\bibinfo  {journal} {J.
  Chem. Theory Comput.}\ }\textbf {\bibinfo {volume} {12}},\ \bibinfo {pages}
  {983} (\bibinfo {year} {2016}{\natexlab{a}})}\BibitemShut {NoStop}%
\bibitem [{\citenamefont {Cotton}\ and\ \citenamefont
  {Miller}(2016{\natexlab{b}})}]{Cotton2016SQC}%
  \BibitemOpen
  \bibfield  {author} {\bibinfo {author} {\bibfnamefont {S.~J.}\ \bibnamefont
  {Cotton}}\ and\ \bibinfo {author} {\bibfnamefont {W.~H.}\ \bibnamefont
  {Miller}},\ }\href {\doibase 10.1063/1.4963914} {\bibfield  {journal}
  {\bibinfo  {journal} {J.~Chem. Phys}\ }\textbf {\bibinfo {volume} {145}},\
  \bibinfo {pages} {144108} (\bibinfo {year} {2016}{\natexlab{b}})}\BibitemShut
  {NoStop}%
\bibitem [{\citenamefont {Liang}\ \emph {et~al.}(2018)\citenamefont {Liang},
  \citenamefont {Cotton}, \citenamefont {Binder}, \citenamefont {Hegger},
  \citenamefont {Burghardt},\ and\ \citenamefont {Miller}}]{Liang2018}%
  \BibitemOpen
  \bibfield  {author} {\bibinfo {author} {\bibfnamefont {R.}~\bibnamefont
  {Liang}}, \bibinfo {author} {\bibfnamefont {S.~J.}\ \bibnamefont {Cotton}},
  \bibinfo {author} {\bibfnamefont {R.}~\bibnamefont {Binder}}, \bibinfo
  {author} {\bibfnamefont {R.}~\bibnamefont {Hegger}}, \bibinfo {author}
  {\bibfnamefont {I.}~\bibnamefont {Burghardt}}, \ and\ \bibinfo {author}
  {\bibfnamefont {W.~H.}\ \bibnamefont {Miller}},\ }\href@noop {} {\bibfield
  {journal} {\bibinfo  {journal} {J. Chem. Phys.}\ }\textbf {\bibinfo {volume}
  {149}},\ \bibinfo {pages} {044101} (\bibinfo {year} {2018})}\BibitemShut
  {NoStop}%
\bibitem [{\citenamefont {Cotton}\ and\ \citenamefont
  {Miller}(2019)}]{Cotton2019}%
  \BibitemOpen
  \bibfield  {author} {\bibinfo {author} {\bibfnamefont {S.~J.}\ \bibnamefont
  {Cotton}}\ and\ \bibinfo {author} {\bibfnamefont {W.~H.}\ \bibnamefont
  {Miller}},\ }\href@noop {} {\bibfield  {journal} {\bibinfo  {journal} {J.
  Chem, Phys.}\ }\textbf {\bibinfo {volume} {150}},\ \bibinfo {pages} {104101}
  (\bibinfo {year} {2019})}\BibitemShut {NoStop}%
\bibitem [{\citenamefont {Runeson}\ and\ \citenamefont
  {Richardson}(2019)}]{spinmap}%
  \BibitemOpen
  \bibfield  {author} {\bibinfo {author} {\bibfnamefont {J.~E.}\ \bibnamefont
  {Runeson}}\ and\ \bibinfo {author} {\bibfnamefont {J.~O.}\ \bibnamefont
  {Richardson}},\ }\href {\doibase 10.1063/1.5100506} {\bibfield  {journal}
  {\bibinfo  {journal} {J. Chem. Phys.}\ }\textbf {\bibinfo {volume} {151}},\
  \bibinfo {pages} {044119} (\bibinfo {year} {2019})},\ \Eprint
  {http://arxiv.org/abs/1904.08293} {arXiv:1904.08293 [physics.chem-ph]}
  \BibitemShut {NoStop}%
\bibitem [{\citenamefont {Runeson}\ and\ \citenamefont
  {Richardson}(2020)}]{multispin}%
  \BibitemOpen
  \bibfield  {author} {\bibinfo {author} {\bibfnamefont {J.~E.}\ \bibnamefont
  {Runeson}}\ and\ \bibinfo {author} {\bibfnamefont {J.~O.}\ \bibnamefont
  {Richardson}},\ }\href {\doibase 10.1063/1.5143412} {\bibfield  {journal}
  {\bibinfo  {journal} {J. Chem. Phys.}\ }\textbf {\bibinfo {volume} {152}},\
  \bibinfo {pages} {084110} (\bibinfo {year} {2020})},\ \Eprint
  {http://arxiv.org/abs/1912.10906} {arXiv:1912.10906 [physics.chem-ph]}
  \BibitemShut {NoStop}%
\bibitem [{\citenamefont {Liu}(2016)}]{Liu2016}%
  \BibitemOpen
  \bibfield  {author} {\bibinfo {author} {\bibfnamefont {J.}~\bibnamefont
  {Liu}},\ }\href@noop {} {\bibfield  {journal} {\bibinfo  {journal} {J. Chem.
  Phys.}\ }\textbf {\bibinfo {volume} {145}},\ \bibinfo {pages} {204105}
  (\bibinfo {year} {2016})}\BibitemShut {NoStop}%
\bibitem [{\citenamefont {He}\ and\ \citenamefont {Liu}(2019)}]{xin2019}%
  \BibitemOpen
  \bibfield  {author} {\bibinfo {author} {\bibfnamefont {X.}~\bibnamefont
  {He}}\ and\ \bibinfo {author} {\bibfnamefont {J.}~\bibnamefont {Liu}},\
  }\href@noop {} {\bibfield  {journal} {\bibinfo  {journal} {J. Chem. Phys.}\
  }\textbf {\bibinfo {volume} {151}},\ \bibinfo {pages} {024105} (\bibinfo
  {year} {2019})}\BibitemShut {NoStop}%
\bibitem [{\citenamefont {Kim}\ and\ \citenamefont
  {Rhee}(2014)}]{Kim2014mapping}%
  \BibitemOpen
  \bibfield  {author} {\bibinfo {author} {\bibfnamefont {H.~W.}\ \bibnamefont
  {Kim}}\ and\ \bibinfo {author} {\bibfnamefont {Y.~M.}\ \bibnamefont {Rhee}},\
  }\href@noop {} {\bibfield  {journal} {\bibinfo  {journal} {J. Chem. Phys.}\
  }\textbf {\bibinfo {volume} {140}},\ \bibinfo {pages} {184106} (\bibinfo
  {year} {2014})}\BibitemShut {NoStop}%
\bibitem [{\citenamefont {Miller}\ and\ \citenamefont
  {White}(1986)}]{Miller1986fermions}%
  \BibitemOpen
  \bibfield  {author} {\bibinfo {author} {\bibfnamefont {W.~H.}\ \bibnamefont
  {Miller}}\ and\ \bibinfo {author} {\bibfnamefont {K.~A.}\ \bibnamefont
  {White}},\ }\href@noop {} {\bibfield  {journal} {\bibinfo  {journal}
  {J.~Chem. Phys.}\ }\textbf {\bibinfo {volume} {84}},\ \bibinfo {pages} {5059}
  (\bibinfo {year} {1986})}\BibitemShut {NoStop}%
\bibitem [{\citenamefont {Li}\ and\ \citenamefont
  {Miller}(2012)}]{Li2012fermions}%
  \BibitemOpen
  \bibfield  {author} {\bibinfo {author} {\bibfnamefont {B.}~\bibnamefont
  {Li}}\ and\ \bibinfo {author} {\bibfnamefont {W.~H.}\ \bibnamefont
  {Miller}},\ }\href@noop {} {\bibfield  {journal} {\bibinfo  {journal}
  {J.~Chem. Phys.}\ }\textbf {\bibinfo {volume} {137}},\ \bibinfo {pages}
  {154107} (\bibinfo {year} {2012})}\BibitemShut {NoStop}%
\bibitem [{\citenamefont {Sun}, \citenamefont {Sasmal},\ and\ \citenamefont
  {Vendrell}(2021)}]{Sun2021}%
  \BibitemOpen
  \bibfield  {author} {\bibinfo {author} {\bibfnamefont {J.}~\bibnamefont
  {Sun}}, \bibinfo {author} {\bibfnamefont {S.}~\bibnamefont {Sasmal}}, \ and\
  \bibinfo {author} {\bibfnamefont {O.}~\bibnamefont {Vendrell}},\ }\href@noop
  {} {\bibfield  {journal} {\bibinfo  {journal} {J. Chem. Phys.}\ }\textbf
  {\bibinfo {volume} {155}},\ \bibinfo {pages} {134110} (\bibinfo {year}
  {2021})}\BibitemShut {NoStop}%
\bibitem [{\citenamefont {Lang}, \citenamefont {Vendrell},\ and\ \citenamefont
  {Hauke}(2021)}]{Lang2021GDTWA}%
  \BibitemOpen
  \bibfield  {author} {\bibinfo {author} {\bibfnamefont {H.}~\bibnamefont
  {Lang}}, \bibinfo {author} {\bibfnamefont {O.}~\bibnamefont {Vendrell}}, \
  and\ \bibinfo {author} {\bibfnamefont {P.}~\bibnamefont {Hauke}},\
  }\href@noop {} {\bibfield  {journal} {\bibinfo  {journal} {arXiv preprint
  arXiv:2104.07139}\ } (\bibinfo {year} {2021})}\BibitemShut {NoStop}%
\bibitem [{\citenamefont {Polley}\ and\ \citenamefont
  {Loring}(2019)}]{Polley2019}%
  \BibitemOpen
  \bibfield  {author} {\bibinfo {author} {\bibfnamefont {K.}~\bibnamefont
  {Polley}}\ and\ \bibinfo {author} {\bibfnamefont {R.~F.}\ \bibnamefont
  {Loring}},\ }\href@noop {} {\bibfield  {journal} {\bibinfo  {journal} {J.
  Chem. Phys.}\ }\textbf {\bibinfo {volume} {150}},\ \bibinfo {pages} {164114}
  (\bibinfo {year} {2019})}\BibitemShut {NoStop}%
\bibitem [{\citenamefont {Polley}\ and\ \citenamefont
  {Loring}(2020)}]{Polley2020vibronic}%
  \BibitemOpen
  \bibfield  {author} {\bibinfo {author} {\bibfnamefont {K.}~\bibnamefont
  {Polley}}\ and\ \bibinfo {author} {\bibfnamefont {R.~F.}\ \bibnamefont
  {Loring}},\ }\href@noop {} {\bibfield  {journal} {\bibinfo  {journal} {J.
  Phys. Chem. B}\ }\textbf {\bibinfo {volume} {124}},\ \bibinfo {pages} {9913}
  (\bibinfo {year} {2020})}\BibitemShut {NoStop}%
\bibitem [{\citenamefont {Polley}\ and\ \citenamefont
  {Loring}(2021)}]{Polley2021}%
  \BibitemOpen
  \bibfield  {author} {\bibinfo {author} {\bibfnamefont {K.}~\bibnamefont
  {Polley}}\ and\ \bibinfo {author} {\bibfnamefont {R.~F.}\ \bibnamefont
  {Loring}},\ }\href@noop {} {\bibfield  {journal} {\bibinfo  {journal} {J.
  Chem. Phys.}\ }\textbf {\bibinfo {volume} {154}},\ \bibinfo {pages} {194110}
  (\bibinfo {year} {2021})}\BibitemShut {NoStop}%
\bibitem [{\citenamefont {Braver}, \citenamefont {Valkunas},\ and\
  \citenamefont {Gelzinis}(2021)}]{Braver2021}%
  \BibitemOpen
  \bibfield  {author} {\bibinfo {author} {\bibfnamefont {Y.}~\bibnamefont
  {Braver}}, \bibinfo {author} {\bibfnamefont {L.}~\bibnamefont {Valkunas}}, \
  and\ \bibinfo {author} {\bibfnamefont {A.}~\bibnamefont {Gelzinis}},\
  }\href@noop {} {\bibfield  {journal} {\bibinfo  {journal} {J. Chem. Theory
  Comput.}\ }\textbf {\bibinfo {volume} {17}},\ \bibinfo {pages} {7157}
  (\bibinfo {year} {2021})}\BibitemShut {NoStop}%
\bibitem [{\citenamefont {Huo}\ and\ \citenamefont {Coker}(2010)}]{Huo2010}%
  \BibitemOpen
  \bibfield  {author} {\bibinfo {author} {\bibfnamefont {P.}~\bibnamefont
  {Huo}}\ and\ \bibinfo {author} {\bibfnamefont {D.~F.}\ \bibnamefont
  {Coker}},\ }\href@noop {} {\bibfield  {journal} {\bibinfo  {journal} {J.
  Chem. Phys.}\ }\textbf {\bibinfo {volume} {133}},\ \bibinfo {pages} {184108}
  (\bibinfo {year} {2010})}\BibitemShut {NoStop}%
\bibitem [{\citenamefont {Huo}\ and\ \citenamefont
  {Coker}(2011)}]{Huo2011densitymatrix}%
  \BibitemOpen
  \bibfield  {author} {\bibinfo {author} {\bibfnamefont {P.}~\bibnamefont
  {Huo}}\ and\ \bibinfo {author} {\bibfnamefont {D.~F.}\ \bibnamefont
  {Coker}},\ }\href {\doibase 10.1063/1.3664763} {\bibfield  {journal}
  {\bibinfo  {journal} {J.~Chem. Phys.}\ }\textbf {\bibinfo {volume} {135}},\
  \bibinfo {pages} {201101} (\bibinfo {year} {2011})}\BibitemShut {NoStop}%
\bibitem [{\citenamefont {Huo}\ and\ \citenamefont
  {Coker}(2012{\natexlab{a}})}]{Huo2012MolPhys}%
  \BibitemOpen
  \bibfield  {author} {\bibinfo {author} {\bibfnamefont {P.}~\bibnamefont
  {Huo}}\ and\ \bibinfo {author} {\bibfnamefont {D.~F.}\ \bibnamefont
  {Coker}},\ }\href@noop {} {\bibfield  {journal} {\bibinfo  {journal} {Mol.
  Phys.}\ }\textbf {\bibinfo {volume} {110}},\ \bibinfo {pages} {1035}
  (\bibinfo {year} {2012}{\natexlab{a}})}\BibitemShut {NoStop}%
\bibitem [{\citenamefont {Huo}\ and\ \citenamefont
  {Miller~III}(2015)}]{Huo2015PLDM}%
  \BibitemOpen
  \bibfield  {author} {\bibinfo {author} {\bibfnamefont {P.}~\bibnamefont
  {Huo}}\ and\ \bibinfo {author} {\bibfnamefont {T.~F.}\ \bibnamefont
  {Miller~III}},\ }\href {\doibase 10.1039/c5cp02517f} {\bibfield  {journal}
  {\bibinfo  {journal} {Phys. Chem. Chem. Phys.}\ }\textbf {\bibinfo {volume}
  {17}},\ \bibinfo {pages} {30914} (\bibinfo {year} {2015})}\BibitemShut
  {NoStop}%
\bibitem [{\citenamefont {Huo}\ and\ \citenamefont
  {Coker}(2012{\natexlab{b}})}]{Huo2012PLDM}%
  \BibitemOpen
  \bibfield  {author} {\bibinfo {author} {\bibfnamefont {P.}~\bibnamefont
  {Huo}}\ and\ \bibinfo {author} {\bibfnamefont {D.~F.}\ \bibnamefont
  {Coker}},\ }\href@noop {} {\bibfield  {journal} {\bibinfo  {journal}
  {J.~Chem. Phys.}\ }\textbf {\bibinfo {volume} {137}},\ \bibinfo {pages}
  {22A535} (\bibinfo {year} {2012}{\natexlab{b}})}\BibitemShut {NoStop}%
\bibitem [{\citenamefont {Huo}\ and\ \citenamefont
  {Coker}(2012{\natexlab{c}})}]{Huo2012_2}%
  \BibitemOpen
  \bibfield  {author} {\bibinfo {author} {\bibfnamefont {P.}~\bibnamefont
  {Huo}}\ and\ \bibinfo {author} {\bibfnamefont {D.~F.}\ \bibnamefont
  {Coker}},\ }\href@noop {} {\bibfield  {journal} {\bibinfo  {journal} {J.
  Chem. Phys.}\ }\textbf {\bibinfo {volume} {136}},\ \bibinfo {pages} {115102}
  (\bibinfo {year} {2012}{\natexlab{c}})}\BibitemShut {NoStop}%
\bibitem [{\citenamefont {Lee}, \citenamefont {Huo},\ and\ \citenamefont
  {Coker}(2016)}]{Lee2016}%
  \BibitemOpen
  \bibfield  {author} {\bibinfo {author} {\bibfnamefont {M.~K.}\ \bibnamefont
  {Lee}}, \bibinfo {author} {\bibfnamefont {P.}~\bibnamefont {Huo}}, \ and\
  \bibinfo {author} {\bibfnamefont {D.~F.}\ \bibnamefont {Coker}},\ }\href@noop
  {} {\bibfield  {journal} {\bibinfo  {journal} {Annu. Rev. Phys. Chem.}\
  }\textbf {\bibinfo {volume} {67}},\ \bibinfo {pages} {639} (\bibinfo {year}
  {2016})}\BibitemShut {NoStop}%
\bibitem [{\citenamefont {Castellanos}\ and\ \citenamefont
  {Huo}(2017)}]{Castellanos2017}%
  \BibitemOpen
  \bibfield  {author} {\bibinfo {author} {\bibfnamefont {M.~A.}\ \bibnamefont
  {Castellanos}}\ and\ \bibinfo {author} {\bibfnamefont {P.}~\bibnamefont
  {Huo}},\ }\href@noop {} {\bibfield  {journal} {\bibinfo  {journal} {J. Phys.
  Chem. Lett.}\ }\textbf {\bibinfo {volume} {8}},\ \bibinfo {pages} {2480}
  (\bibinfo {year} {2017})}\BibitemShut {NoStop}%
\bibitem [{\citenamefont {Mandal}\ and\ \citenamefont
  {Huo}(2019)}]{Mandal2019}%
  \BibitemOpen
  \bibfield  {author} {\bibinfo {author} {\bibfnamefont {A.}~\bibnamefont
  {Mandal}}\ and\ \bibinfo {author} {\bibfnamefont {P.}~\bibnamefont {Huo}},\
  }\href@noop {} {\bibfield  {journal} {\bibinfo  {journal} {J. Phys. Chem.
  Lett.}\ }\textbf {\bibinfo {volume} {10}},\ \bibinfo {pages} {5519} (\bibinfo
  {year} {2019})}\BibitemShut {NoStop}%
\bibitem [{\citenamefont {Hsieh}\ and\ \citenamefont
  {Kapral}(2012)}]{Hsieh2012FBTS}%
  \BibitemOpen
  \bibfield  {author} {\bibinfo {author} {\bibfnamefont {C.-Y.}\ \bibnamefont
  {Hsieh}}\ and\ \bibinfo {author} {\bibfnamefont {R.}~\bibnamefont {Kapral}},\
  }\href {\doibase 10.1063/1.4736841} {\bibfield  {journal} {\bibinfo
  {journal} {J.~Chem. Phys.}\ }\textbf {\bibinfo {volume} {137}},\ \bibinfo
  {pages} {22A507} (\bibinfo {year} {2012})}\BibitemShut {NoStop}%
\bibitem [{\citenamefont {Hsieh}\ and\ \citenamefont
  {Kapral}(2013)}]{Hsieh2013FBTS}%
  \BibitemOpen
  \bibfield  {author} {\bibinfo {author} {\bibfnamefont {C.-Y.}\ \bibnamefont
  {Hsieh}}\ and\ \bibinfo {author} {\bibfnamefont {R.}~\bibnamefont {Kapral}},\
  }\href {\doibase 10.1063/1.4798221} {\bibfield  {journal} {\bibinfo
  {journal} {J.~Chem. Phys.}\ }\textbf {\bibinfo {volume} {138}},\ \bibinfo
  {pages} {134110} (\bibinfo {year} {2013})}\BibitemShut {NoStop}%
\bibitem [{\citenamefont {Kelly}(2020)}]{Kelly2020}%
  \BibitemOpen
  \bibfield  {author} {\bibinfo {author} {\bibfnamefont {A.}~\bibnamefont
  {Kelly}},\ }\href@noop {} {\bibfield  {journal} {\bibinfo  {journal} {Faraday
  Discuss.}\ }\textbf {\bibinfo {volume} {221}},\ \bibinfo {pages} {547}
  (\bibinfo {year} {2020})}\BibitemShut {NoStop}%
\bibitem [{\citenamefont {Mannouch}\ and\ \citenamefont
  {Richardson}(2020{\natexlab{a}})}]{Mannouch2020a}%
  \BibitemOpen
  \bibfield  {author} {\bibinfo {author} {\bibfnamefont {J.~R.}\ \bibnamefont
  {Mannouch}}\ and\ \bibinfo {author} {\bibfnamefont {J.~O.}\ \bibnamefont
  {Richardson}},\ }\href@noop {} {\bibfield  {journal} {\bibinfo  {journal} {J.
  Chem. Phys.}\ }\textbf {\bibinfo {volume} {153}},\ \bibinfo {pages} {194109}
  (\bibinfo {year} {2020}{\natexlab{a}})}\BibitemShut {NoStop}%
\bibitem [{\citenamefont {Mannouch}\ and\ \citenamefont
  {Richardson}(2020{\natexlab{b}})}]{Mannouch2020b}%
  \BibitemOpen
  \bibfield  {author} {\bibinfo {author} {\bibfnamefont {J.~R.}\ \bibnamefont
  {Mannouch}}\ and\ \bibinfo {author} {\bibfnamefont {J.~O.}\ \bibnamefont
  {Richardson}},\ }\href@noop {} {\bibfield  {journal} {\bibinfo  {journal} {J.
  Chem. Phys.}\ }\textbf {\bibinfo {volume} {153}},\ \bibinfo {pages} {194110}
  (\bibinfo {year} {2020}{\natexlab{b}})}\BibitemShut {NoStop}%
\bibitem [{\citenamefont {Sun}\ and\ \citenamefont {Miller}(1997)}]{Sun1997}%
  \BibitemOpen
  \bibfield  {author} {\bibinfo {author} {\bibfnamefont {X.}~\bibnamefont
  {Sun}}\ and\ \bibinfo {author} {\bibfnamefont {W.~H.}\ \bibnamefont
  {Miller}},\ }\href@noop {} {\bibfield  {journal} {\bibinfo  {journal} {J.
  Chem. Phys.}\ }\textbf {\bibinfo {volume} {106}},\ \bibinfo {pages} {916}
  (\bibinfo {year} {1997})}\BibitemShut {NoStop}%
\bibitem [{\citenamefont {Provazza}\ and\ \citenamefont
  {Coker}(2018)}]{Provazza2018_lin}%
  \BibitemOpen
  \bibfield  {author} {\bibinfo {author} {\bibfnamefont {J.}~\bibnamefont
  {Provazza}}\ and\ \bibinfo {author} {\bibfnamefont {D.~F.}\ \bibnamefont
  {Coker}},\ }\href@noop {} {\bibfield  {journal} {\bibinfo  {journal} {J.
  Chem. Phys.}\ }\textbf {\bibinfo {volume} {148}},\ \bibinfo {pages} {181102}
  (\bibinfo {year} {2018})}\BibitemShut {NoStop}%
\bibitem [{\citenamefont {Provazza}\ \emph {et~al.}(2018)\citenamefont
  {Provazza}, \citenamefont {Segatta}, \citenamefont {Garavelli},\ and\
  \citenamefont {Coker}}]{Provazza2018_nonlin}%
  \BibitemOpen
  \bibfield  {author} {\bibinfo {author} {\bibfnamefont {J.}~\bibnamefont
  {Provazza}}, \bibinfo {author} {\bibfnamefont {F.}~\bibnamefont {Segatta}},
  \bibinfo {author} {\bibfnamefont {M.}~\bibnamefont {Garavelli}}, \ and\
  \bibinfo {author} {\bibfnamefont {D.~F.}\ \bibnamefont {Coker}},\ }\href@noop
  {} {\bibfield  {journal} {\bibinfo  {journal} {J. Chem. Theory Comput.}\
  }\textbf {\bibinfo {volume} {14}},\ \bibinfo {pages} {856} (\bibinfo {year}
  {2018})}\BibitemShut {NoStop}%
\bibitem [{\citenamefont {Barford}(2013)}]{BardfordBook}%
  \BibitemOpen
  \bibfield  {author} {\bibinfo {author} {\bibfnamefont {W.}~\bibnamefont
  {Barford}},\ }\href@noop {} {\emph {\bibinfo {title} {Electronic and optical
  properties of conjugated polymers}}},\ \bibinfo {edition} {2nd}\ ed.\
  (\bibinfo  {publisher} {Oxford University Press},\ \bibinfo {address}
  {Oxford},\ \bibinfo {year} {2013})\BibitemShut {NoStop}%
\bibitem [{\citenamefont {May}\ and\ \citenamefont {K\"uhn}(2011)}]{KuehnBook}%
  \BibitemOpen
  \bibfield  {author} {\bibinfo {author} {\bibfnamefont {V.}~\bibnamefont
  {May}}\ and\ \bibinfo {author} {\bibfnamefont {O.}~\bibnamefont {K\"uhn}},\
  }\href@noop {} {\emph {\bibinfo {title} {Charge and Energy Transfer Dynamics
  in Molecular Systems}}},\ \bibinfo {edition} {3rd}\ ed.\ (\bibinfo
  {publisher} {Wiley},\ \bibinfo {year} {2011})\BibitemShut {NoStop}%
\bibitem [{Note1()}]{Note1}%
  \BibitemOpen
  \bibinfo {note} {In principle, the approach outlined in this paper can also
  treat systems containing a manifold of ground states, where $\protect
  \mathaccentV {hat}05E{V}_{\protect \text {g}}(x)$ simply becomes a matrix
  within this subspace.}\BibitemShut {Stop}%
\bibitem [{\citenamefont {Saller}, \citenamefont {Runeson},\ and\ \citenamefont
  {Richardson}()}]{NRPMDChapter}%
  \BibitemOpen
  \bibfield  {author} {\bibinfo {author} {\bibfnamefont {M.~A.~C.}\
  \bibnamefont {Saller}}, \bibinfo {author} {\bibfnamefont {J.~E.}\
  \bibnamefont {Runeson}}, \ and\ \bibinfo {author} {\bibfnamefont {J.~O.}\
  \bibnamefont {Richardson}},\ }\enquote {\bibinfo {title} {Path-integral
  approaches to non-adiabatic dynamics},}\ in\ \href {\doibase
  10.1002/9781119417774.ch20} {\emph {\bibinfo {booktitle} {Quantum Chemistry
  and Dynamics of Excited States: Methods and Applications}}},\ \bibinfo
  {editor} {edited by\ \bibinfo {editor} {\bibfnamefont {L.}~\bibnamefont
  {Gonz\'alez}}\ and\ \bibinfo {editor} {\bibfnamefont {R.}~\bibnamefont
  {Lindh}}}\ (\bibinfo  {publisher} {Wiley})\ pp.\ \bibinfo {pages}
  {629--653}\BibitemShut {NoStop}%
\bibitem [{\citenamefont {Stock}\ and\ \citenamefont
  {Thoss}(2005)}]{Stock2005nonadiabatic}%
  \BibitemOpen
  \bibfield  {author} {\bibinfo {author} {\bibfnamefont {G.}~\bibnamefont
  {Stock}}\ and\ \bibinfo {author} {\bibfnamefont {M.}~\bibnamefont {Thoss}},\
  }\href {\doibase 10.1002/0471739464.ch5} {\bibfield  {journal} {\bibinfo
  {journal} {Adv. Chem. Phys.}\ }\textbf {\bibinfo {volume} {131}},\ \bibinfo
  {pages} {243} (\bibinfo {year} {2005})}\BibitemShut {NoStop}%
\bibitem [{Note2()}]{Note2}%
  \BibitemOpen
  \bibinfo {note} {In many cases, we may choose a smaller value of $\omega
  _{\protect \text {shift}}$, which reduces the oscillation frequency of the
  correlation functions but in practice is not expected to alter the shape of
  the spectrum.}\BibitemShut {Stop}%
\bibitem [{\citenamefont {Craig}, \citenamefont {Thoss},\ and\ \citenamefont
  {Wang}(2007)}]{Craig2007condensed}%
  \BibitemOpen
  \bibfield  {author} {\bibinfo {author} {\bibfnamefont {I.~R.}\ \bibnamefont
  {Craig}}, \bibinfo {author} {\bibfnamefont {M.}~\bibnamefont {Thoss}}, \ and\
  \bibinfo {author} {\bibfnamefont {H.}~\bibnamefont {Wang}},\ }\href {\doibase
  10.1063/1.2772265} {\bibfield  {journal} {\bibinfo  {journal} {J.~Chem.
  Phys.}\ }\textbf {\bibinfo {volume} {127}},\ \bibinfo {pages} {144503}
  (\bibinfo {year} {2007})}\BibitemShut {NoStop}%
\bibitem [{\citenamefont {Breuer}\ and\ \citenamefont
  {Petruccione}(2002)}]{OpenQuantum}%
  \BibitemOpen
  \bibfield  {author} {\bibinfo {author} {\bibfnamefont {H.-P.}\ \bibnamefont
  {Breuer}}\ and\ \bibinfo {author} {\bibfnamefont {F.}~\bibnamefont
  {Petruccione}},\ }\href@noop {} {\emph {\bibinfo {title} {The Theory of Open
  Quantum Systems}}}\ (\bibinfo  {publisher} {Oxford University Press},\
  \bibinfo {address} {Oxford},\ \bibinfo {year} {2002})\BibitemShut {NoStop}%
\bibitem [{\citenamefont {Berkelbach}, \citenamefont {Markland},\ and\
  \citenamefont {Reichman}(2012)}]{Berkelbach2012}%
  \BibitemOpen
  \bibfield  {author} {\bibinfo {author} {\bibfnamefont {T.~C.}\ \bibnamefont
  {Berkelbach}}, \bibinfo {author} {\bibfnamefont {T.~E.}\ \bibnamefont
  {Markland}}, \ and\ \bibinfo {author} {\bibfnamefont {D.~R.}\ \bibnamefont
  {Reichman}},\ }\href@noop {} {\bibfield  {journal} {\bibinfo  {journal} {J.
  Chem. Phys.}\ }\textbf {\bibinfo {volume} {136}},\ \bibinfo {pages} {084104}
  (\bibinfo {year} {2012})}\BibitemShut {NoStop}%
\bibitem [{\citenamefont {Berkelbach}, \citenamefont {Reichman},\ and\
  \citenamefont {Markland}(2012)}]{Berkelbach2012hybrid}%
  \BibitemOpen
  \bibfield  {author} {\bibinfo {author} {\bibfnamefont {T.~C.}\ \bibnamefont
  {Berkelbach}}, \bibinfo {author} {\bibfnamefont {D.~R.}\ \bibnamefont
  {Reichman}}, \ and\ \bibinfo {author} {\bibfnamefont {T.~E.}\ \bibnamefont
  {Markland}},\ }\href {\doibase 10.1063/1.3671372} {\bibfield  {journal}
  {\bibinfo  {journal} {J.~Chem. Phys.}\ }\textbf {\bibinfo {volume} {136}},\
  \bibinfo {pages} {034113} (\bibinfo {year} {2012})}\BibitemShut {NoStop}%
\bibitem [{\citenamefont {Montoya-Castillo}\ and\ \citenamefont
  {Reichman}(2017)}]{Montoya2017}%
  \BibitemOpen
  \bibfield  {author} {\bibinfo {author} {\bibfnamefont {A.}~\bibnamefont
  {Montoya-Castillo}}\ and\ \bibinfo {author} {\bibfnamefont {D.~R.}\
  \bibnamefont {Reichman}},\ }\href@noop {} {\bibfield  {journal} {\bibinfo
  {journal} {J. Chem. Phys.}\ }\textbf {\bibinfo {volume} {146}},\ \bibinfo
  {pages} {024107} (\bibinfo {year} {2017})}\BibitemShut {NoStop}%
\bibitem [{\citenamefont {Berkelbach}()}]{pyrho}%
  \BibitemOpen
  \bibfield  {author} {\bibinfo {author} {\bibfnamefont {T.~C.}\ \bibnamefont
  {Berkelbach}},\ }\href@noop {} {\enquote {\bibinfo {title} {Pyrho: A python
  package for reduced density matrix techniques},}\ }\bibinfo {howpublished}
  {\url{https://github.com/berkelbach-group/pyrho}}\BibitemShut {NoStop}%
\bibitem [{\citenamefont {Thoss}, \citenamefont {Wang},\ and\ \citenamefont
  {Miller}(2001)}]{Thoss2001hybrid}%
  \BibitemOpen
  \bibfield  {author} {\bibinfo {author} {\bibfnamefont {M.}~\bibnamefont
  {Thoss}}, \bibinfo {author} {\bibfnamefont {H.}~\bibnamefont {Wang}}, \ and\
  \bibinfo {author} {\bibfnamefont {W.~H.}\ \bibnamefont {Miller}},\ }\href
  {\doibase 10.1063/1.1385562} {\bibfield  {journal} {\bibinfo  {journal}
  {J.~Chem. Phys.}\ }\textbf {\bibinfo {volume} {115}},\ \bibinfo {pages}
  {2991} (\bibinfo {year} {2001})}\BibitemShut {NoStop}%
\bibitem [{\citenamefont {Jeon}, \citenamefont {Park},\ and\ \citenamefont
  {Cho}(2010)}]{Jeon2010}%
  \BibitemOpen
  \bibfield  {author} {\bibinfo {author} {\bibfnamefont {J.}~\bibnamefont
  {Jeon}}, \bibinfo {author} {\bibfnamefont {S.}~\bibnamefont {Park}}, \ and\
  \bibinfo {author} {\bibfnamefont {M.}~\bibnamefont {Cho}},\ }\enquote
  {\bibinfo {title} {Two-dimensional optical spectroscopy: Theory and
  experiment},}\ in\ \href@noop {} {\emph {\bibinfo {booktitle} {Encyclopedia
  of Analytical Chemistry}}}\ (\bibinfo  {publisher} {American Cancer
  Society},\ \bibinfo {year} {2010})\BibitemShut {NoStop}%
\bibitem [{\citenamefont {Gelin}, \citenamefont {Borrelli},\ and\ \citenamefont
  {Domcke}(2017)}]{Gelin2017}%
  \BibitemOpen
  \bibfield  {author} {\bibinfo {author} {\bibfnamefont {M.~F.}\ \bibnamefont
  {Gelin}}, \bibinfo {author} {\bibfnamefont {R.}~\bibnamefont {Borrelli}}, \
  and\ \bibinfo {author} {\bibfnamefont {W.}~\bibnamefont {Domcke}},\
  }\href@noop {} {\bibfield  {journal} {\bibinfo  {journal} {J. Chem. Phys.}\
  }\textbf {\bibinfo {volume} {147}},\ \bibinfo {pages} {044114} (\bibinfo
  {year} {2017})}\BibitemShut {NoStop}%
\bibitem [{\citenamefont {Miller}(1978)}]{Miller1978radiation}%
  \BibitemOpen
  \bibfield  {author} {\bibinfo {author} {\bibfnamefont {W.~H.}\ \bibnamefont
  {Miller}},\ }\href@noop {} {\bibfield  {journal} {\bibinfo  {journal}
  {J.~Chem. Phys.}\ }\textbf {\bibinfo {volume} {69}},\ \bibinfo {pages} {2188}
  (\bibinfo {year} {1978})}\BibitemShut {NoStop}%
\bibitem [{\citenamefont {Kim}\ \emph {et~al.}(2012)\citenamefont {Kim},
  \citenamefont {Kelly}, \citenamefont {Park},\ and\ \citenamefont
  {Rhee}}]{Kim2012FMO}%
  \BibitemOpen
  \bibfield  {author} {\bibinfo {author} {\bibfnamefont {H.~W.}\ \bibnamefont
  {Kim}}, \bibinfo {author} {\bibfnamefont {A.}~\bibnamefont {Kelly}}, \bibinfo
  {author} {\bibfnamefont {J.~W.}\ \bibnamefont {Park}}, \ and\ \bibinfo
  {author} {\bibfnamefont {Y.~M.}\ \bibnamefont {Rhee}},\ }\href {\doibase
  10.1021/ja303025q} {\bibfield  {journal} {\bibinfo  {journal} {J. Am. Chem.
  Soc.}\ }\textbf {\bibinfo {volume} {134}},\ \bibinfo {pages} {11640}
  (\bibinfo {year} {2012})}\BibitemShut {NoStop}%
\end{thebibliography}%

\end{document}

\renewcommand\theequation{S\arabic{equation}} 
\renewcommand\thefigure{S\arabic{figure}} 
\renewcommand\thetable{S\arabic{table}} 

\title{A partially linearized spin-mapping approach for simulating nonlinear optical spectra: supplementary material}% Force line breaks with \\

\author{Jonathan R. Mannouch}
\email{jonathan.mannouch@phys.chem.ethz.ch}
\author{Jeremy O. Richardson}%
\email{jeremy.richardson@phys.chem.ethz.ch}
\affiliation{Laboratory of Physical Chemistry, ETH Z\"{u}rich, 8093 Z\"{u}rich, Switzerland}

\date{\today}% It is always \today, today,
 
\maketitle

In this supplementary material, we present additional results for both linear and nonlinear spectra, which were not included within the main paper. While these results give rise to the same overall conclusions with regards to the relative accuracy of each of the considered methods, the discrepancies between the obtained results are not as large. They are nevertheless included here for completeness. We also present the results for a modified spin-PLDM approach for calculating the excited-state absorption correlation functions used to compute 2D and pump--probe spectra. Through application to the Fenna--Matthews--Olsen (FMO) complex, we show that this new approach is just as accurate as the full spin-PLDM method, but requires significantly fewer trajectories to reach convergence. Data files containing all of the response functions corresponding to the spectra presented in this paper for standard PLDM, spin-LSC and spin-PLDM are also included. We also provide python scripts which can be used to perform the necessary Fourier transforms to generate the associated spectra.
\section{Circular Dichroism Spectra}
In addition to electronic absorption and fluorescence spectra, an array of other linear spectroscopic quantities exist, which can be obtained as the Fourier transform of a single-time optical response function. An example is the circular dichroism spectrum, which measures the difference in absorption of right- and left-circularly polarized light. The associated correlation function is defined as:\cite{Bosnich1969,Kramer2018}
\begin{subequations}
\begin{align}
J_{\text{CD}}(t)&=\Tr\left[\eu{i\hat{H}_{\text{g}}t}\hat{m}^{-}\eu{-i\hat{H}_{\text{e}}t}\hat{\mu}^{+}\hat{\rho}_{\text{g}}\right], \\
\hat{m}^{-}&=\sum_{n=1}^{F}\left(\vec{R}_{n}\times\vec{\mu}_{n}\right)\hat{a}_{n} , \\
\hat{\mu}^{+}&=\sum_{n=1}^{F}\vec{\mu}_{n}\hat{a}^{\dagger}_{n} ,
\end{align}
\end{subequations}
where $\hat{m}=\hat{m}^{+}+\hat{m}^{-}$ and $\hat{\mu}=\hat{\mu}^{+}+\hat{\mu}^{-}$ are the magnetic and electric dipole operators respectively and $\hat{a}_{n}$ is the exciton destruction operator for chromophore $n$. Additionally $\hat{H}_{\text{g}}$ and $\hat{H}_{\text{e}}$ are the Hamiltonians associated with the ground and single-exciton subspaces, $\vec{R}_{n}$ is the position vector associated with the centre of chromophore $n$ and $\times$ corresponds to the vector cross-product.
From these expressions, the fully linearized and partially linearized mapping expressions for the circular dichroism optical response function can be obtained in accordance with the expressions outlined in the main paper for absorption spectra. Additionally, the rotational averaging of the linear spectra over all possible electric field orientations can be achieved % by taking the average %of three single-time correlation functions, which correspond 
by considering the components of the two observable operators along each of the three Cartesian axes.\cite{Hein2012}
This is implemented as a post-processing step after each trajectory is computed to determine its contribution to the ensemble average.
%In the main paper, the linear optical spectra were only calculated for the Frenkel biexciton model, for which the transition dipole moments, $\mu_{n}$, on each of the two chromophores are antiparallel. Therefore for this model, $\mu_{n}$ is a scalar. For the FMO complex, $\vec{\mu}_{n}$ is a three-dimensional vector and so the single-time correlation functions must be rotationally averaged. 

In order to calculate the circular dichroism spectrum, the Fourier transform of the optical response function is taken as follows:
\begin{subequations}
\begin{align}
I_{\text{CD}}(\omega)&\propto\text{Im}\left[\int_{0}^{\infty}\rd t\,S_{\text{CD}}^{(1)}(t)\,\eu{i\omega t}\right] \label{eq:spec_lin} , \\
S_{\text{CD}}^{(1)}(t)&=-2\theta(t)\text{Im}[J_{\text{CD}}(t)] ,
\end{align}
\end{subequations}
where the positive part of the spectrum is normalized for positive frequencies. We now analyse the electronic absorption, circular dichroism and fluorescence spectra for the FMO complex at a range of temperatures, calculated using the quantum master equation and mapping-based methods considered in the main paper.
\section{The Fenna--Matthews--Olsen complex}
For the FMO complex, we consider the same model as in Ref.~\onlinecite{Kramer2018}, where complete details of the model and all the required parameters can be found. For all of the classical trajectory techniques, $f=60$ nuclear degrees of freedom per chromophore were used. The numerically exact results were computed using the hierarchical equations of motion (HEOM); for the $T=30$\,K linear spectra, these were obtained from Ref.~\onlinecite{Kramer2018} while the other spectra were calculated using the open-source \texttt{pyrho} package.\cite{pyrho}
\subsection{Linear Spectra}
\begin{figure*}
\includegraphics[scale=.85]{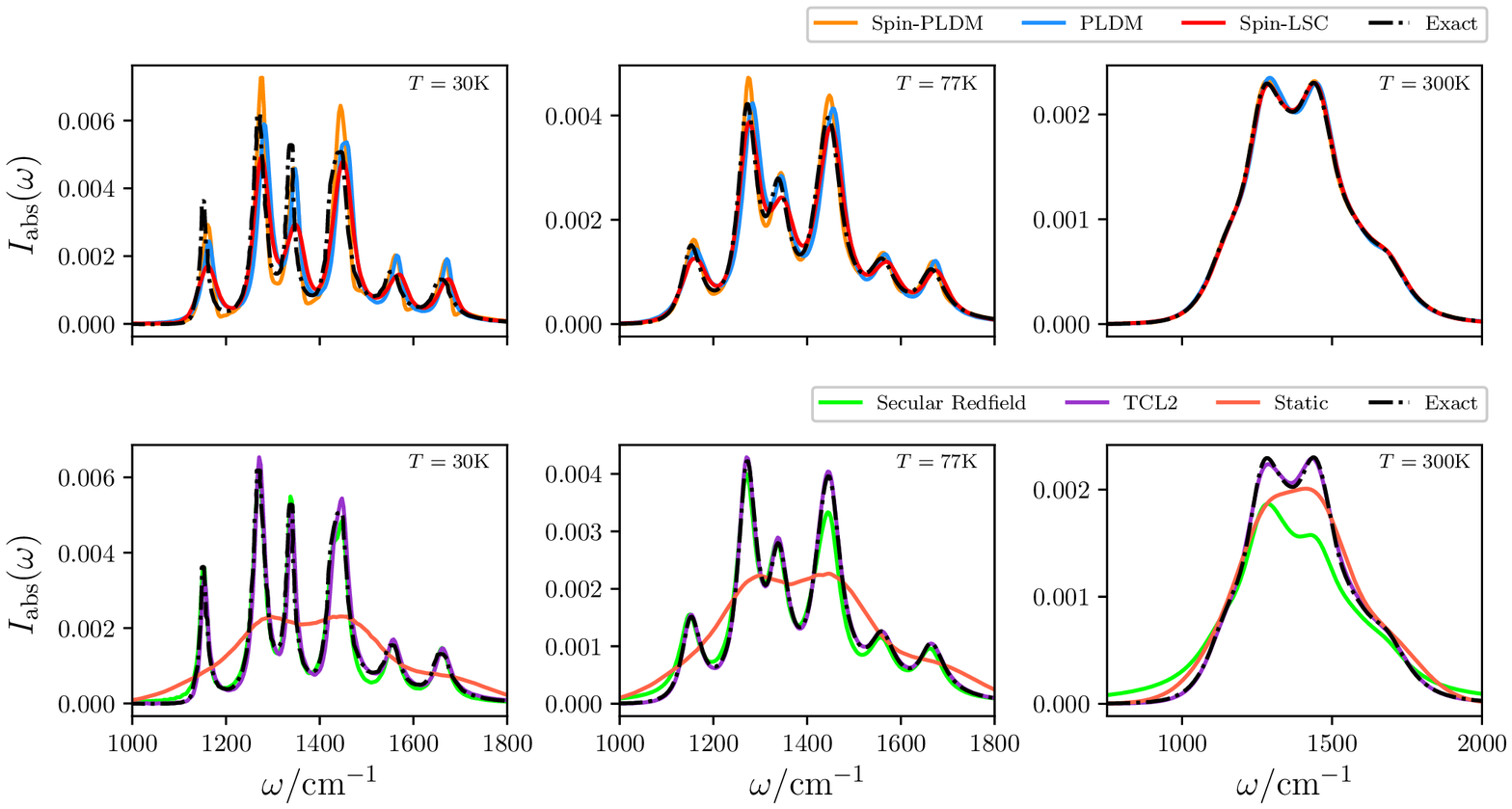}
\caption{The absorption spectrum for a seven-state FMO model, calculated at various temperatures as indicated. We use $\omega_{\text{shift}}=1430\,\text{cm}^{-1}$, with all other model parameters given in Ref.~\onlinecite{Kramer2018}. Exact HEOM results\cite{pyrho,Kramer2018} are given by the dashed black lines. These results may be compared with those calculated using Ehrenfest (Figs.~5, 6 and 7 from Ref.~\onlinecite{Gao2020_lin}) and LSC-IVR, PBME and traceless MMST (Figs.~8, 9 and 10 from Ref.~\onlinecite{Gao2020_lin}). Our Redfield results are in agreement with those already published (Fig.~4 from Ref.~\onlinecite{Kramer2018}).}\label{fig:abs-fmo}
\end{figure*}
\begin{figure*}
\includegraphics[scale=.85]{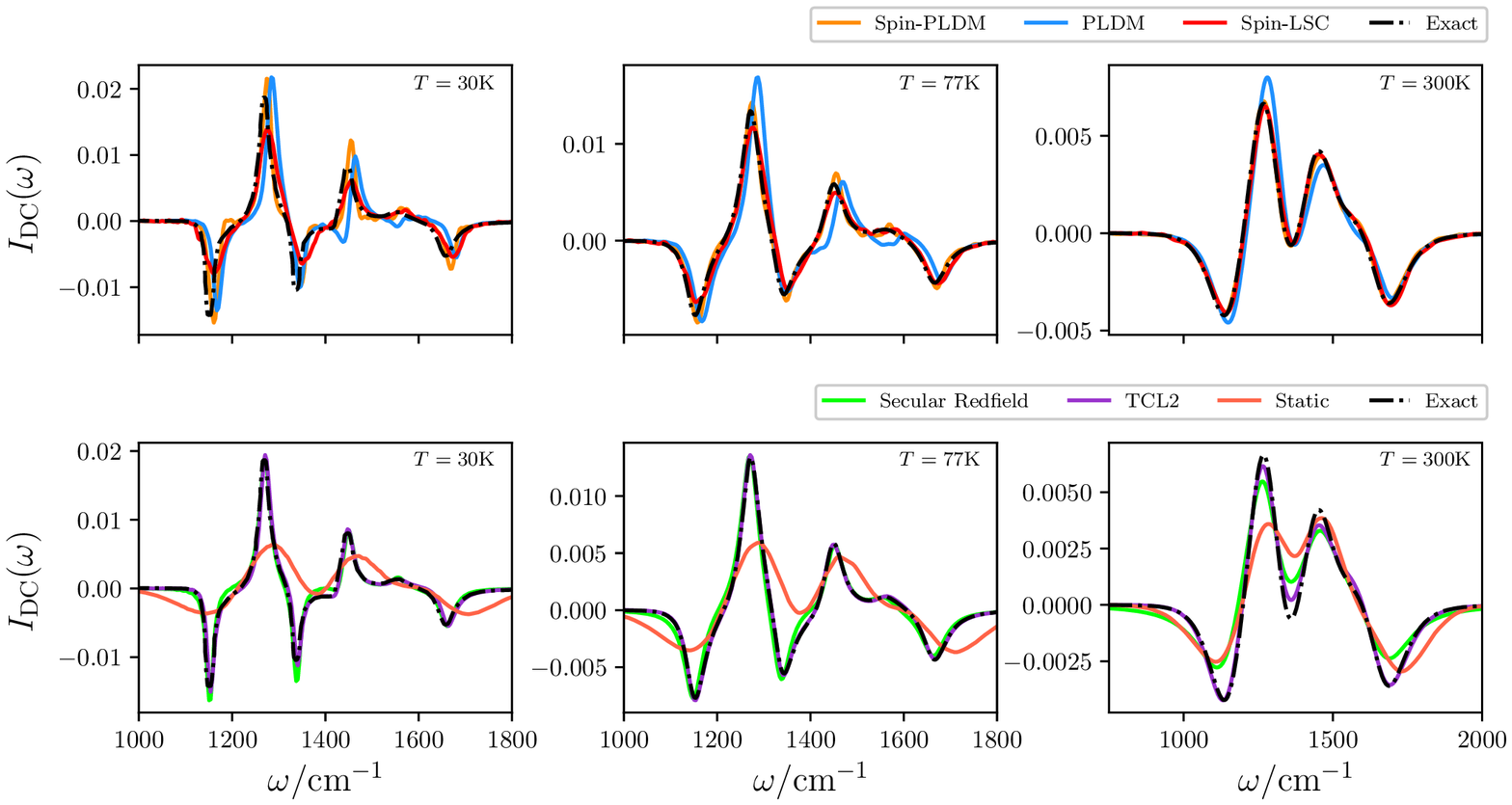}
\caption{The circular dichroism spectrum for a seven-state FMO model, calculated at various temperatures. We use $\omega_{\text{shift}}=1430\,\text{cm}^{-1}$, with all other model parameters given in Ref.~\onlinecite{Kramer2018}. Exact HEOM results\cite{pyrho,Kramer2018} are given by the dashed black lines. Our Redfield results are in agreement with those already published (Fig.~6 from Ref.~\onlinecite{Kramer2018}).}\label{fig:dichrom-fmo}
\end{figure*}
\begin{figure*}
\includegraphics[scale=.85]{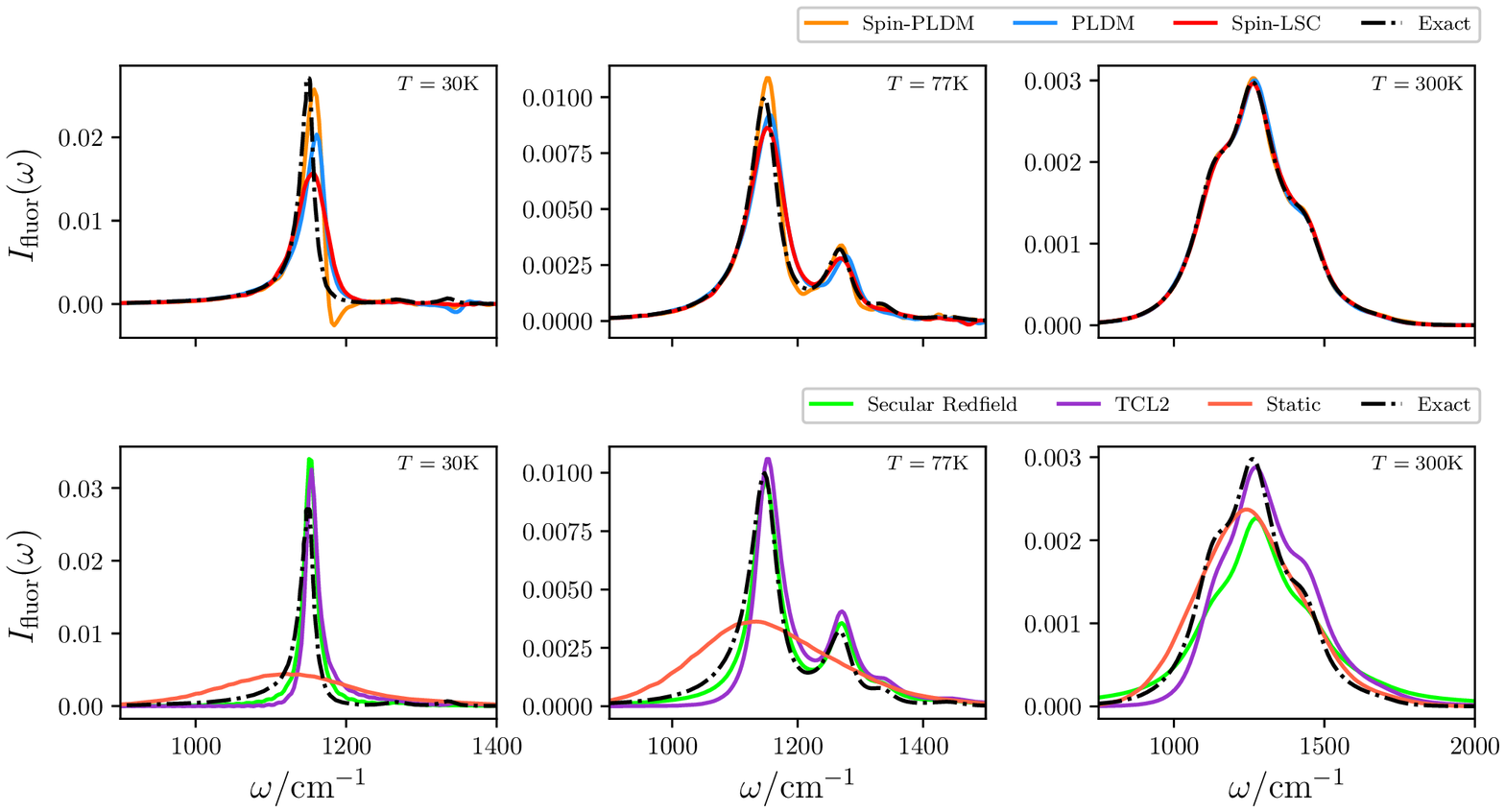}
\caption{The fluorescence spectrum for a seven-state FMO model, calculated at various temperatures as indicated. We use $\omega_{\text{shift}}=1430\,\text{cm}^{-1}$, with all other model parameters given in Ref.~\onlinecite{Kramer2018}. Exact HEOM results\cite{pyrho,Kramer2018} are given by the dashed black lines. Our Redfield results are in agreement with those already published (Fig.~6 from Ref.~\onlinecite{Kramer2018}).}\label{fig:fluor-fmo}
\end{figure*}
The associated electronic absorption, circular dichroism and fluorescence spectra are given in Figs.~\ref{fig:abs-fmo}, \ref{fig:dichrom-fmo} and \ref{fig:fluor-fmo} respectively. Considering that the error of the static nuclear approximation is large in comparison with the numerically exact results suggests that homogeneous broadening dominates. Such a regime is thus challenging and requires a dynamical theory for the nuclei in order to correctly reproduce the qualitative features of the spectra.

For the mapping-based methods (given in the first row of all three figures), we find that all of the generated spectra in the high-temperature limit ($T=300$\,K) are essentially indistinguishable from the exact results. This is in agreement with previous work, which has found that for the standard partially linearized density matrix (PLDM), spin-PLDM and spin-LSC methods, the relatively short-time dynamics become exact for high-temperature harmonic models.\cite{spinmap,multispin,Mannouch2020a} However for the circular dichroism spectrum (Fig.~\ref{fig:dichrom-fmo}) at this temperature, the standard PLDM result does exhibit a noticeably greater error than spin-PLDM\@.

As the temperature is decreased, the peaks associated with the spin-LSC linear spectra become more and more over-broadened compared to all of the numerically exact results. This is in agreement with the findings in the main paper, where this phenomenon was attributed to the fact that the fully linearized methods are unable to correctly describe the dynamical coherence between the ground and single-exciton subspaces. %, where the dynamics associated with the forward and backward excitonic paths should be treated separately.
Although also not exact, the partially linearized results (standard and spin-PLDM) achieve closer agreement to the numerically exact spectra at low temperatures. In general though for all linear spectra and parameter regimes considered, spin-PLDM is found to be just as good or even better than standard PLDM at correctly reproducing the important spectral features.

\begin{figure*}
\includegraphics[scale=.81]{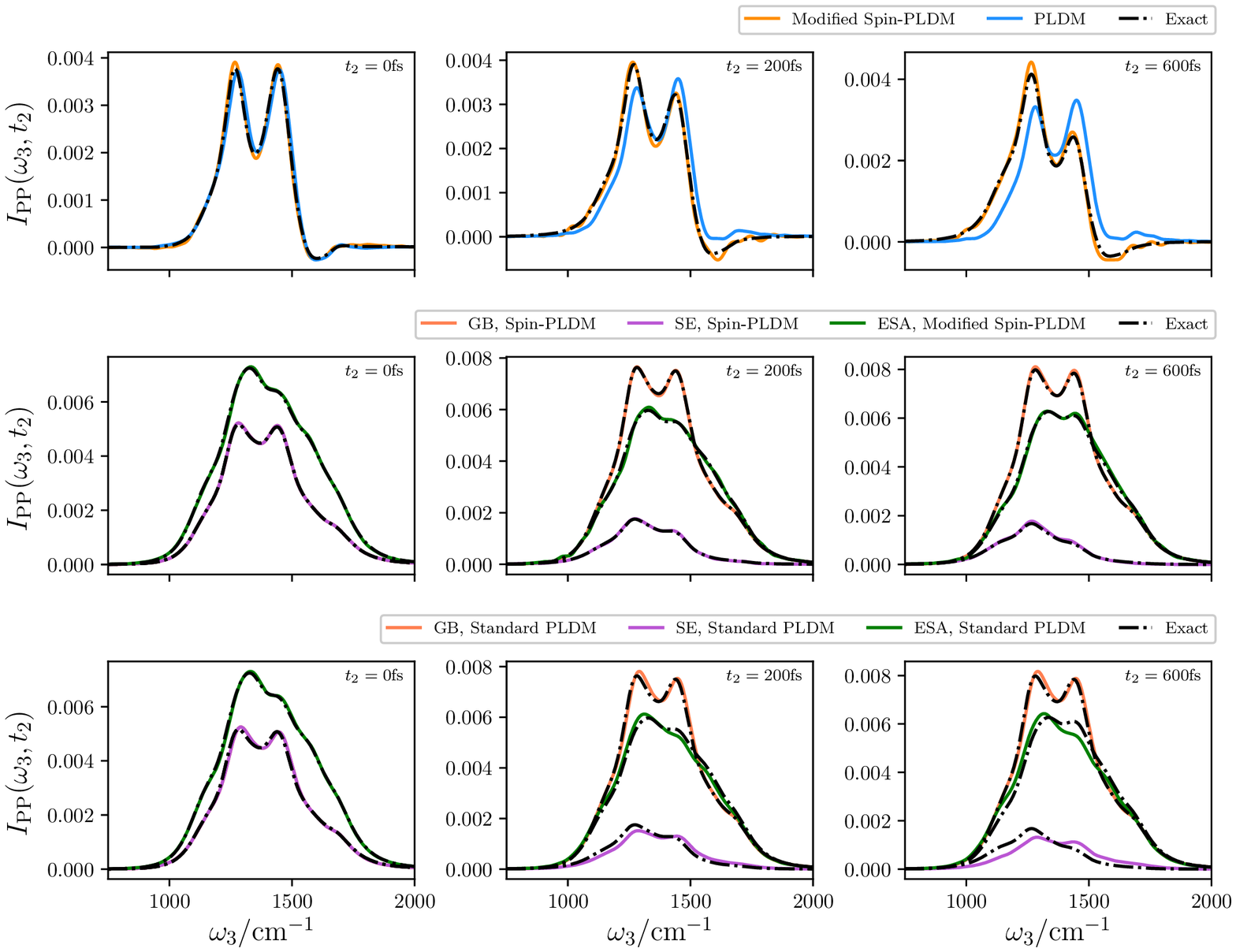}
\caption{The pump--probe spectra for a seven-state FMO model at $T=300\,$K, calculated for various $t_{2}$ delay times as indicated. We use $\omega_{\text{shift}}=1430\,\text{cm}^{-1}$, with all other model parameters given in Ref.~\onlinecite{Kramer2018}. The top row of figures gives the full pump--probe spectra, whereas the middle row gives the individual contributions calculated with spin-PLDM, except that the excited-state absorption term was obtained with our modified version. Finally, the bottom row of figures gives the individual contributions calculated with standard PLDM\@. Exact HEOM results\cite{pyrho} are given by the dashed black lines.}\label{fig:pump-fmo-approx}
\end{figure*}
For the quantum master equations (given in the second row of all three figures), the second-order time-convolutionless (TCL2) quantum master equation is by far the most accurate and is able to almost exactly reproduce the electronic absorption and circular dichroism spectra at all temperatures. This was explained in the main paper by noting that the TCL2 approach is also able to correctly describe the dynamical coherence between the ground and single-exciton subspaces, as demonstrated by the fact that as for the partially linearized approaches, the method is exact for harmonic models without diabatic couplings.\cite{Fetherolf2017} This is not true for Redfield theory, which although seemingly accurate at low temperatures, exhibits large errors in the high-temperature limit, as has been noted in previous work.\cite{Kramer2018} In general, all quantum master equations are unable to correctly reproduce electronic fluorescence spectra because the initial coupled exciton--nuclear state cannot be correctly described by reduced densities for which the nuclear degrees of freedom have been integrated out.
\subsection{Pump--Probe spectra}
\begin{figure*}
\includegraphics[scale=.86]{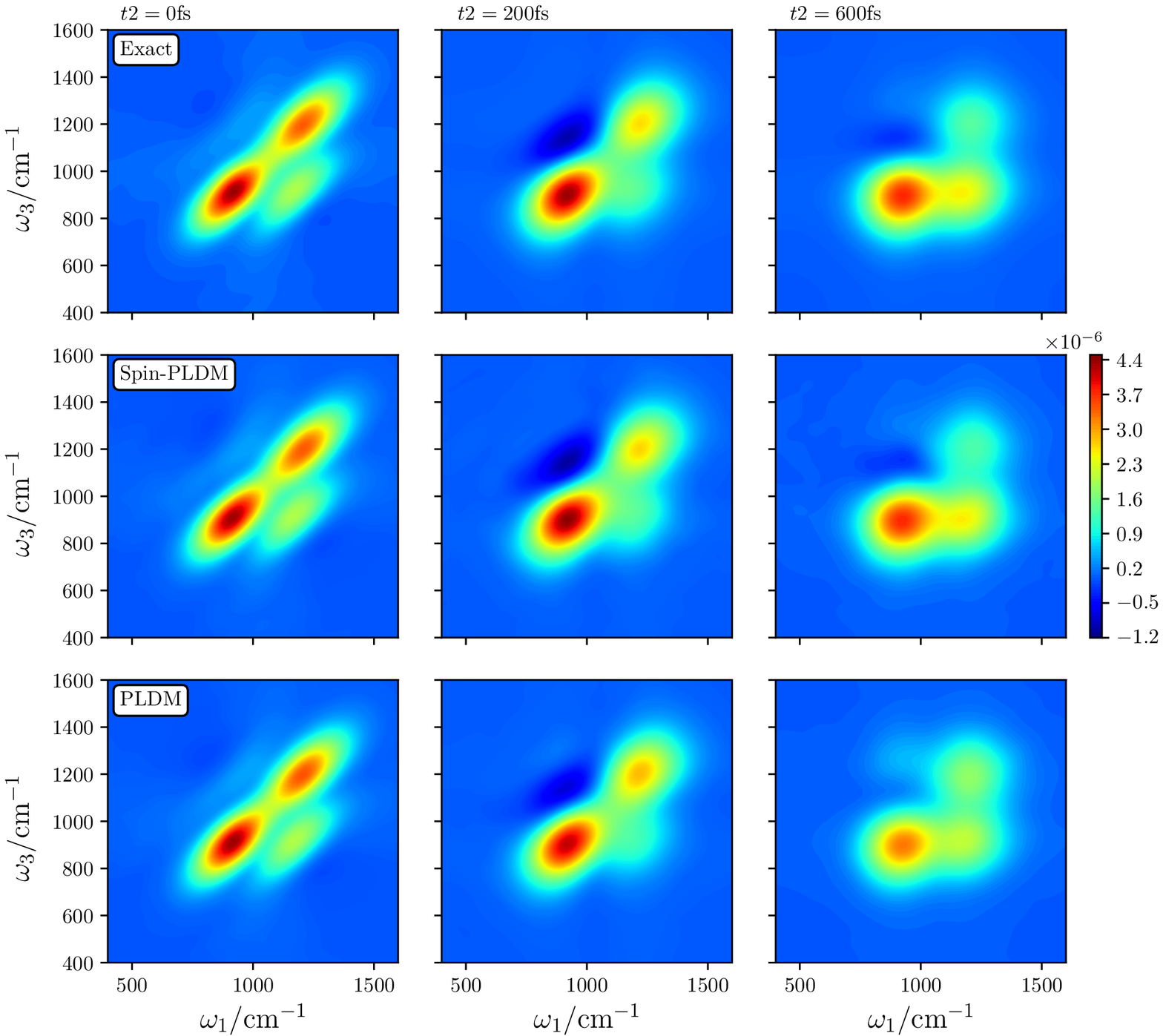}
\caption{The 2D electronic spectra for a relatively slow-bath and high-temperature Frenkel biexciton model ($\epsilon_{1}-\epsilon_{2}=100\,\text{cm}^{-1}$, $\omega_{\text{shift}}=1050\,\text{cm}^{-1}$, $\Delta_{12}=100\,\text{cm}^{-1}$, $\Lambda=50\,\text{cm}^{-1}$, $\omega_{c}=18\,\text{cm}^{-1}$ and $T=300\,$K), calculated for various $t_{2}$ delay times as indicated. Exact HEOM results\cite{pyrho} correspond to the first row of figures. We recommend that the reader compares these results with those calculated using Ehrenfest, LSC-IVR and PBME (Fig.~5 from Ref.~\onlinecite{Gao2020_nonlin}), traceless MMST (Fig.~4 from from Ref.~\onlinecite{Gao2020_nonlin}) and TCL2 (Fig.~5 from Ref.~\onlinecite{Fetherolf2017}). Both our PLDM (Fig.~4 from Ref.~\onlinecite{Provazza2018_nonlin}) and HEOM (Fig.~5 from Ref.~\onlinecite{Fetherolf2017}, Fig.~4 from Ref.~\onlinecite{Provazza2018_nonlin} and Figs.~2, 4 and 5 from Ref.~\onlinecite{Gao2020_nonlin}) results are in agreement with those already published.}\label{fig:2d-exciton}
\end{figure*}
Here we present the pump--probe spectra associated with the FMO complex at $T=300\,$K, calculated using our modified spin-PLDM approach described in the main paper. Within this approach the ground-state bleaching (GB) and stimulated emission (SE) contributions to the spectra are calculated using the original spin-PLDM approach and are hence identical to the results presented in the main paper. The calculation of the excited state absorption (ESA) contribution to the spectra is however different, because the mapping variables in the double-exciton subspace ($\mathcal{Z}_{\text{ee}}$) are now treated with standard PLDM ($\gamma=0$), while the mapping variables in the single-exciton subspace ($\mathcal{Z}_{\text{e}}$ and $\mathcal{Z}_{\text{e}}'$) are treated with the same zero-point energy parameter used in the original spin-PLDM method. This modification requires significantly fewer trajectories to reach convergence, because the standard PLDM focused initial conditions are more efficient than those for spin-PLDM, especially when sampling the large double-exciton space. The reason for this is explained in Ref.~\onlinecite{Mannouch2020b}. By comparing the modified (Fig.~\ref{fig:pump-fmo-approx}), original spin-PLDM (given by Fig.~8 in the main paper) and standard PLDM results (Fig.~\ref{fig:pump-fmo-approx}), we see that the superior accuracy of the spin-PLDM approach is still retained in the modified approach, as desired.
\section{The Frenkel Biexciton Model}
The Frenkel biexciton model is a two chromophore system commonly used to benchmark methods used to compute electronic spectroscopic quantities.\cite{Provazza2018_lin,Provazza2018_nonlin,Gao2020_lin,Gao2020_nonlin,Fetherolf2017} In particular, the relatively slow-bath and high-temperature version of the model ($T=300$\,K) has been used numerous times as a test for the calculation of 2D electronic spectra.\cite{Provazza2018_nonlin,Gao2020_nonlin,Fetherolf2017} Even though this regime is less challenging than the relatively fast-bath and low-temperature case considered in the main paper, we include this here for completeness and in order to aid comparison with other results published in the literature. Further details of the model, including all the required parameters, can be found in Ref.~\onlinecite{Gao2020_lin}. For all of the classical trajectory techniques, $f=100$ nuclear degrees of freedom per chromophore were used. Exact results can also be computed using HEOM, where we have used the open source \texttt{pyrho} package.\cite{pyrho} 

The spin-PLDM 2D electronic spectra (given by the second row in Fig.~\ref{fig:2d-exciton}) are essentially indistinguishable from the exact results (given by the first row in Fig.~\ref{fig:2d-exciton}). This is again to be expected, as the dynamics produced by spin-PLDM have been observed to be numerically exact for high-temperature harmonic models.\cite{Mannouch2020a} While the standard PLDM results (given by the third row in Fig.~\ref{fig:2d-exciton}) are also practically perfect for $t_{2}=0$, a small error can clearly be seen when $t_{2}$ is increased. This is most noticeable for the $t_{2}=600$\,fs results, where even though standard PLDM is still able to qualitatively reproduce the correct features of the spectrum, the relative peak heights are different in comparison to the exact results. This error arises because standard PLDM is known to often be unable to correctly reproduce the long-time behaviour of dynamical observables, even in the high-temperature limit.\cite{Mannouch2020a}   

\bibliography{jonathan,references}